\documentclass[twocolumn]{aastex631}

\usepackage{amsmath}
\usepackage{rotating}

\shorttitle{Gravitational lensing of elongated early galaxies}
\shortauthors{Pandya \& Loeb et al.}

\begin{document}

\title{Preliminary Evidence for Lensing-Induced Alignments of High-Redshift Galaxies in JWST-CEERS}

\correspondingauthor{Viraj Pandya}
\email{vgp2108@columbia.edu}

\author[0000-0002-2499-9205]{Viraj Pandya}
\altaffiliation{Hubble Fellow}
\affiliation{Columbia Astrophysics Laboratory, Columbia University, 550 West 120th Street, New York, NY 10027, USA}

\author[0000-0003-4330-287X]{Abraham Loeb}
\affiliation{Center for Astrophysics $\vert$ Harvard $\&$ Smithsonian, Cambridge, MA 02138, USA}
\affiliation{Black Hole Initiative, Harvard University, Cambridge, MA 02138, USA}

\author[0000-0001-8688-2443]{Elizabeth J.\ McGrath}
\affiliation{Department of Physics and Astronomy, Colby College, Waterville, ME 04901, USA}

\author[0000-0001-6813-875X]{Guillermo Barro}
\affiliation{Department of Physics, University of the Pacific, Stockton, CA 90340 USA}

\author[0000-0001-8519-1130]{Steven L. Finkelstein}
\affiliation{Department of Astronomy, The University of Texas at Austin, Austin, TX, USA}

\author[0000-0002-6610-2048]{Henry C. Ferguson}
\affiliation{Space Telescope Science Institute, 3700 San Martin Dr., Baltimore, MD 21218, USA}

\author[0000-0002-6610-2048]{Norman A. Grogin}
\affiliation{Space Telescope Science Institute, 3700 San Martin Dr., Baltimore, MD 21218, USA}

\author[0000-0001-9187-3605]{Jeyhan S. Kartaltepe}
\affiliation{Laboratory for Multiwavelength Astrophysics, School of Physics and Astronomy, Rochester Institute of Technology, 84 Lomb Memorial Drive, Rochester, NY 14623, USA}

\author[0000-0002-6610-2048]{Anton M. Koekemoer}
\affiliation{Space Telescope Science Institute, 3700 San Martin Dr., Baltimore, MD 21218, USA}

\author[0000-0003-3382-5941]{Casey Papovich}
\affiliation{Department of Physics and Astronomy, Texas A\&M University, College Station, TX 77843-4242, USA}
\affiliation{George P. and Cynthia Woods Mitchell Institute for Fundamental Physics and Astronomy, Texas A\&M University, College Station, TX 77843-4242, USA}

\author[0000-0003-3382-5941]{Nor Pirzkal}
\affiliation{Space Telescope Science Institute, 3700 San Martin Dr., Baltimore, MD 21218, USA}

\author[0000-0003-3466-035X]{{L. Y. Aaron} {Yung}}
\affiliation{Space Telescope Science Institute, 3700 San Martin Dr., Baltimore, MD 21218, USA}

\begin{abstract}
The majority of low-mass ($\log_{10} M_*/M_{\odot}=9-10$) galaxies at high redshift ($z>1$) appear elongated in projection. We use JWST-CEERS observations to explore the role of gravitational lensing in this puzzle. The typical galaxy-galaxy lensing shear $\gamma\sim1\%$ is too low to explain the predominance of elongated early galaxies with ellipticity $e\approx0.6$. However, non-parametric quantile regression with Bayesian Additive Regression Trees reveals hints of an excess of tangentially-aligned source-lens pairs with $\gamma>10\%$. On larger scales, we also find evidence for weak lensing shear. We rule out the null hypothesis of randomly oriented galaxies at $\gtrsim99\%$ significance in multiple NIRCam chips, modules and pointings. The number of such regions is small and attributable to chance, but coherent alignment patterns suggest otherwise. On the chip scale, the average complex ellipticity $\langle e\rangle\sim10\%$ is non-negligible and beyond the level of our PSF uncertainties. The shear variance $\langle\overline{\gamma}^2\rangle\sim10^{-3}$ is an order of magnitude above the conventional weak lensing regime but is more sensitive to PSF systematics, intrinsic alignments, cosmic variance and other biases. Taking it as an upper limit, the maximum implied ``cosmic shear'' is only a few percent and cannot explain the elongated shapes of early galaxies. The alignments themselves may arise from lensing by a protocluster or filament at $z\sim0.75$ where we find an overabundance of massive lens galaxies. We recommend a weak lensing search for overdensities in ``blank'' deep fields with JWST and the Roman Space Telescope.
\end{abstract}


\section{Introduction}

Almost thirty years ago, Hubble Space Telescope (HST) observations revealed that faint early galaxies preferentially appear elongated in projection \citep{cowie95,vandenbergh96}. This has since been confirmed with newer instruments and statistical samples from larger surveys with HST \citep{elmegreen05,ravindranath06,vanderwel14,zhang19}. Most recently, \citet{pandya24} revived interest in this puzzle by showing that low-mass galaxies with $\log_{10} M_*/M_{\odot}\sim9-10$ at $z>1$ continue to appear preferentially elongated with axis ratios $b/a\sim0.3-0.6$ even in deeper observations with the James Webb Space Telescope (JWST) at rest-frame optical wavelengths \citep[see also][]{kartaltepe23,robertson23,vegaferrero23}. 

These early elongated galaxies have continued to defy a clear explanation. The safest bet is that there is a surface brightness selection effect against detecting rounder face-on disks \citep{dalcanton96,elmegreen05,loeb24} or that we are preferentially seeing an elongated star-forming ``proto-bar'' but not the extended stellar disk. \citet{zhang19} showed that at most $\sim20\%$ of elongated galaxies would be missed by HST if their light profiles were re-projected into the face-on view, which is not enough to explain the high inferred elongated fractions of $\sim50-70\%$. \citet{pandya24} demonstrated that JWST should be even more complete to disks with all orientations at $z\sim1-8$ down to $\log_{10} (M_*/M_{\odot})\sim9$ over a reasonable range of sizes, but stressed the need to scrutinize even deeper surveys. Alternatively, we may be seeing tidal debris from ongoing interactions since merger rates are expected to increase towards high redshift \citep{conselice14,somervilledave15}. Finally, these objects may truly have prolate or triaxial 3D shapes that reflect their formation via mergers along cosmic web filaments \citep{ceverino15,tomassetti16}. 

Here we explore a possibility not previously considered in the literature: are early galaxies preferentially elongated because they have a higher probability of being gravitationally lensed? It is already well known that the optical depth to lensing increases for higher redshift systems \citep[e.g.,][]{barkana00} and that this results in a ``magnification bias'' which modifies the number counts of high-redshift galaxies \citep{turner84,wyithe11,mason15}. The larger covering fraction of possible foreground lenses combined with bigger angular diameter distances to high-redshift sources can also help maximize the Einstein radius for strong lensing \citep{treu10}. JWST has already identified multiple strongly lensed background galaxies in massive clusters missed by Hubble \citep[e.g.,][]{pascale22,mowla22,frye23,mowla24,bradley24}, with roughly $\sim25-65$ cases expected in ``blank'' JWST deep fields \citep{holloway23,casey23}. One new Einstein ring has already been detected in COSMOS-Web based on serendipitous visual inspection \citep{mercier23,vandokkum24}.

Perhaps the most relevant connection is to weak lensing which is sensitive to the assumed intrinsic shapes of galaxies. JWST is uniquely enabling the use of lower mass, higher redshift sources for weak lensing studies around massive clusters \citep{finner23,harvey24} but the intrinsic shapes and intrinsic alignments of these preferentially elongated background galaxies remain poorly understood. \citet{pandya19} proposed that if most early galaxies are indeed nearly prolate and forming along cosmic web filaments, then they are expected to show very strong intrinsic alignments which may be an under-appreciated source of bias for weak lensing. On the other hand, lensing by foreground large-scale structure may itself contribute a non-negligible amount of ``cosmic shear'' that leads to some net elongation and alignments of distant sources. Of course, the fact that it is the lower mass systems at high-redshift that preferentially appear elongated suggests that this phenomenon is due to their intrinsic shapes rather than lensing, but it is still important to quantify the magnitude of the effect and possible implications.

With these fundamental questions in mind, here we pursue a systematic search for both galaxy-galaxy lensing candidates and large-scale alignments in ``blank'' JWST deep fields. This is complementary to traditional lensing studies around bright foreground clusters and involves looking for the statistical correlations between ellipticity, orientation and shear that are the hallmarks of gravitational lensing \citep[e.g.,][]{tyson90,brainerd96}. For this pilot study, we will use the JWST Cosmic Evolution Early Release Science survey \citep[CEERS; Program ID 1345;][]{finkelstein23}, which was selected to not have an obvious bright cluster in the foreground but may still have overdensities at higher redshift. By averaging over the orientations of background galaxies on multiple scales, we will demonstrate the potential of JWST for constraining the presence of any such foreground mass concentration even if it is ``dark'' \citep[e.g.,][]{kaiser00,bacon00,wittman00,maoli01,rhodes01,refregier03,kilbinger14}. With that said, it is not our goal to measure the precise amplitude of any lensing signal but rather to place an upper limit on its contribution to the elongation of early galaxies.

This paper is organized as follows. In section \ref{sec:data} we describe the data and in section \ref{sec:methods} we detail our methods. We present our results on galaxy-galaxy lensing in section \ref{sec:results1} and on large-scale alignments in section \ref{sec:results2}. We discuss possible explanations and implications in \ref{sec:discussion}. Finally, we summarize in section \ref{sec:summary}. We assume a standard \citet{planck15} cosmology throughout with $h=0.6774$, $\Omega_{\rm m,0}=0.3075$, $\Omega_{\rm \Lambda,0}=0.691$ and $\Omega_{\rm b,0}=0.0486$.

\section{Data}\label{sec:data}

We use data from the JWST-CEERS survey \citep[Program ID 1345;][]{finkelstein23}.\footnote{These data can be found on MAST: \dataset[10.17909/z7p0-8481]{http://dx.doi.org/10.17909/z7p0-8481}} CEERS covers a $\sim100$ arcmin$^2$ portion of the Extended Groth Strip \citep[EGS;][]{groth94,davis07} with NIRCam imaging in 10 pointings. Data reduction details are given in \citet{bagley23}. Here we use a source catalog derived by \citet{finkelstein23} with the original SourceExtractor code \citep{bertin96}. We use photometric redshifts and stellar masses derived from 13-filter spectral energy distribution (SED) fitting by \citet[][]{barro23} using the EAZY code \citep{brammer08}. The 13 filters include six broadband ones from NIRCam (F115W, F150W, F200W, F277W, F356W, F444W), one medium-band NIRCam filter (F410M), six broadband filters from HST-ACS/WFC3 (F606W, F814W, F105W, F125W, F140W, F160W). The multi-wavelength SEDs give reasonably well constrained photometric redshifts and stellar masses for our selected sample as described below in subsection \ref{sec:errorprop}. 

Our galaxy shape measurements are based on single-component S\'ersic fits with \texttt{galfit} \citep{peng02} by McGrath et al. (in prep.). This provides the key quantities that we need for our lensing analysis: effective (half-light) radius, projected axis ratio, and position angle of the major axis measured for each galaxy independently in all six broadband NIRCam filters. The shape catalog also includes empirical errors for these quantities that are derived by matching each observed galaxy to 100 simulated sources which were inserted into blank regions of the mosaic and which have similar magnitude, size and S\'ersic index. These empirical errors help quantify the systematic uncertainties that generally dominate galaxy shape measurements and are not accounted for by the formal statistical uncertainties from \texttt{galfit}. 
During the S\'ersic model fitting process, an empirical, filter-dependent global PSF was used to recover ``intrinsic'' galaxy sizes, axis ratios and position angles before convolution with a global empirical PSF. These global PSFs were created by stacking stars throughout the CEERS footprint \citep[see section 3.2][]{finkelstein23}. In Appendix \ref{sec:psf}, we investigate the spatial dependence of the PSF in CEERS by measuring the quadrupole moments of individual stars and quantify the level of bias expected.

\section{Methods}\label{sec:methods}
Here we describe our methods to study galaxy-galaxy lensing and large-scale alignments (in that order).

\subsection{Lens and Source Selection}\label{sec:sel}
We select possible foreground massive lens galaxies by requiring $z<1$, $\log_{10}(M_*/M_{\odot})>10.5$ and good S\'ersic fits (\texttt{galfit} flag of zero or one) in the F115W filter which traces the rest-frame optical/near-infrared. Of the 53885 galaxies available in the catalog, this leads to 77 massive lens galaxies. Based on visual inspection, we discard one that is clearly a deblending artifact resulting in 76 lenses.

For background source galaxies, we impose $z>1$ and $\log_{10}(M_*/M_{\odot})>9$ which is above the CEERS galaxy completeness limit to at least $z=8$ for a reasonable range of sizes and apparent magnitudes \citep[Appendix B of][]{pandya24}. We do not make any cut on color or star formation rate. Of the 53385 sources in the catalog, this yields 6648 galaxies. We further require that the galaxies have good S\'ersic fits (\texttt{galfit} flag of zero or one) in the filter that most closely tracks the rest-frame optical ($\sim5000$\AA) at their redshift. This means F115W for $z=1.0-1.5$, F150W for $z=1.5-2.0$, F200W for $z=2-3$, F356W for $z=3-6$ and F444W for $z>6$. We only include galaxies whose intrinsic (i.e., PSF-deconvolved) S\'ersic effective radius is greater than the PSF FWHM of their assigned rest-optical filter. These cuts yield 4135 source galaxies but we discard 267 that are clearly artifacts based on visual inspection, and another 20 that have catastrophically high uncertainty estimates of $\Delta(b/a)>1$, $\Delta(\rm{PA})>30^{\circ}$ or $\Delta(r_{\rm e})>1''$. This leaves 3848 sources of which 3073 ($\sim80\%$) are low-mass with $\log_{10}(M_*/M_{\odot})=9-10$. 

\subsection{Lens-Source Pair Selection}
For every background galaxy, we start by computing its on-sky separation to all 76 possible foreground lenses. We define a subset of these as ``nearest'' source-lens pairs by assigning the closest on-sky lens to each background galaxy. We verified that our results would be identical if we had instead assigned to each source the lens with the highest predicted shear.

\begin{figure*}
\centering
\includegraphics[width=\hsize]{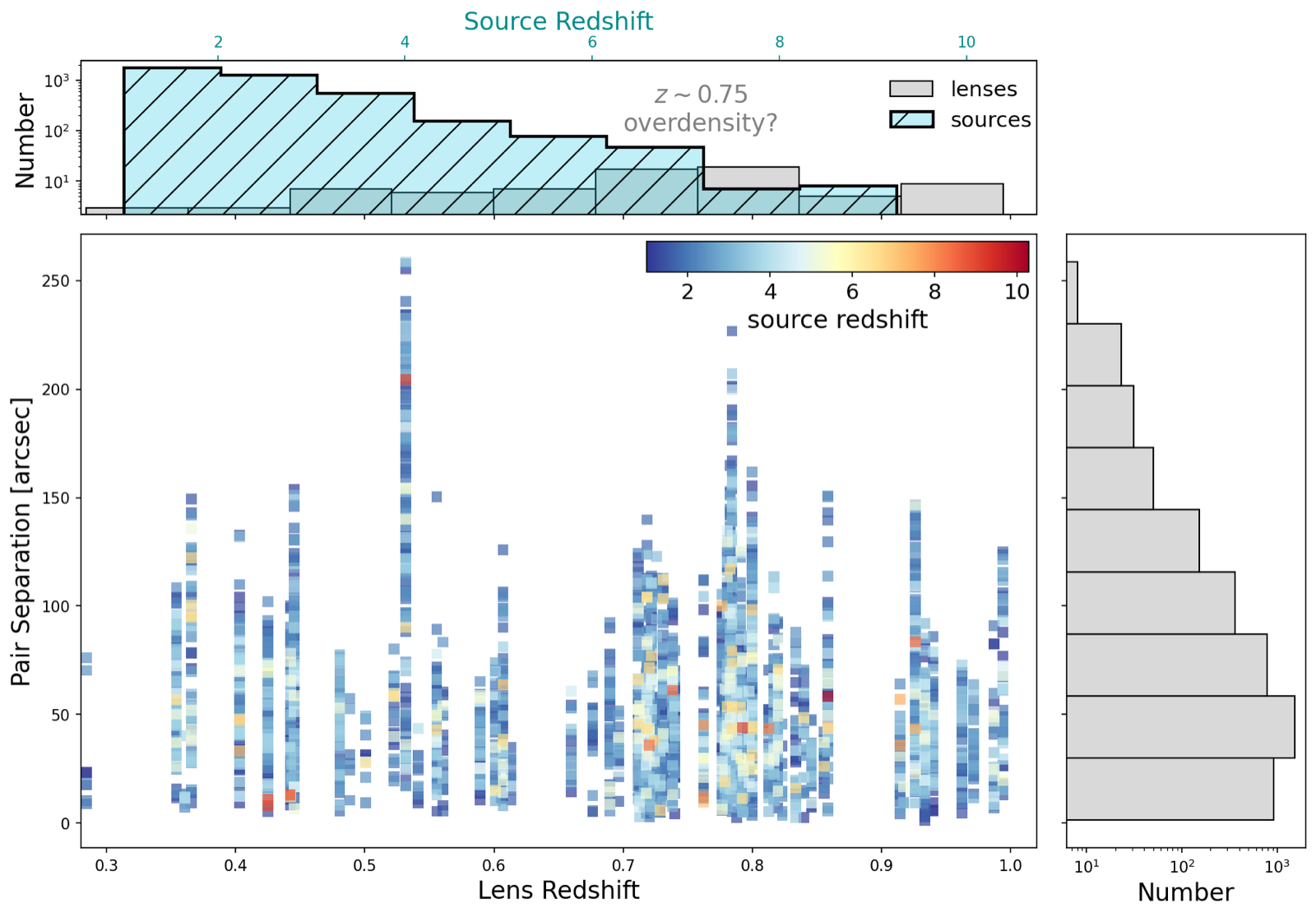}
\caption{Joint distribution of lens redshift, source redshift and pair separation for all background galaxies paired with their nearest on-sky massive foreground lens. The lens redshift distribution is roughly uniform between $z\sim0.3-1.0$ with an overdensity at $z\sim0.75$ (top gray histogram). The source redshift distribution drops off rapidly at $z>1$ (top cyan histogram). At fixed lens redshift, background galaxies span a range of source redshifts and pair separations with $\theta\lesssim250$ arcsec. On average, background galaxies are $\sim1$ arcmin from their nearest on-sky massive foreground lens (right gray histogram).}
\label{fig:pairs}
\end{figure*}

Figure \ref{fig:pairs} shows the demographics of our source-lens pairs in terms of the joint distribution of lens redshift, source redshift and on-sky pair separation. The lenses are roughly uniformly distributed over $z\sim0.3-1.0$ but there appears to be an overdensity at $z\sim0.75$ (top gray histogram). The background sources span a range of redshifts but drop off steeply from $z\sim1-10$ (top cyan histogram). Interestingly, the highest redshift background galaxies (orange/red points) are preferentially found near lenses with $z\sim0.75$. It is unclear if this is due to magnification bias, a selection effect, cosmic variance or some combination thereof. At a given lens redshift, background sources span a range of on-sky pair separations $\theta\lesssim250$ arcsec. For the average background galaxy, the nearest massive foreground lens is $\sim1$ arcmin away. 

\subsection{Lens Model} 
We assume a singular isothermal sphere (SIS) lens model for all 76 massive foreground galaxies. For this initial exploratory study, we believe this strikes a balance between the simplicity of point lens models and the complex shear analysis that is standard in weak lensing \citep{schneider92}.

Since we are interested in calculating the galaxy-galaxy lensing shear, we begin by computing the angular diameter distances to the lens ($D_{\rm L}$), to the source ($D_{\rm S}$), and between the lens and source ($D_{\rm LS}$). Then we can compute the Einstein radius as \citep[e.g., see][]{narayan96,treu10,schneider15}: 
\begin{equation}\label{eqn:einstein}
\theta_{\rm E} = 4\pi\frac{D_{\rm LS}}{D_{\rm S}}\left(\frac{\sigma_{\rm SIS}}{c}\right)^2
\end{equation}
where $c$ is the speed of light and $\sigma_{\rm SIS}$ is the (radially-constant) 1D velocity dispersion of the SIS halo. For massive early-type lenses, the central stellar velocity dispersion $\sigma_*$ has been shown to correlate well with $\sigma_{\rm SIS}$ \citep[i.e., being roughly equal;][]{bolton08a,zahid16,zahid18}. We thus approximate the central stellar velocity dispersion of all lenses assuming their centers are baryon-dominated via
\begin{equation}\label{eqn:sigmastar}
\sigma_* = \sqrt{\frac{GM_*}{R_{\rm eff}^{\rm F115W}}}\;.
\end{equation}
This yields a roughly log-normal distribution of $\sigma_*$ with shape parameter $s\sim0.3$ and scale parameter $275$ km/s, which peaks at $\sim250$ km/s and has a tail out to $\sim580$ km/s. The distribution is reasonable and compares well with, e.g., the range of $\sigma_*$ measured for luminous red galaxies (LRGs) at $z<0.7$ in the SDSS-III/BOSS survey \citep{thomas13}. In detail, since our lens sample extends down to $\log_{10} M_*/M_{\odot}>10.5$, it likely includes many ``fast rotator'' early-type galaxies \citep[e.g.,][]{cappellari16}. Equation \ref{eqn:sigmastar} may need a correction to account for the greater fraction of rotation in these stellar systems. We forgo any such correction here, leaving it as a systematic that is only relevant for our galaxy-galaxy lensing analysis but not cosmic shear calculations (the latter is described next in subsection \ref{sec:complex}).

Taking $\sigma_{\rm SIS}=\sigma_*$, we can thus compute $\theta_{\rm E}$. The shear experienced by a background galaxy a projected distance $\theta$ away from the center of a foreground lens is then simply 
\begin{equation}\label{eqn:gamma}
\gamma = \frac{1}{2}\frac{\theta_{\rm E}}{\theta}\;.
\end{equation}
This shear can be thought of as a differential change in projected ellipticity. It is larger for smaller on-sky source-lens separations and for more massive lenses which have larger Einstein radii. For a given on-sky separation $\theta$ and SIS lens, the shear is maximized as $D_{\rm LS}\to D_{\rm S}$.

\begin{figure*}
\centering
\includegraphics[width=0.75\hsize]{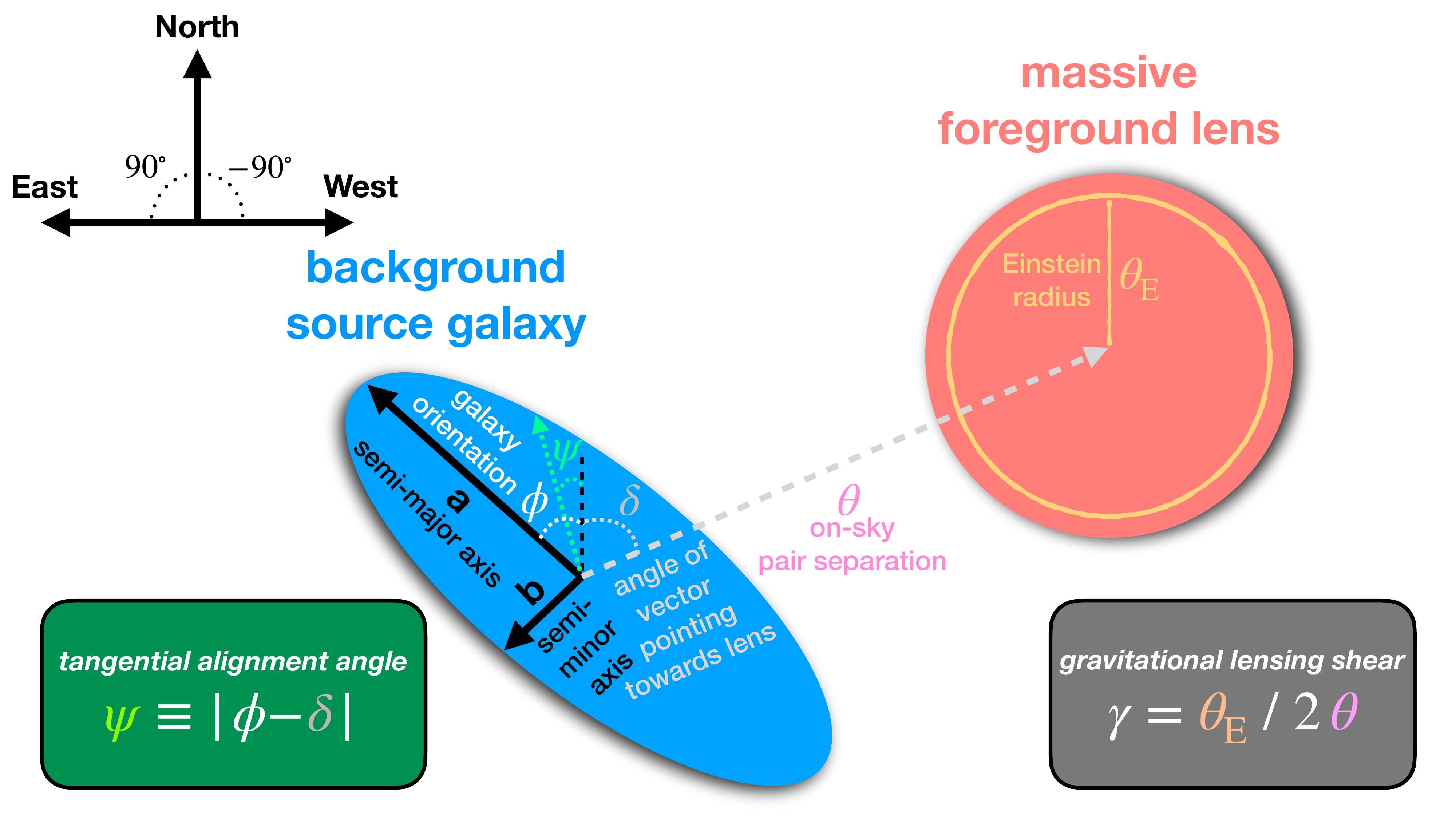}
\caption{Illustration of the various vectors and angles computed in this work. We are primarily interested in the position angle $\phi$ of the major axis of the background galaxy and its tangential alignment angle $\psi$ with respect to the position of the lens. The on-sky angular pair separation is $\theta$. All angles are defined between $-90^{\circ}$ to $90^{\circ}$ east of north.}
\label{fig:illustration}
\end{figure*}

In addition to shear, we also compute the tangential alignment angle for every source-lens pair as illustrated in Figure \ref{fig:illustration}. Every background galaxy already has a position angle $\phi$ for its major axis from the best-fitting S\'ersic model. We compute another position angle $\delta$ from the vector that connects the centers of the background and lens galaxies. Both position angles are defined with the same astronomical convention where $0^{\circ}$ is north (up) and we limit the range between $-90^{\circ}$ and $90^{\circ}$ east of north. The absolute difference between the two is $\psi\equiv|\phi-\delta|$ and lies between $0-90^{\circ}$. When $\psi\approx0^{\circ}$, it means the major axis of the background galaxy is pointing towards its nearest lens. When $\psi\approx90^{\circ}$, it instead means the major axis of the background galaxy is perpendicular to the vector connecting the source-lens pair, i.e., we say the background galaxy is tangentially aligned with the lens.


\subsection{Bayesian Quantile Regression with BART}
Since we are dealing with small sample sizes, we opt for a Bayesian analysis of galaxy-galaxy lensing. Specifically, we will perform Bayesian ``quantile regression'' which predicts the conditional distribution of different quantiles of some property and their dependence on other variables. This provides more information than linear regression which instead fits for a single mean relation and is susceptible to outliers and non-linearities. Unlike simple linear regression, quantile regression can also capture heteroscedasticity in the data such as the scatter in $y$ changing with $x$. While quantile regression can be done in a parametric way (e.g., fitting different lines for different quantiles of $y$), here we opt for a non-parametric approach using Bayesian Additive Regression Trees \citep[BART;][]{chipman10,hill20,martin21}.
BART approximates a function by summing over many small regression trees whose individual sizes and depths are subject to priors that avoid overfitting.\footnote{In the limit of infinite regression trees, BART approaches a Gaussian process but without the need to choose and tune a covariance function (i.e., kernel). BART can also be thought of as a Bayesian cousin of machine learning techniques like random forests and gradient-boosted decision trees.} In addition to the sum-of-trees, there is an error term that is fit as part of the overall model. 

We use the implementation of BART in the probabilistic programming language PyMC \citep{quiroga23,pymc2023}, which utilizes Hamiltonian Monte Carlo for accelerated gradient-based sampling of the error term, and particle Gibbs sampling for BART. BART has a number of hyper-parameters that characterize its internal priors for the depths, splitting rules and values of its trees to minimize overfitting. The default values have been shown to work well for a variety of problems so we do not vary those. We use 100 trees to fit our thousands of points in the  $\log \gamma-e$ and $\log\gamma-\psi$ planes, but using 50 or 200 trees did not change our results. Note that $\gamma$ has to be inferred while $e$ and $\psi$ are direct observables so we treat the latter as ``independent'' variables. 

We assume an asymmetric Laplace likelihood which is naturally well-suited for quantile regression \citep{yu01,koenker01,kozumi11,benoit17}. It has three parameters: one controlling the mean, another controlling the scale, and yet another controlling the asymmetry. The location parameter of this likelihood is the BART random variable with its default internal priors as described above. For the scale parameter (which quantifies uncertainty around the BART-based mean), we use a half-normal prior with standard deviation arbitrarily set to five. We verified that our results are not sensitive to wider and narrower choices of half-normal or exponential priors. We sum this scale parameter in quadrature with the shear uncertainty estimated for each source as described in subsection \ref{sec:errorprop} below. Finally, the asymmetry parameter is set to the quantile we are trying to fit. In this work, we will fit five quantiles: 0.1, 0.5, 0.9, 0.95 and 0.997. 

We ran four chains with 3000 tuning (burn-in) and 3000 sampling draws.
We verified the convergence of the chains as follows. Trace plots revealed that the posteriors of the scale parameter for the asymmetric Laplace likelihood were similar from all four independent chains. Since the BART random variable is really a collection of predictions from multiple trees for each of our thousands of observed points, the recommended convergence check involves the cumulative distribution function of the effective sample size and Gelman-Rubin statistic for each chain. We verified that the effective sample size was $\gg1000$ and the Gelman-Rubin statistic was $\lesssim1.02$ for all BART components of each chain, thus implying convergence.

\subsection{Complex Ellipticity Analysis}\label{sec:complex}
We will average over the orientations of background galaxies both in bins of galaxy-galaxy shear and more generally on the scale of individual NIRCam chips ($64''\times64''$), modules ($2.2'\times2.2'$) and pointings ($2.2'\times5.1'$). For this, we will follow standard practice in gravitational lensing studies \citep[e.g.,][]{schneider92} and assign each galaxy a complex number
\begin{equation}\label{eqn:complex}
e = |e| \exp(2i\phi) = |e|\cos(2\phi) + i|e|\sin(2\phi)
\end{equation}
where $|e|\equiv (1-b/a)$ is the usual projected ellipticity and $\phi$ is the position angle measured east of north (up).\footnote{In weak lensing, it is common practice to instead define $|e|=(a-b)/(a+b)$ or $|e|=(a^2-b^2)/(a^2+b^2)$. Our results are not sensitive to the choice of a particular definition for the norm $|e|$.} The use of $2\phi$ accounts for the symmetry of rotating an ellipse by $\pi$, i.e., galaxy shape vectors do not point in one or the other direction for a given orientation. This formalism makes it easy to compute the average ellipticity as the magnitude of the average complex number $\langle e \rangle$ and the average orientation as its phase.\footnote{It is common practice to divide individual ellipticities by a ``shear polarizability tensor'' $P^{\gamma}$ or ``shear responsivity factor'' $\mathcal{R}$ \citep{massey07}. These are of order unity and may depend on galaxy morphology so we neglect them here for simplicity.}

We will also compute the ``shear variance'' $\langle \overline{\gamma}^2\rangle$ which quantifies the degree of correlation between the complex ellipticities of all possible pairs of galaxies in some region. This summary statistic is frequently used in cosmic shear studies \citep{refregier03,kilbinger14} and can be thought of as a simpler alternative to more sophisticated shear-shear correlation functions, which otherwise depend continuously on pair separation. Specifically, we adapt equation (8) of \citet{hammerle02}: 
\begin{equation}\label{eqn:shearvar}
\overline{\gamma}^2_{\rm reg} = \frac{1}{N_{\rm gal}(N_{\rm gal}-1)}\sum_{i=1}^{N_{\rm gal}}\sum_{j=i+1}^{N_{\rm gal}} e_{i} e_{j}^*
\end{equation}
where the normalization accounts for the number of unique pairs. Note that the inner summation is staggered by one to prevent double counting pairs and that the product involves the complex conjugate $e_j^*$. This $\overline{\gamma}^2_{\rm reg}$ is the shear variance for a single chip, module or pointing, but we want the average over all regions of a given type: 
\begin{equation}\label{eqn:avgshearvar}
\langle \overline{\gamma}^2\rangle = \frac{1}{N_{\rm reg}}\sum_{n=1}^{N_{\rm reg}} \vert \overline{\gamma}^2_{\rm reg,n}\vert
\end{equation}
where the sum involves the norm of $\overline{\gamma}^2_{\rm reg,n}$, which in general is a complex number. We will compute this shear variance for both galaxies and our PSF stars (Appendix \ref{sec:psf}) with the latter serving as null tests for systematics.

\subsection{Monte Carlo Error Propagation \& Systematics}\label{sec:errorprop}
We use an efficient vectorized Monte Carlo method to estimate uncertainties on the galaxy-galaxy lensing shear. First, we assume negligible astrometric uncertainties which allows us to fix the combination of source-lens pairs for simplicity. This is sensible since we have so few possible lens galaxies and the nearest one in projection to each source would maximize its shear. Then, for each source-lens pair, we draw 1000 random realizations of source redshift, lens redshift, lens stellar mass and lens effective radius using the uncertainty estimates for those quantities. The median fractional uncertainties are $\sigma_z/(1+z_{\rm phot})\approx0.014$ for source redshifts, $\sigma_z/(1+z_{\rm phot})\approx0.012$ for lens redshifts, $\sigma_{\rm \log M_*}/\log M_*\approx0.002$ for lens stellar masses and $\sigma_{\rm R_{\rm eff}^{\rm{F115W}}}/R_{\rm eff}^{\rm{F115W}}\approx0.074$ for lens effective radius.\footnote{The redshift and mass uncertainties reflect the difference between the 84th and 16th percentile estimates from \citet{barro23}. The size uncertainty from McGrath et al. (in prep.) accounts for the formal \texttt{Galfit} error as well as systematics using the approach of \citet{vanderwel12}.} Finally, we compute the distribution of galaxy-galaxy lensing shear over all realizations for a given source-lens pair. The standard deviation around the mean of that shear distribution gives the uncertainty on the shear for that source-lens pair.

We use a similarly efficient Monte Carlo approach to propagate errors on $\langle e\rangle$, $\langle\overline{\gamma}^2\rangle$ and the star-galaxy cross-correlation. When computing these summary statistics for any sample of $N$ galaxies or stars, we first create 1000 random realizations of size $N$ each. For every observed galaxy, we randomly draw an ellipticity, position angle and any other quantity of interest from a Gaussian with mean equal to the fiducial catalog value and standard deviation equal to the empirical error from McGrath et al. (in prep.). For stars, we do the same thing but in lieu of errors on quadrupole moments, we use the standard deviation of $e_1$ and $e_2$ from the star sample itself as a measure of uncertainty on those quantities. For simplicity, we do not propagate errors on galaxy redshift or mass since our analysis is not done in fine $z$ or $M_*$ intervals. Then we can compute our summary statistics in all realizations and take the mean and standard deviation with the latter quantifying the uncertainty due to galaxy/star shape error propagation. It is almost always the case that the summary statistics are well constrained, e.g., $\mu_{\langle e\rangle}/\sigma_{\langle e\rangle}\gg1$. However, this cannot, by itself, be used to assess the significance of an alignment signal. For that, we need a null hypothesis test to rule out randomly oriented galaxies, which we describe in subsection \ref{sec:null}.

Our Monte Carlo error propagation method gives a formal uncertainty for summary statistics computed on the scale of the entire survey. However, we only have one survey limited to a small ($\sim0.028$ deg$^2$) part of the sky so our shear variance $\langle\overline{\gamma}^2\rangle$ is almost certainly dominated by cosmic variance. It is not our goal in this paper to constrain cosmological parameters or perform detailed comparisons to theory so we do not attempt to estimate the cosmic variance uncertainty on $\langle\overline{\gamma}^2\rangle$. Instead, we will simply place an upper limit on the weak lensing shear experienced by high-redshift galaxies. We defer an estimation of the cosmic variance uncertainty to future weak lensing analyses which can empirically constrain the field-to-field variance using multiple ``blank'' JWST deep fields.

We can adapt the shear variance calculation of equation \ref{eqn:shearvar} to compute the star-galaxy cross-correlation as a check of systematics: 
\begin{equation}\label{eqn:xcorr}
e_{\rm gal}\star e_{\rm star}^* = \frac{1}{N_{\rm gal}N_{\rm star}}\sum_{i=1}^{N_{\rm gal}}\sum_{j=1}^{N_{\rm star}} e_{\rm{gal},i} e_{\rm{star},j}^*\;.
\end{equation}
For this, we measured the complex ellipticity of individual stars throughout CEERS using quadrupole moments as described in Appendix \ref{sec:psf}. This provides an important null test for baselining possible PSF systematics but we caution that we only have $\sim130$ stars throughout the survey footprint and typically only $\sim1-2$ stars per NIRCam chip, with $\sim30\%$ of chips having no star. To ensure that galaxies are only paired up with stars in their corresponding rest-frame optical filter, we will compute this separately for each NIRCam filter. Since we want to know the fractional contribution of the star-galaxy cross-correlation to the total shear variance, we record the ratio of the norms $|e_{\rm gal}\star e_{\rm star}^*|/|\overline{\gamma}^2_{\rm reg}|$ for all individual regions before taking the average as in \ref{eqn:avgshearvar}. This approach of averaging ratios should be more robust especially on smaller scales where ``shot noise'' from the galaxy ellipticity distribution can dominate the variance between regions of the same size. Note that our galaxy shapes were measured by fitting a S\'ersic model convolved with a global empirical PSF and that we do not attempt to perform local PSF corrections here (but see Appendix \ref{sec:psf}). The star-galaxy cross-correlation thus reflects any additional local PSF contamination unaccounted for by the global PSF.

\subsection{Null Hypothesis Test for Alignments}\label{sec:null}
For any sample of galaxies that show elevated $\langle e\rangle$ suggestive of alignments, we need to compare their observed $\langle e\rangle$ to the distribution of $\langle e\rangle$ expected under the null hypothesis that they are randomly oriented. For $N$ observed galaxies, we create 1 million realizations, each of which is assigned $N$ random complex ellipticity vectors. We randomly draw orientations $\phi$ from a uniform distribution $\mathcal{U}(-90^{\circ},90^{\circ})$ which should ``break'' any alignments (if present). Since there is no good, universal distribution from which to randomly draw ellipticities, we fix the distribution of complex magnitudes $e$ to the actual observed ellipticities of the galaxies.\footnote{A uniform distribution is clearly not a good choice for high-redshift, low-mass bins which are biased towards high $e$ \citep{pandya24} or for high-mass spheroid-dominated bins which are biased towards low $e$ \citep[e.g.][]{chang13}.} This lets us compute the distribution of the average $\langle e\rangle$ under the null hypothesis that the $N$ galaxies are randomly oriented. We then calculate the $p$-value for a given chip as the fraction of its random realizations where the null $\langle e \rangle$ exceeds the observed $\langle e\rangle$. This procedure is used both for our galaxy-galaxy lensing candidates and when averaging over the orientations of all background galaxies in a given NIRCam chip, module, pointing and the scale of the entire survey.

\section{Galaxy-Galaxy Lensing}\label{sec:results1}
In this section we present our results on galaxy-galaxy lensing in JWST-CEERS.

\subsection{Shear versus ellipticity}
\begin{figure*}
\centering
\includegraphics[width=\hsize]{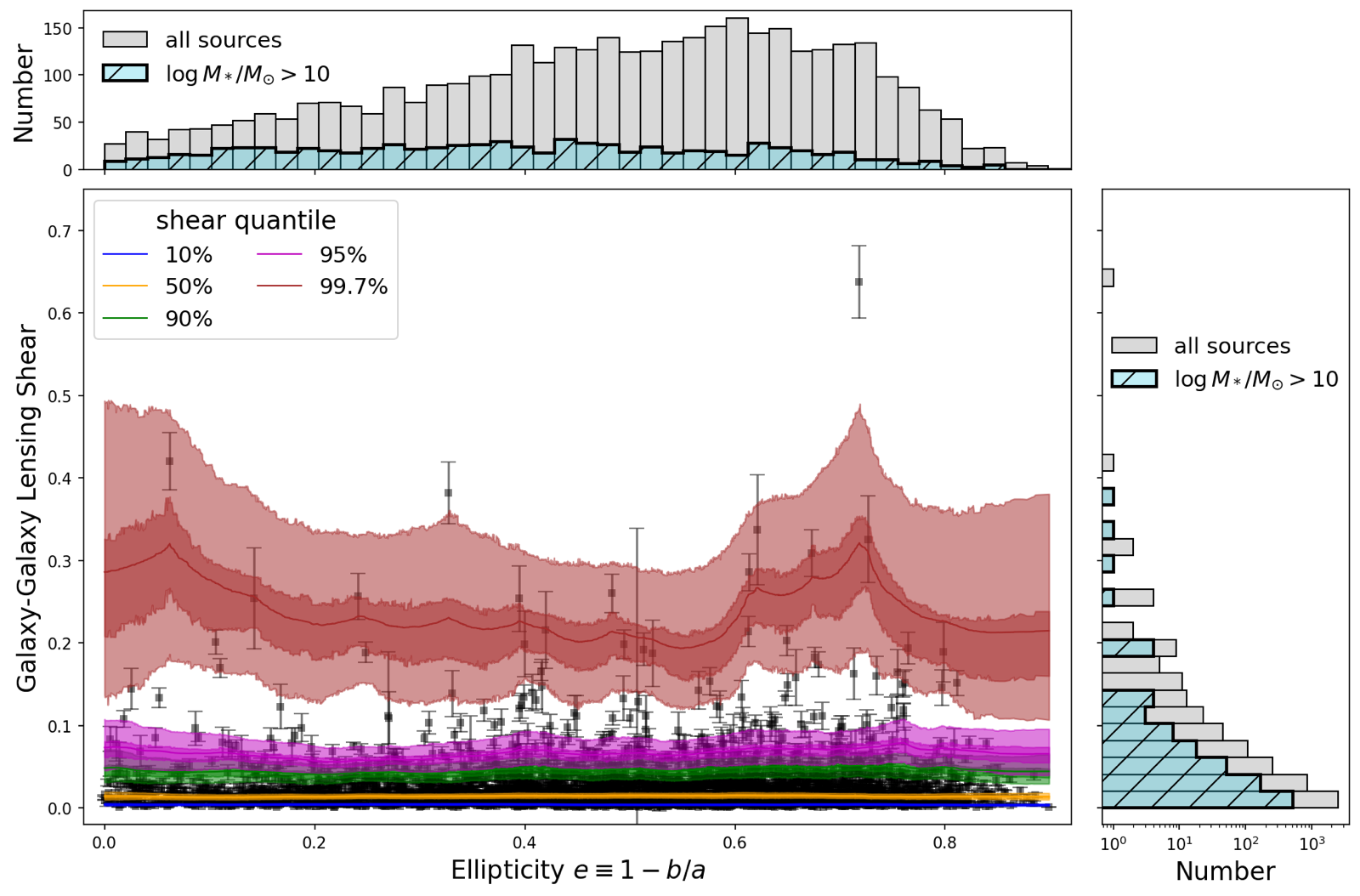}
\caption{Joint distribution of shear and ellipticity for all nearest source-lens pairs (black points). The errorbars reflect our Monte Carlo shear estimates (see subsection \ref{sec:errorprop}). The colored solid lines show the conditional distribution of shear on ellipticity in five quantiles from our non-parametric quantile regression with Bayesian Additive Regression Trees: 10\% (blue), 50\% (orange), 90\% (green), 95\% (magenta) and 99.7\% (red). The light and dark shaded regions reflect the 50\% and 95\% credible intervals, respectively. Most background galaxies are predicted to have very small shear but there is a strong tail towards $\gamma>0.1$. The data does not show a strong correlation between shear and ellipticity regardless of shear quantile. The projected ellipticity distribution of background galaxies is non-uniform and biased towards high values (top gray histogram). If we restrict to high-mass background galaxies, they show a flatter distribution (top cyan histogram) as expected for randomly oriented disks, meaning that it is the low-mass galaxies that are preferentially elongated.}
\label{fig:shear}
\end{figure*}

Figure \ref{fig:shear} shows the joint distribution of galaxy-galaxy lensing shear\footnote{Recall that $\gamma=0.5\theta_{\rm E}/\theta$. The distribution of $\theta_{\rm E}$ itself for all source-lens pairs is roughly a log-normal with mean $\mu\sim0.4''$ and scale $\sigma\sim1.1$. It peaks at $\theta_{\rm E}\approx0.7''$ and extends to $7''$.\label{lognormal}} and source ellipticity. The ellipticity distribution of background galaxies is clearly non-uniform and biased towards elongated objects with $e\sim0.6$ on average. The ellipticity distribution of massive sources alone is flatter, consistent with them being randomly oriented disks, but this implies that the low-mass sources are the preferentially elongated ones. Most pairs have very small shear with the mean being $\langle\gamma\rangle\approx0.02$ consistent with what is expected for weak lensing \citep[e.g.,][]{oguri12}. Because the mean ellipticity of background galaxies is much larger than the typical expected shear, galaxy-galaxy lensing cannot be the primary driver for the preferential elongation of low-mass high-redshift galaxies.

However, there is a non-negligible tail towards larger shears in excess of $0.1$ which is beyond the conventional weak lensing regime. One object that is highly elongated with $e\approx0.7$ has an unusually large shear of $\gamma\approx0.6$. Our Bayesian non-parametric quantile regression with BART reveals that shear does not correlate with ellipticity for the majority of pairs below the 95\% quantile of shear. Even for shears in the $99.7$\% quantile, the data does not show a strong correlation with ellipticity and the Bayesian credible interval is large. At very high $e>0.7$, we see a downturn in the median $\gamma$ of the 99.7\% shear quantile, but the uncertainty is large due to the small number of such extreme objects. We confirmed that the $19$ objects with $e>0.85$ and $\gamma<0.1$ are in fact highly elongated galaxies and not artifacts \citep[coincidentally, two of them are shown as prolate candidates in Figure 17 of][]{pandya24}. The tail of large shear pairs motivates looking for correlations with the source-lens tangential alignment angle, which we turn to next.

\subsection{Shear versus tangential alignment angle}
\begin{figure*}
\centering
\includegraphics[width=\hsize]{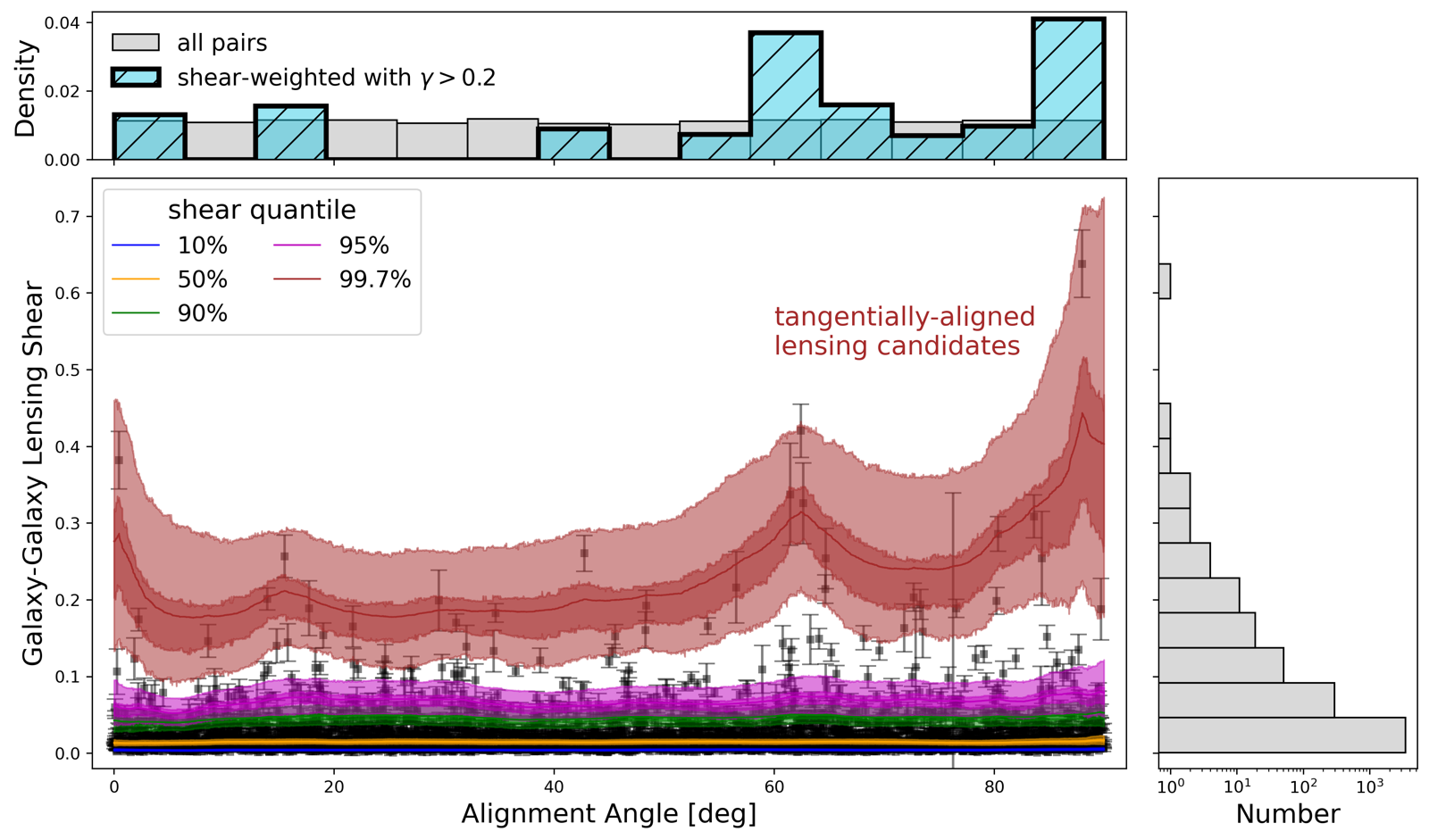}
\caption{Joint distribution of shear and tangential alignment angle for all nearest source-lens pairs (black points). The errorbars reflect our Monte Carlo shear estimates (see subsection \ref{sec:errorprop}). The colored lines show the conditional distribution of shear on tangential alignment angle in five quantiles from our non-parametric quantile regression with Bayesian Additive Regression Trees: 10\% (blue), 50\% (orange), 90\% (green), 95\% (magenta) and 99.7\% (red). The light and dark shaded regions reflect the 50\% and 95\% Bayesian credible intervals, respectively. Most pairs have small shear (right gray histogram) and there is no correlation with $\psi$ even in the $95\%$ quantile of shear (magenta curve). In contrast, there is a hint of an excess of tangentially-aligned pairs at high shear with $\gamma>0.2$ (top cyan histogram). There is also a hint of a positive correlation in the 99.7\% shear quantile (red curve) wherein $\gamma$ tends to be $\sim5\%$ higher for tangentially-aligned pairs but the uncertainties are large. We will refer to pairs with $\gamma>0.1$ and $\psi>75^{\circ}$ as tangentially-aligned lensing candidates.}
\label{fig:alignments}
\end{figure*}

Figure \ref{fig:alignments} plots the joint distribution of shear $\gamma$ and tangential alignment angle $\psi$. Most pairs follow a broad, relatively uniform distribution of $\psi$ (top gray histogram). There is a hint that high-shear pairs with $\gamma>0.2$ tend to be tangentially aligned with larger $\psi$, but this is only significant at the $\sim96.6\%$ level ($p=0.0337$) based on a two-sample Kolmogorov-Smirnov test using the unbinned, unweighted distributions of $\psi$. While this hypothesis test based on the 1D marginalized distribution of $\psi$ is useful, it is not entirely appropriate since we are dealing with multi-dimensional data and really looking for correlations \citep{feigelson12}.

Our Bayesian non-parametric quantile regression confirms that most pairs have low shear (the 95\% quantile is $\gamma\lesssim0.1$), and that those low-shear pairs have no correlation with alignment angle. The conditional distribution of the extreme $99.7$\% quantile of shear shows hints of a positive correlation with alignment angle as expected, but our Bayesian credible intervals (i.e., uncertainties) are large. At the low-$\psi$ end, there is one source-lens pair with high $\gamma\sim0.38$. Visual inspection reveals that this is clearly a high-redshift dropout galaxy that only appears in redder NIRCam filters and is almost perfectly radially-aligned with a massive foreground galaxy $\sim2.8''$ away. For pairs in the $99.7\%$ quantile of shear, when $\psi$ is large, the shear tends to be larger by $\sim5\%$. There is also a hint of heteroscedasticity in the data such that the scatter in shear increases with $\psi$. In other words, at small $\psi$, the 99.7\% quantile of shear may simply be a tail of the log-normal distribution of shear of all pairs, but at large $\psi$, the extremes of the conditional shear distribution may comprise a broader tail. This would imply that among the many tangentially-aligned pairs, there may be some that are lensed, precisely those with high shear. With that said, our sample size is small and this heteroscedasticity appears to be driven by one tangentially-aligned source with high shear $\gamma\sim0.63$. We next inspect this and other sources with $\gamma>0.1$ and $\psi>75^{\circ}$, but note that our simple quantile regression method should be tested in the future on larger surveys with known strong lenses.

\subsection{Images of lensing candidates}
\begin{figure*}
\centering
\includegraphics[width=\hsize]{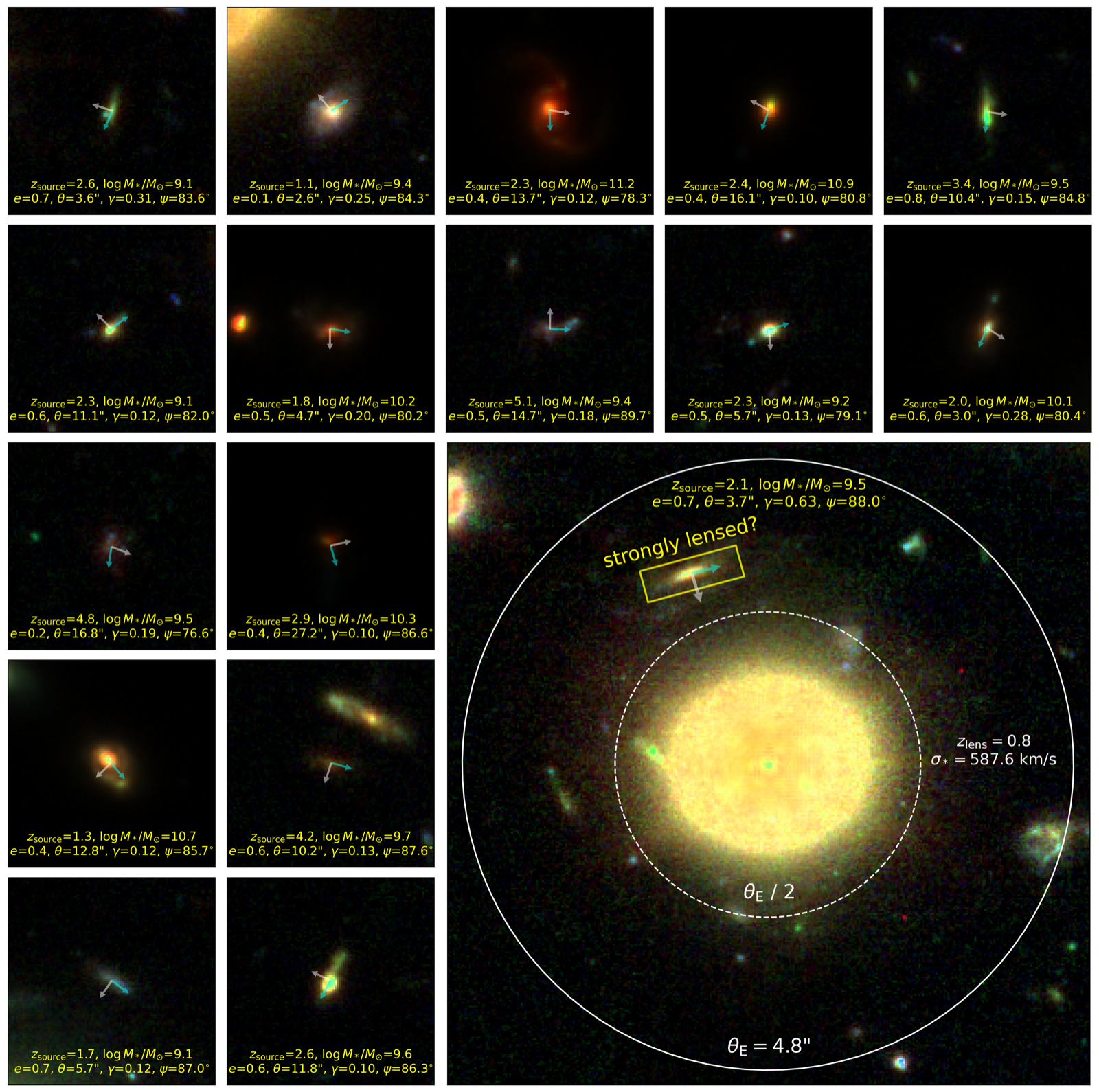}
\caption{Example $3''\times3''$ false-color RGB (F115W+F200W+F356W) postage stamps of our 17 tangentially-aligned lensing candidates with $\gamma>0.1$ and $\psi>75^{\circ}$. The gray arrows point in the direction of their closest foreground massive lens (typically beyond the cutout) and cyan arrows denote the major axis position angle. The large panel shows a $10''\times10''$ cutout around our one strong lensing candidate. The solid white circle shows our fiducial Einstein radius with the source clearly inside it but there is no obvious second image on the opposite (lower-right) side of the lens. If the Einstein radius was halved (dashed white circle), a second image would not be expected and the shear would drop to $\gamma\sim0.3$.}
\label{fig:sources}
\end{figure*}

Figure \ref{fig:sources} shows images of the 17 source-lens pairs that have $\gamma>0.1$ and $\psi>75^{\circ}$. We will refer to these as our tangentially-aligned galaxy-galaxy lensing candidates. Table \ref{tab:pairs} lists their properties to facilitate follow-up observations. The cut on $\gamma>0.1$ places these candidates in a shear regime that is an order of magnitude above the conventional weak lensing limit. Many of them are elongated and some even show arc-like distortions that may be suggestive of intermediate lensing. None of these are truly in the strong lensing regime (i.e., all have $\theta>\theta_{\rm E}$) except possibly one, which we now discuss.

The highest shear candidate is a $\sim10^{9.5}M_{\odot}$ galaxy that is clearly elongated with $e\sim0.72$ and has a photometric redshift $z\sim2.1$. It is nearly perfectly tangentially aligned ($\psi\approx88^{\circ}$) and lies within the Einstein radius of its associated lens ($\theta\sim3.75''$, $\theta_{\rm E}\sim4.75''$), making it our only strong lensing candidate within the JWST-CEERS footprint. If its associated lens, which happens to be our most massive one with the highest $\sigma_*$, satisfies our SIS assumption, then we expect another brighter image on the opposite side outside the Einstein radius with an image separation $\Delta\theta=2\theta_{\rm E}=9.5''$. We do not find any obvious source there with a similar redshift as the main image. The massive foreground lens galaxy itself is spectroscopically-confirmed to be at $z\sim0.78$ and our formal Monte Carlo uncertainty on the shear is only $\gamma=0.638\pm0.044$. If we overestimated its stellar velocity dispersion by a factor of two, then with all else fixed, $\theta_{\rm E}$ would be halved and the source would no longer lie within the Einstein radius so a second image would not be expected (it would still have $\gamma\sim0.3$ in this case). Of course, another possibility is that the photometric redshift of our lensing candidate is very wrong and it is instead physically associated with the lens as a satellite. Note that the rest of our results in this paper do not hinge on whether or not this candidate is confirmed (namely, the large-scale alignments in section \ref{sec:results2} are independent of this galaxy-galaxy lensing issue).

\begin{table*}
\footnotesize 
\centering 
\begin{tabular}{|c|c|c|c|c|c|c|c|h|h|h|h|h|h|h|h|h|h|h|h|}\hline 
Source ID & Source RA & Source Dec & Source $z$ & Source $\log_{10} (M_*/M_{\odot})$ & Source $R_{\rm e}$ & Source $e$ & Source $\phi$ & Rest Filter & $\theta$ & $\theta_{\rm E}$ & $\gamma$ & $\psi$ & Lens ID & Lens RA & Lens Dec & Lens $z$ & Lens $\log_{10} M_*/M_{\odot}$ & Lens $R_{\rm e}$ & Lens $\sigma_*$ \\\hline
23392 & 214.80777 & 52.86992 & $2.619\pm0.044$ & $9.07\pm0.03$ & $1.39\pm0.10$ & $0.673\pm0.020$ & $-69.171\pm2.636$ & F200W & $3.643$ & $2.249\pm0.206$ & $0.309\pm0.028$ & $83.628\pm2.636$ & 23335 & 214.80843 & 52.87069 & $0.823\pm0.023$ & $10.94\pm0.01$ & $2.528\pm0.194$ & $387.05\pm16.47$ \\
24257 & 214.8167 & 52.86969 & $1.118\pm0.061$ & $9.43\pm0.05$ & $1.90\pm0.17$ & $0.142\pm0.028$ & $78.415\pm3.154$ & F115W & $2.605$ & $1.323\pm0.320$ & $0.254\pm0.061$ & $84.312\pm3.154$ & 25049 & 214.81657 & 52.8704 & $0.840\pm0.029$ & $11.27\pm0.03$ & $3.310\pm0.330$ & $492.55\pm28.20$ \\
24459 & 214.81947 & 52.87288 & $2.338\pm0.023$ & $11.23\pm0.01$ & $6.67\pm1.01$ & $0.397\pm0.016$ & $-43.107\pm1.603$ & F200W & $13.734$ & $3.408\pm0.423$ & $0.124\pm0.015$ & $78.321\pm1.603$ & 25049 & 214.81657 & 52.8704 & $0.840\pm0.029$ & $11.27\pm0.03$ & $3.310\pm0.330$ & $492.59\pm28.89$ \\
24638 & 214.83818 & 52.88739 & $2.359\pm0.029$ & $10.89\pm0.07$ & $1.14\pm0.03$ & $0.377\pm0.008$ & $-64.522\pm0.402$ & F200W & $16.117$ & $3.351\pm0.509$ & $0.104\pm0.016$ & $80.828\pm0.402$ & 12631 & 214.84013 & 52.89141 & $0.780\pm0.015$ & $11.12\pm0.01$ & $2.534\pm0.356$ & $473.64\pm35.12$ \\
28283 & 214.85403 & 52.87866 & $3.382\pm0.049$ & $9.45\pm0.01$ & $1.60\pm0.12$ & $0.811\pm0.018$ & $-48.633\pm2.200$ & F356W & $10.399$ & $3.150\pm0.312$ & $0.151\pm0.015$ & $84.810\pm2.200$ & 28072 & 214.85181 & 52.87682 & $0.817\pm0.020$ & $10.75\pm0.03$ & $1.279\pm0.083$ & $435.19\pm20.87$ \\
29037 & 214.85188 & 52.87375 & $2.349\pm0.133$ & $9.06\pm0.06$ & $1.04\pm0.17$ & $0.583\pm0.024$ & $81.108\pm5.014$ & F200W & $11.072$ & $2.728\pm0.286$ & $0.123\pm0.013$ & $81.966\pm5.014$ & 28072 & 214.85181 & 52.87682 & $0.817\pm0.020$ & $10.75\pm0.03$ & $1.279\pm0.083$ & $436.10\pm21.40$ \\
51412 & 214.98434 & 52.89845 & $1.805\pm0.052$ & $10.23\pm0.11$ & $3.30\pm0.22$ & $0.494\pm0.016$ & $34.954\pm1.771$ & F150W & $4.736$ & $1.881\pm0.166$ & $0.199\pm0.018$ & $80.206\pm1.771$ & 51197 & 214.98547 & 52.89777 & $0.719\pm0.009$ & $10.99\pm0.01$ & $3.072\pm0.247$ & $370.27\pm15.64$ \\
54273 & 214.81218 & 52.81713 & $5.119\pm0.052$ & $9.43\pm0.02$ & $1.31\pm0.14$ & $0.522\pm0.030$ & $43.650\pm2.122$ & F356W & $14.683$ & $5.515\pm1.179$ & $0.188\pm0.040$ & $89.716\pm2.122$ & 53095 & 214.80865 & 52.81919 & $0.820\pm0.024$ & $11.24\pm0.01$ & $2.588\pm0.490$ & $545.67\pm55.79$ \\
65925 & 215.1386 & 52.98904 & $2.311\pm0.014$ & $9.22\pm0.02$ & $0.70\pm0.05$ & $0.508\pm0.019$ & $61.166\pm2.291$ & F200W & $5.695$ & $1.470\pm0.129$ & $0.129\pm0.011$ & $79.061\pm2.291$ & 66073 & 215.13988 & 52.98812 & $0.689\pm0.015$ & $10.93\pm0.02$ & $4.039\pm0.289$ & $302.59\pm12.81$ \\
70412 & 215.09525 & 52.93305 & $1.995\pm0.032$ & $10.06\pm0.03$ & $2.21\pm0.20$ & $0.613\pm0.011$ & $-67.699\pm1.035$ & F150W & $2.990$ & $1.708\pm0.135$ & $0.286\pm0.023$ & $80.356\pm1.035$ & 70465 & 215.09496 & 52.93227 & $0.720\pm0.017$ & $10.91\pm0.02$ & $2.983\pm0.190$ & $342.92\pm12.80$ \\
86621 & 214.86714 & 52.79953 & $2.121\pm0.041$ & $9.47\pm0.03$ & $5.90\pm1.26$ & $0.718\pm0.022$ & $60.037\pm5.622$ & F200W & $3.746$ & $4.779\pm0.329$ & $0.638\pm0.044$ & $88.000\pm5.622$ & 86622 & 214.86783 & 52.79875 & $0.800\pm0.020$ & $11.28\pm0.01$ & $2.352\pm0.134$ & $589.05\pm18.28$ \\
86695 & 214.87056 & 52.80254 & $4.764\pm0.067$ & $9.49\pm0.07$ & $1.61\pm0.15$ & $0.248\pm0.018$ & $-53.020\pm2.234$ & F356W & $16.823$ & $6.347\pm0.406$ & $0.189\pm0.012$ & $76.554\pm2.234$ & 86622 & 214.86783 & 52.79875 & $0.800\pm0.020$ & $11.28\pm0.01$ & $2.352\pm0.134$ & $589.09\pm17.98$ \\
87402 & 214.87488 & 52.80144 & $2.850\pm0.094$ & $10.27\pm0.06$ & $2.43\pm0.20$ & $0.371\pm0.020$ & $-28.896\pm1.915$ & F200W & $27.162$ & $5.488\pm0.367$ & $0.101\pm0.007$ & $86.647\pm1.915$ & 86622 & 214.86783 & 52.79875 & $0.800\pm0.020$ & $11.28\pm0.01$ & $2.352\pm0.134$ & $588.60\pm18.22$ \\
94812 & 214.84508 & 52.79176 & $4.247\pm0.146$ & $9.70\pm0.07$ & $1.60\pm0.18$ & $0.606\pm0.028$ & $29.231\pm3.247$ & F356W & $10.169$ & $2.741\pm0.403$ & $0.135\pm0.020$ & $87.646\pm3.247$ & 94939 & 214.84778 & 52.79094 & $0.712\pm0.013$ & $11.07\pm0.02$ & $3.538\pm0.440$ & $380.89\pm27.33$ \\
95439 & 214.86063 & 52.80203 & $1.654\pm0.078$ & $9.07\pm0.04$ & $1.91\pm0.75$ & $0.749\pm0.069$ & $7.401\pm22.201$ & F150W & $5.748$ & $1.396\pm0.147$ & $0.121\pm0.013$ & $87.037\pm22.201$ & 85548 & 214.86221 & 52.80185 & $0.560\pm0.019$ & $10.98\pm0.03$ & $4.727\pm0.314$ & $296.80\pm14.25$ \\
95498 & 214.84607 & 52.78814 & $2.622\pm0.145$ & $9.59\pm0.03$ & $1.11\pm0.11$ & $0.635\pm0.013$ & $-73.424\pm1.987$ & F200W & $11.786$ & $2.402\pm0.343$ & $0.102\pm0.015$ & $86.267\pm1.987$ & 94939 & 214.84778 & 52.79094 & $0.712\pm0.013$ & $11.07\pm0.02$ & $3.538\pm0.440$ & $380.60\pm26.35$ \\
95846 & 214.86427 & 52.79873 & $1.316\pm0.066$ & $10.66\pm0.05$ & $3.49\pm0.19$ & $0.361\pm0.007$ & $-4.828\pm0.416$ & F115W & $12.799$ & $3.026\pm0.344$ & $0.118\pm0.013$ & $85.681\pm0.416$ & 86622 & 214.86783 & 52.79875 & $0.800\pm0.020$ & $11.28\pm0.01$ & $2.352\pm0.134$ & $588.07\pm18.57$ \\\hline
\end{tabular}
\label{tab:pairs}
\caption{Properties of our 17 tangentially-aligned lensing candidates and their main associated lenses. All IDs and positions are from the CEERS photometric catalog with the latter in decimal degrees. Photometric redshifts and stellar masses are from SED fitting, effective S\'ersic radii are in proper kpc, and the lens stellar velocity dispersions are in km/s. Position angles are defined between $-90^{\circ}<\phi<90^{\circ}$ east of north, and alignment angles are between $0^{\circ}<\psi<90^{\circ}$ with $0^{\circ}$ for radial and $90^{\circ}$ for tangential alignment. Pair separations and Einstein radii are in arcsec. Uncertainties on physical properties and on the galaxy-galaxy lensing shear are described in subsection \ref{sec:errorprop}. The full table is available for download from the journal.} 
\end{table*}


\subsection{Source magnifications and demographics}\label{sec:magnif}

\begin{figure}
\centering
\includegraphics[width=\hsize]{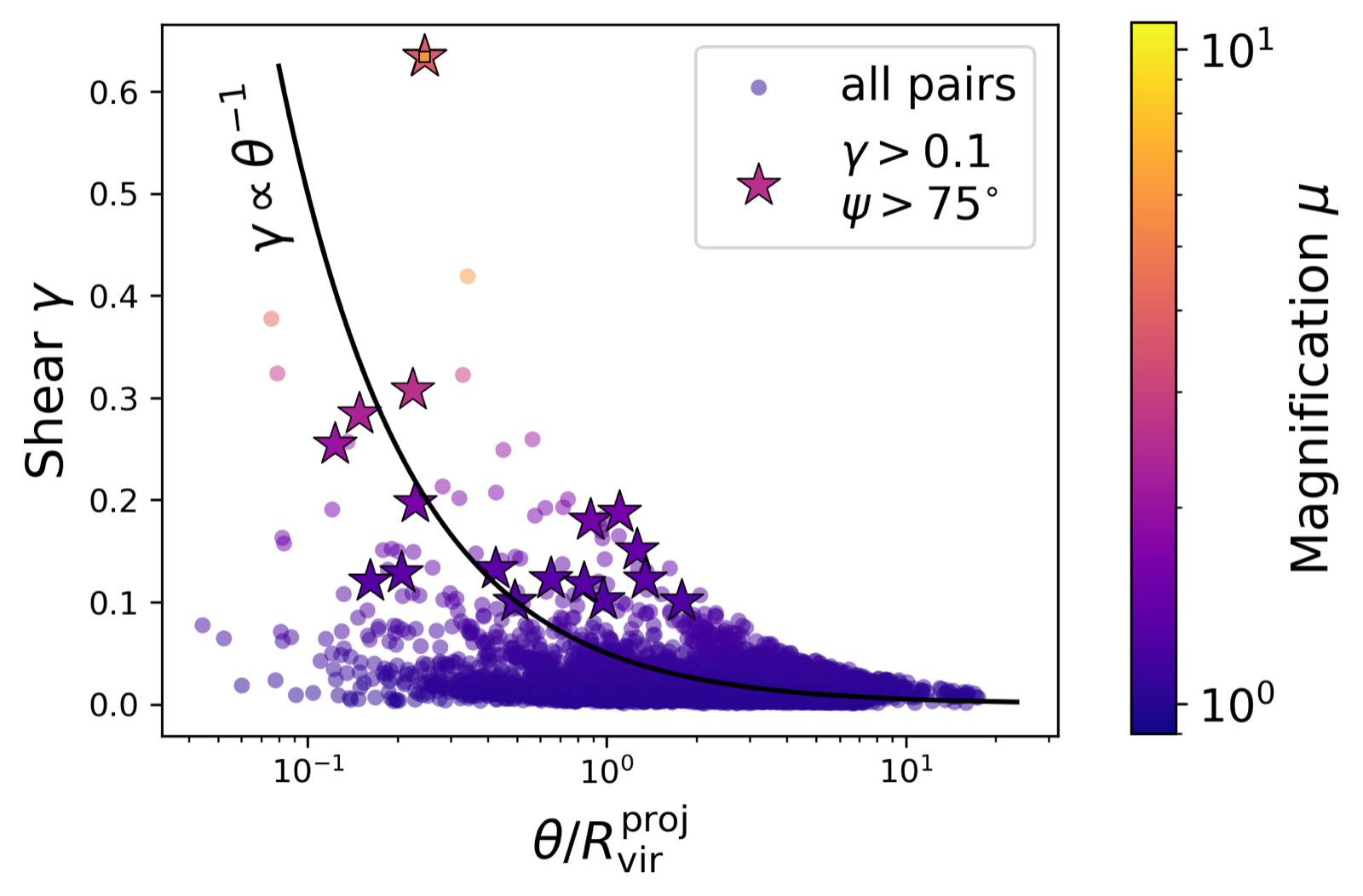}
\caption{Radial dependence of shear on ``impact parameter'' for all source-lens pairs (small circles) and for our tangentially-aligned lensing candidates with $\gamma>0.1$ and $\psi>75^{\circ}$ (stars). All of our lensing candidates fall within twice the projected virial radius of their associated foreground lens and their shear drops off as $\gamma\propto\theta^{-1}$. The colorbar denotes the lensing magnification $\mu$ which is $\sim1-2$ for most candidates. If our highest shear object is truly strongly lensed, the inner image has $|\mu|\sim3.7$ and the outer image is predicted to have $\mu\sim5.7$ (small inset square).}
\label{fig:radial}
\end{figure}

Figure \ref{fig:radial} shows that all of our lensing candidates fall within twice the projected virial radius of their associated lens. Here we have estimated the projected halo virial radius assuming $R_{\rm eff}/R_{\rm vir}=0.02$ \citep{kravtsov13,somerville18}. Importantly, the shear of our lensing candidates falls off with ``impact parameter'' $\theta/R_{\rm vir}^{\rm proj}$ as $\gamma\propto\theta^{-1}$, as it should. The scatter around this relation must come from additional $\theta_{\rm E}$ dependence in Equation \ref{eqn:gamma} or due to our simple scatter-free approximation that $R_{\rm eff}/R_{\rm vir}=0.02$. We can estimate the magnification due to an SIS lens as 
\begin{equation}
\mu_{\pm} = \theta_{\pm}/\beta
\end{equation}
where $\beta=\theta_{\pm}-\theta_{\rm E}$ is the ``true'' source position, $\theta_{\pm}$ are the two image positions and $\mu_{\pm}$ are their corresponding magnifications \citep{narayan96}. All of our sources, except possibly the one strong lensing candidate, lie outside their Einstein radii so there is only one image $\theta_+$ and these all have $\mu\sim1-2$. Our strong lensing candidate has $\mu_+\sim-3.7$ for the inner image (where the negative sign means the image parity is flipped with respect to the true source) and the second image is predicted to be $\sim1.5\times$ brighter with $\mu_-\sim5.7$.

\begin{figure*}
\centering
\includegraphics[width=\hsize]{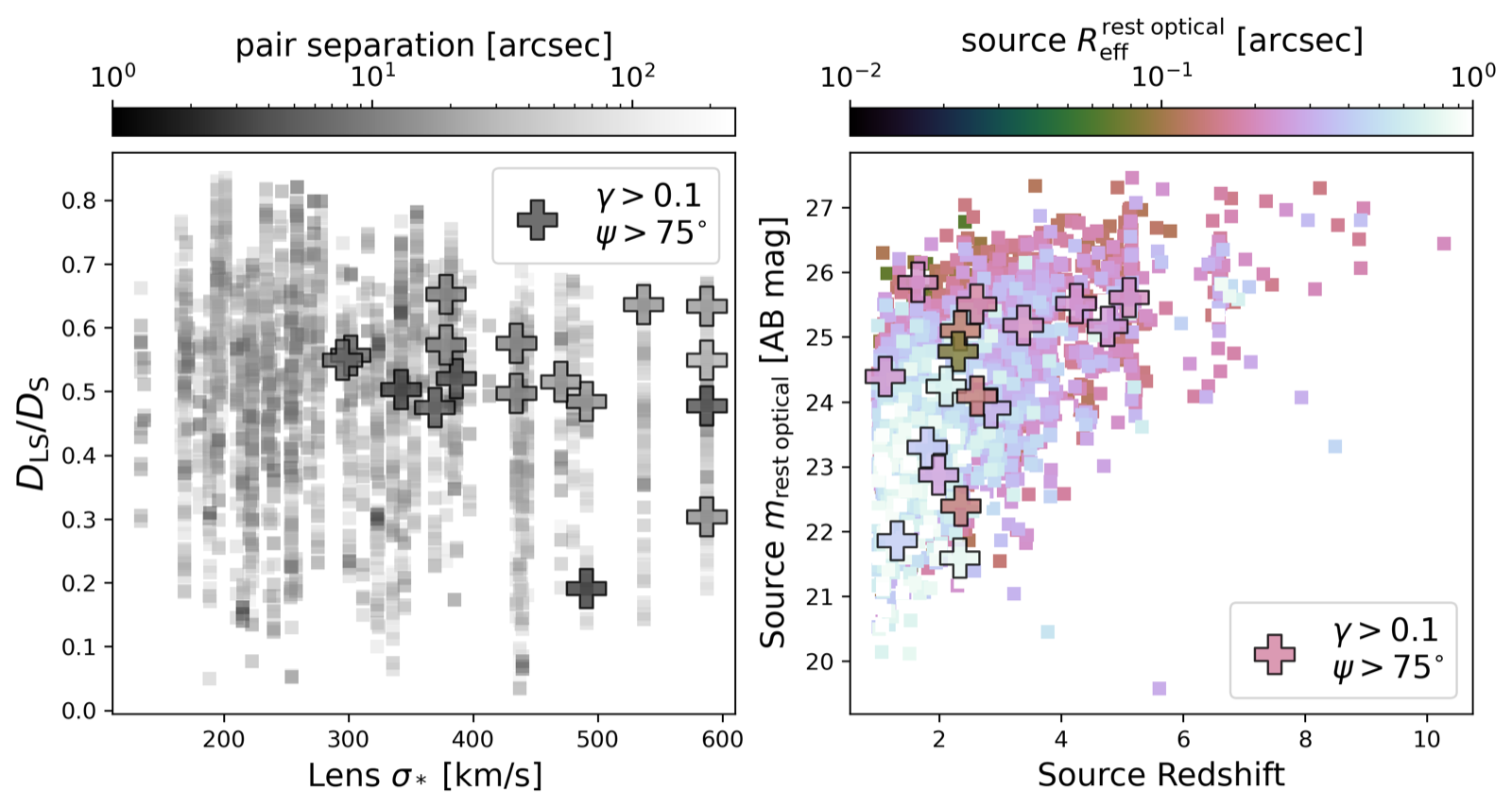}
\caption{Demographics of our lensing candidates relative to all source-lens pairs. \textit{Left:} Joint distribution of SIS lens velocity dispersion, angular diameter distance ratio $D_{\rm LS}/D_{\rm S}$ and on-sky pair separation $\theta$ for all source-lens pairs (small gray squares). These three quantities set the level of shear. Darker colors correspond to smaller pair separations. The large crosses are for our lensing candidates with $\gamma>0.1$ and $\psi>60^{\circ}$. Lensing candidates are preferentially found around more massive lenses (higher $\sigma_*$), at smaller pair separations (darker colors), and at greater angular diameter distance from their nearest lens which maximizes the Einstein radius $\theta_{\rm E}$. \textit{Right:} Joint distribution of source redshift, source apparent magnitude and source S\'ersic effective radius with the latter two measured in the NIRCam filter that most closely tracks the rest-frame optical. Lensing candidates roughly track the underlying population of source-lens pairs except they are biased towards $z\gtrsim2$ to maximize the angular diameter distance to our lenses at $z<1$. Their rest-optical sizes are $\sim0.1-0.5$ arcsec and apparent magnitudes $\sim22-26$ AB mag.}
\label{fig:demographics}
\end{figure*}

Figure \ref{fig:demographics} shows the joint distribution of lens $\sigma_*$, the angular diameter distance ratio $D_{\rm LS}/D_{\rm S}$ and the pair separation $\theta$, all of which go into predicting the shear (equation \ref{eqn:gamma}). The lensing candidates are associated with preferentially more massive lenses, smaller on-sky separations and higher $D_{\rm LS}/D_{\rm S}$ since that maximizes the Einstein radius $\theta_{\rm E}$ for an SIS lens with a given $\sigma_*$. In other words, the distribution of $\theta_{\rm E}$ is biased towards larger values ($\sim1.5-6''$) for lensing candidates compared to the overall log-normal distribution given in Footnote \ref{lognormal}.

Figure \ref{fig:demographics} also plots the joint distribution of source redshift, source apparent magnitude and source S\'ersic effective radius with the latter two in the NIRCam filter that most closely tracks the rest-frame optical. The lensing candidates appear well mixed with the general population of sources except that they are biased towards $z\gtrsim2$ which just reflects the need for some minimal source-lens angular diameter distance so that $\theta_{\rm E}$ is maximized. The highest redshift candidate is at $z\sim5$, and all candidates have apparent magnitudes in the range $\sim22-26$ AB mag in their rest-optical filters. The candidates have rest-optical $R_{\rm eff}\sim0.1-0.5$ arcsec which requires adaptive optics or space-based telescopes for observational follow-up. 

\subsection{Correlated orientations of lensing candidates}\label{sec:orientations}
\begin{figure*}
\centering
\includegraphics[width=\hsize]{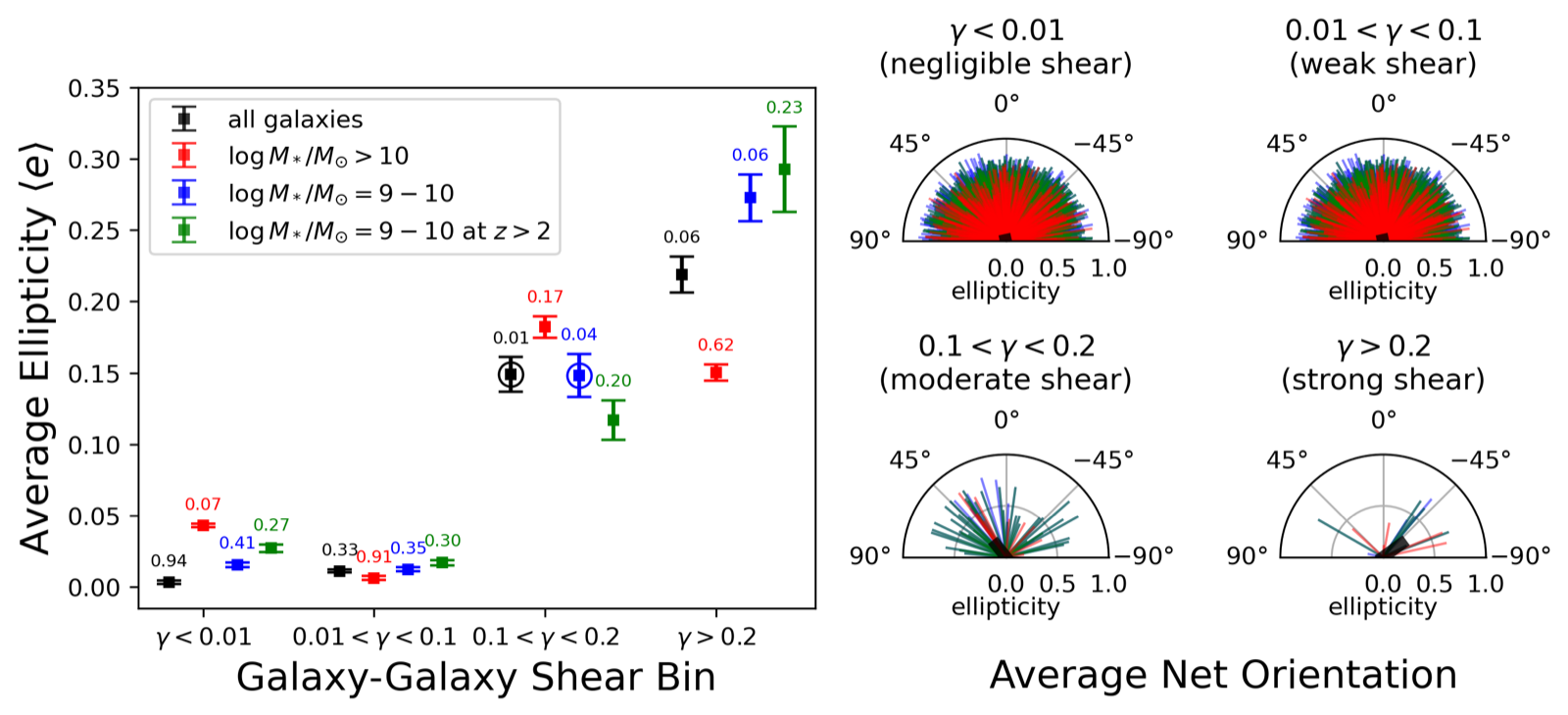}
\caption{Average ellipticity and orientation as a function of galaxy-galaxy lensing shear. \textit{Left:} Magnitude of the average complex ellipticity vector in increasing bins of shear from left to right: negligible ($\gamma<0.01$), weak ($0.01<\gamma<0.1$), moderate ($0.1<\gamma<0.2$) and strong ($\gamma>0.2$). All points reflect the mean and standard error from our Monte Carlo error propagation method (subsection \ref{sec:errorprop}). Our combined background sample is shown in black, high-mass only in red, low-mass only in blue, and $z>2$ low-mass in green. At small shear, the average ellipticity is very small as it should be for randomly oriented galaxies. At larger shear, we find a significantly larger average residual ellipticity. The $p$-value from our null hypothesis test is written above each point. Given the small sample sizes, only the two moderate shear points enclosed in circles are significant at the $>95\%$ level. \textit{Right:} Complex ellipticity vectors in bins of shear for individual galaxies and the complex average of the combined sample (black lines). For visualization, the symmetric complex ellipticity vectors have been ``folded'' back onto the range $-90^{\circ}$ to $90^{\circ}$ measured east of north (up). The magnitude of the average complex ellipticity vector is very small for the negligible and weak shear bins, but is clearly non-zero with a preferred orientation for the two higher shear bins.}
\label{fig:vectors}
\end{figure*}

The possible excess of tangentially-aligned source-lens pairs in the 99.7\% quantile of shear in Figure \ref{fig:alignments} motivates averaging over the orientations of background galaxies in bins of galaxy-galaxy lensing shear. For galaxies with negligible ($\gamma<0.01$) and weak ($0.01<\gamma<0.1$) shear, we expect the magnitude of their average complex ellipticity vector to be negligible, assuming they are randomly oriented and there are no systematics. Figure \ref{fig:vectors} confirms this even when splitting by mass. However, at moderate ($0.1<\gamma<0.2$) and high ($\gamma>0.2$) shear, we start to see evidence for a non-zero average complex ellipticity and hence preferred orientation. The standard errors come from our Monte Carlo error propagation method (subsection \ref{sec:errorprop}) and show that the trend is well constrained modulo any systematics. We compare these observed averages to the null distribution of $\langle e\rangle$ assuming the galaxies are randomly oriented (subsection \ref{sec:null}). Given the small sample sizes, only the combined and low-mass subsamples in the moderate shear bin remain statistically significant at the $>95\%$ level (the $p$-values are reported above each point in Figure \ref{fig:vectors}).

This follows the expectation that as shear increases, there must be some residual orientation dictated by lensing. The complex average $\langle e \rangle\sim10-30\%$ is at least an order of magnitude above the conventional weak lensing regime and unlikely to be due to PSF systematics (Appendix \ref{sec:psf}). These coherent alignments are surprising because our sources trace a number of different foreground lens galaxies. Of course, if this were repeated using galaxy-galaxy lensing candidates from multiple widely-spaced surveys, we would expect $\langle e\rangle\to0$ even in the moderate and strong shear regimes. But the fact that our galaxy-galaxy lensing candidates have correlated orientations within the relatively small ($\sim0.028$ deg$^2$) CEERS footprint motivates a more detailed spatial analysis of alignments on larger scales, which we turn to next. 

\section{Large-Scale Galaxy Alignments}\label{sec:results2}
In this section, we present results on large-scale alignments by averaging over the complex ellipticities of background galaxies on multiple scales defined by the NIRCam instrument. We start with individual NIRCam chips ($64''\times64''$), then individual modules ($2.2'\times2.2'$), followed by entire pointings ($2.2'\times5.1'$ with a $\sim0.7'$ gap), and finally the scale of the entire survey ($\sim30'\times6'$). For this pilot study, restricting ourselves to the on-sky survey geometry imposed by NIRCam mitigates systematics from detector gaps and otherwise arbitrary binning. Appendix \ref{sec:psf} quantifies the expected bias from PSF systematics.

\subsection{Alignments on NIRCam chip scales $(64''\times64'')$}
\begin{figure*}
\centering
\includegraphics[width=0.5\hsize]{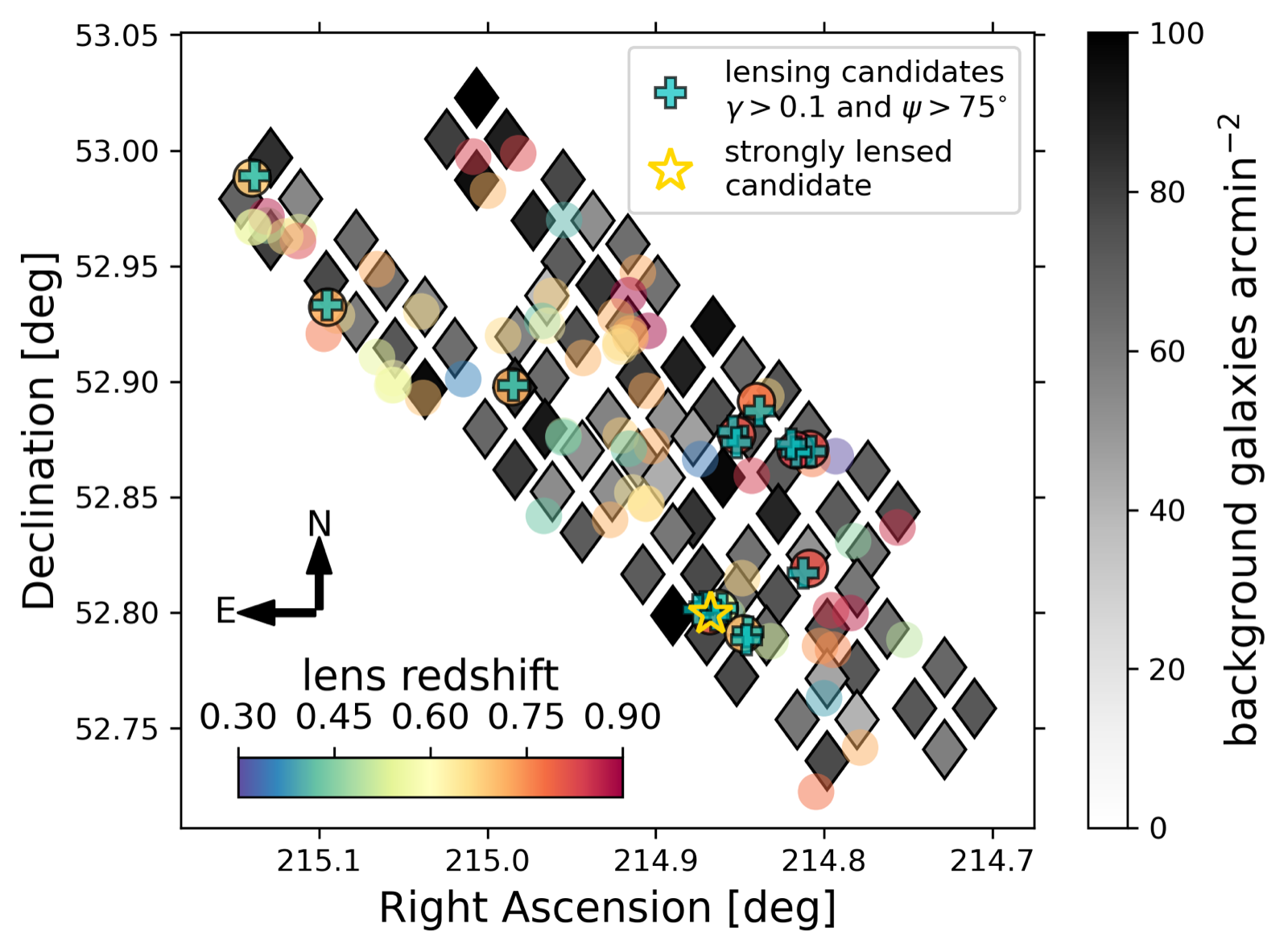}\includegraphics[width=0.5\hsize]{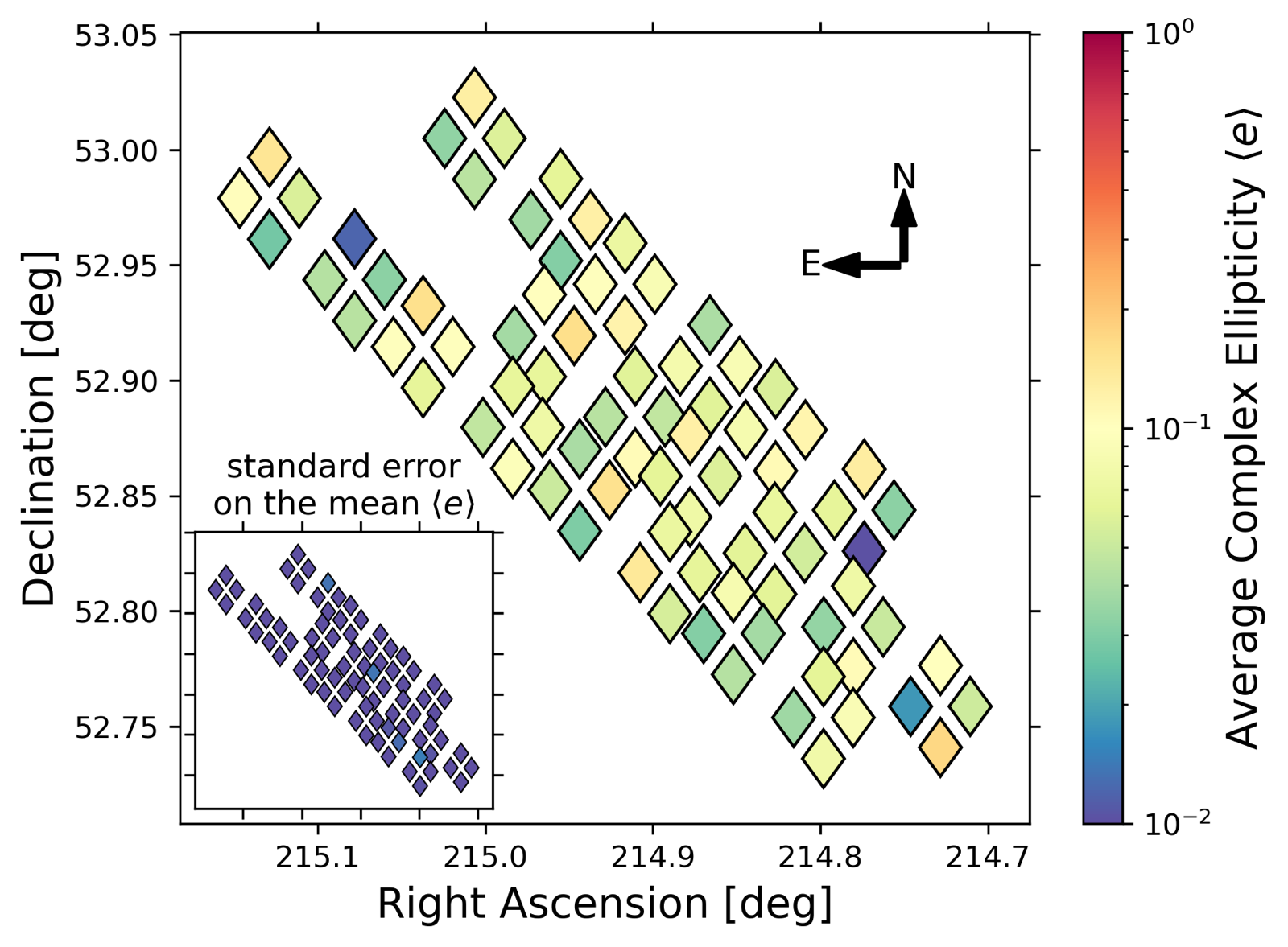}
\caption{Projected on-sky lensing maps for CEERS in the standard ``north-up'' frame. \textit{Left:} Positions of the galaxy-galaxy lensing candidates (cyan pluses), foreground lenses (circles colored by redshift) and our overall background galaxy sample (grey histogram). The gray boxes show that there are $\sim50-100$ per arcmin$^2$ in each NIRCam chip ($64''\times64''\sim0.0003$ deg$^2$) across this $\sim100$ arcmin$^2$ survey. The lensing candidates, including the one possible strong lens (yellow star), appear clustered near the southwest. The foreground lenses are roughly uniformly distributed but Figure \ref{fig:pairs} suggests an overdensity at $z\sim0.75$. There is generally at least one $z\sim0.75$ lens near each lensing candidate. \textit{Right:} Magnitude of the average complex ellipticity $\langle e \rangle$ when averaging over the orientations of background galaxies in individual NIRCam chips. The average complex ellipticity is non-negligible and reaches values in excess of $10\%$ in some parts of the map. The standard error on the mean (inset panel) is generally quite low but we are likely dominated by systematics.}
\label{fig:onsky}
\end{figure*}

Figure \ref{fig:onsky} shows that, on average, there are $\sim50-100$ background galaxies per arcmin$^2$ that satisfy our $\log_{10} M_*/M_{\odot}>9$ cut in CEERS, which covers a total area of about $\sim100$ arcmin$^2$ ($\sim0.028$ deg$^2$). This is sufficient for weak lensing and cosmic shear studies. First, we show the on-sky positions of our galaxy-galaxy lensing candidates with $\gamma>0.1$ and $\psi>75^{\circ}$ compared to the foreground lenses and our overall background source sample. The galaxy-galaxy lensing candidates appear clustered near the southwest, which may help explain why we see correlated orientations in Figure \ref{fig:vectors}. The lenses themselves appear roughly uniformly distributed throughout the field, though we showed in Figure \ref{fig:pairs} that there is an overabundance of lenses at $z\sim0.75$. There is generally at least one, if not multiple, $z\sim0.75$ foreground lenses near each lensing candidate. The projected $R_{\rm vir}$ of our lenses are $\sim0.1'-3.5'$, i.e., extending over a single NIRCam chip or entire module.

Figure \ref{fig:onsky} also shows the magnitude of the average complex ellipticity when averaging over background galaxies in individual NIRCam chips. The average complex ellipticity is non-negligible and clearly in excess of $\langle e \rangle\sim10\%$ in many bins. This is an order of magnitude above the conventional weak lensing regime and implies strong alignments in the ellipticities and orientations of background galaxies. The standard error on the mean $\langle e \rangle$ from our Monte Carlo error propagation is quite low so in the next subsection we will use statistical null tests to quantify the significance of these alignments. 

\begin{figure}
\centering
\includegraphics[width=\hsize]{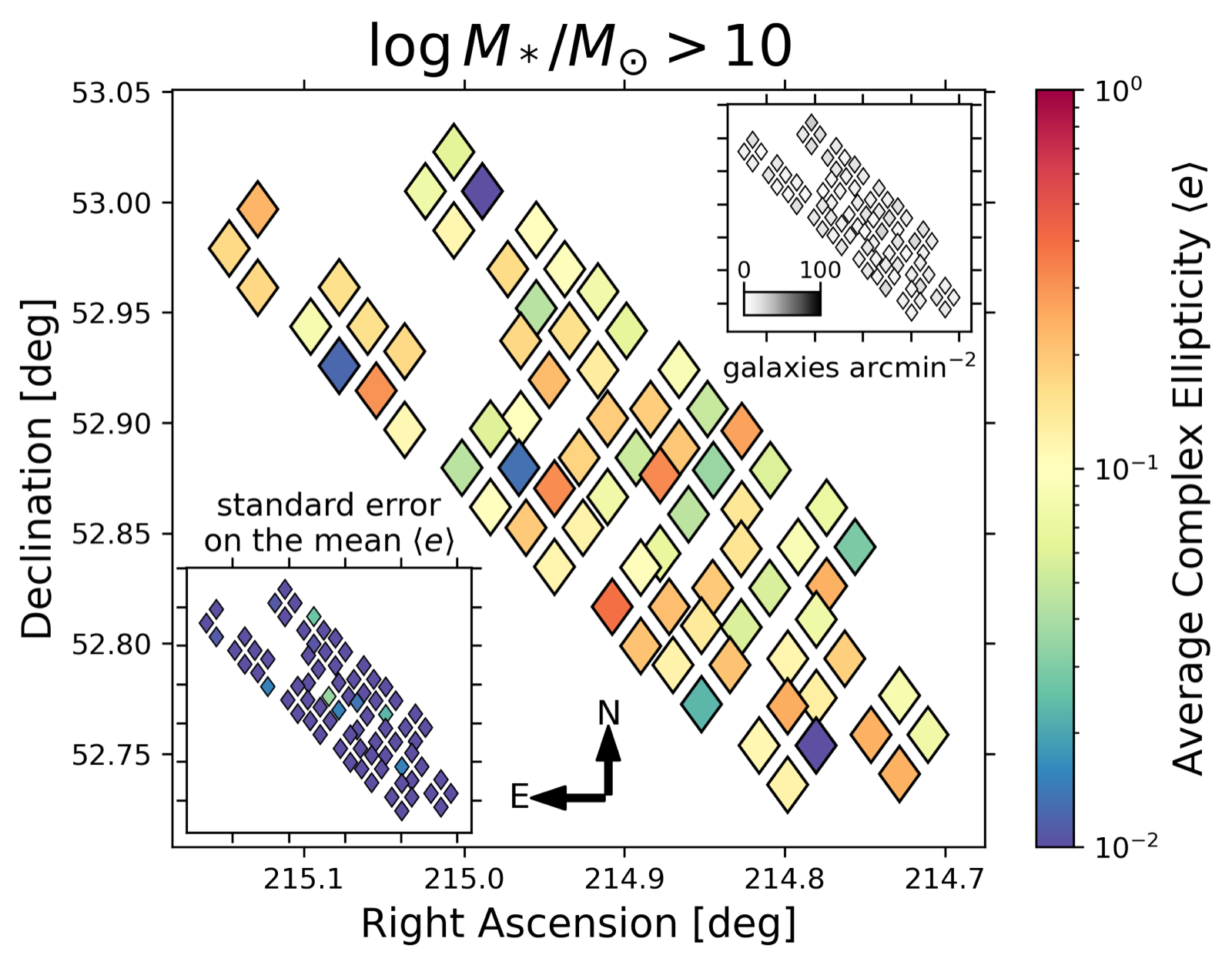}

\includegraphics[width=\hsize]{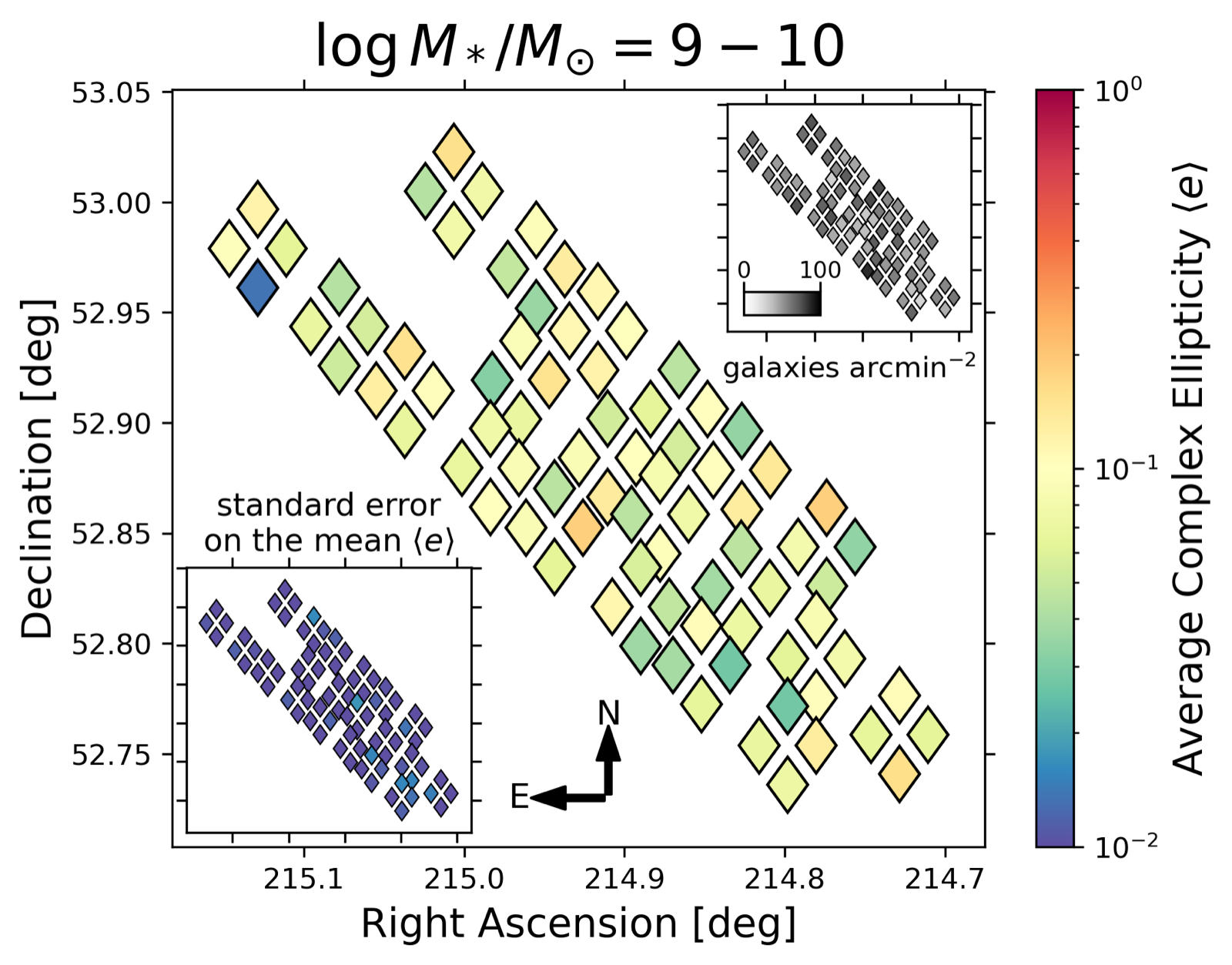}

\includegraphics[width=\hsize]{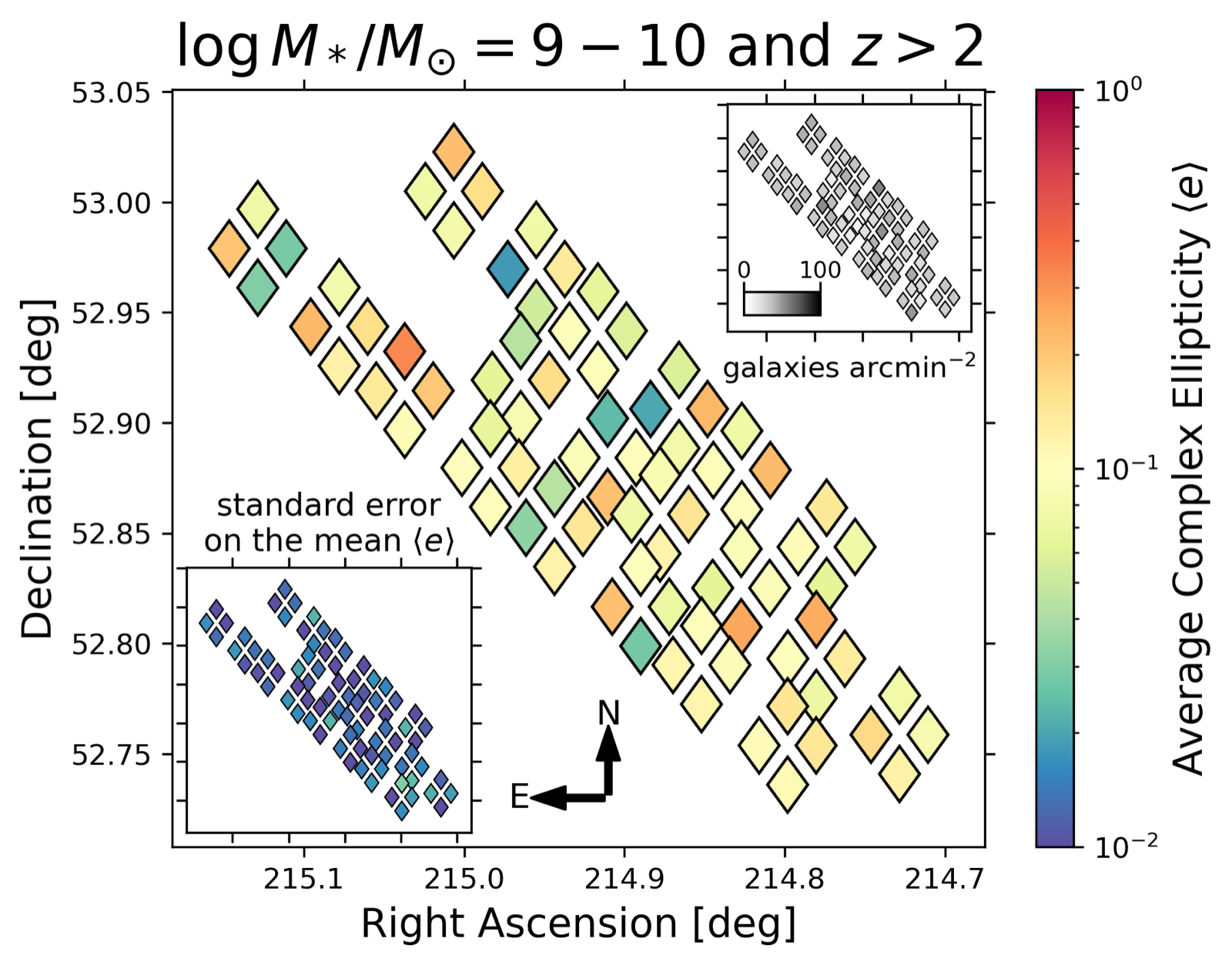}
\caption{Same chip-scale ``shear map'' as in Figure \ref{fig:onsky} but now split into high-mass (top), low-mass (middle), and high-redshift low-mass (bottom) background galaxies. Only bins with $\geq5$ galaxies are used. The top-right inset panels show the number of galaxies arcmin$^{-2}$ in each NIRCam chip. \textit{Top:} Massive sources show an elevated $\langle e\rangle\sim0.3$ in many chips. \textit{Middle:} Low-mass sources at any $z>1$ also show non-negligible $\langle e \rangle\gg0.01$ in many regions. \textit{Bottom:} Low-mass sources at $z>2$ show larger $\langle e \rangle\gtrsim10\%$ in more chips compared to the $z<2$ low-mass sources.}
\label{fig:fields_mass}
\end{figure}

Figure \ref{fig:fields_mass} splits the chip-scale ``shear map'' into three separate ones for high-mass, low-mass, and high-redshift ($z>2$) low-mass background galaxies. 
There are large number of NIRCam chips where high-mass galaxies show strong alignments with $\langle e \rangle>10\%$. Despite the relative scarcity of high-mass galaxies, the standard error on $\langle e\rangle$ from our Monte Carlo error propagation is quite low, though we are likely dominated by systematics. The low-mass sample down to $z>1$ shows non-negligible $\langle e\rangle\gg0.01$ similar to the combined map from Figure \ref{fig:onsky}. Interestingly, restricting to just low-mass galaxies at $z>2$ \citep[in the predominantly elongated regime;][]{vanderwel14,zhang19,pandya24} reveals preferentially larger $\langle e \rangle\sim20-30\%$ in multiple regions compared to the $z<2$ low-mass sample. Whether this is telling us something about lensing by a large foreground mass concentration, intrinsic alignments or systematics requires us to first assess its statistical significance, which we turn to next.

\subsection{Statistical significance from null hypothesis tests}
\begin{figure*}
\centering
\includegraphics[width=\hsize]{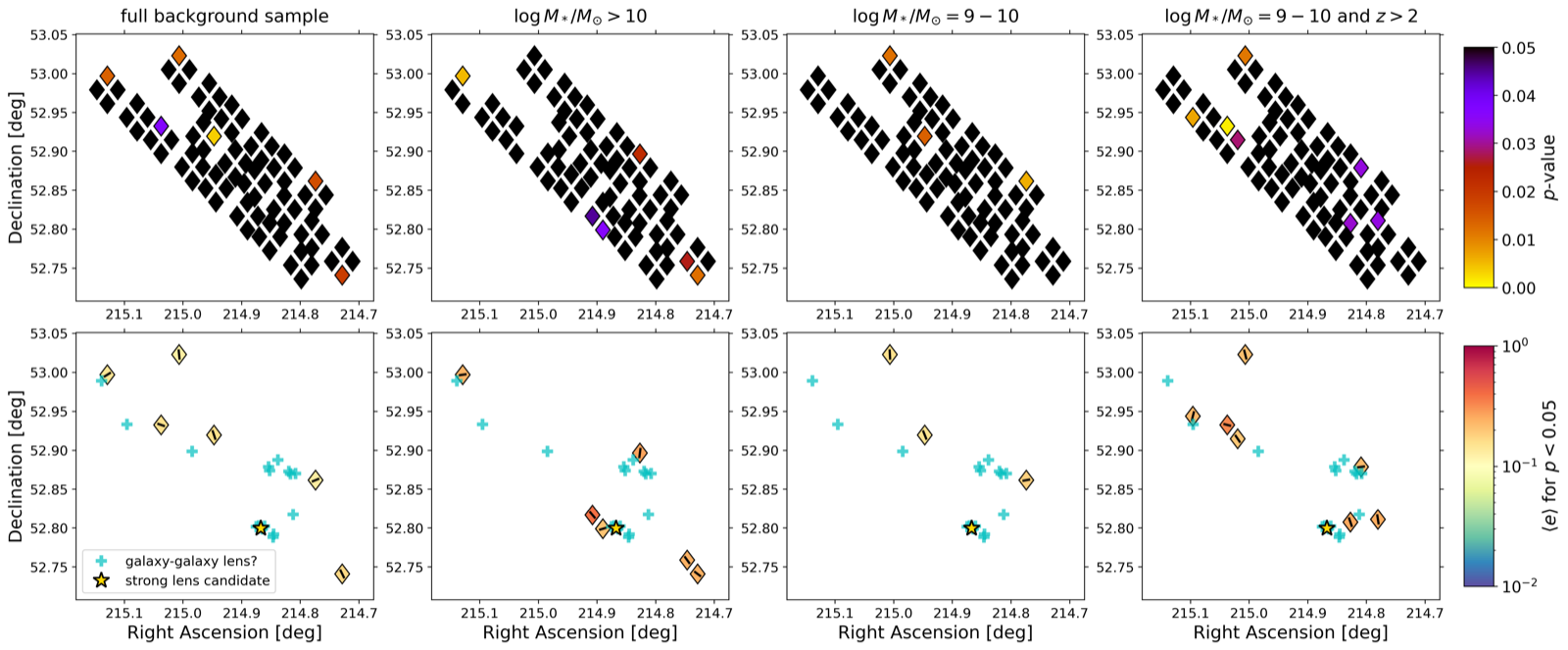}
\caption{Statistical significance of chip-scale ``shear maps'' for all background galaxies combined (left column), high-mass sources (second-from-left column), low-mass sources (second-from-right column), and $z>2$ low-mass sources (right column). The top row shows the $p$-value in individual $64''\times64''$ chips calculated as described in the text. There are multiple regions with ``internal'' chip-scale alignments at the $>95\%$ confidence level. The middle row shows the average complex ellipticity magnitude $\langle e \rangle$ (colors) and phase (lines in the ``north-up'' frame) for these statistically significant bins alone. Note how $\langle e \rangle\gtrsim0.1$ in all significant cases suggestive of strong alignments. Our galaxy-galaxy lensing candidates (cyan crosses and gold star) tend to be clustered near the southwest where there are multiple chips with coherent alignments.}
\label{fig:pvals}
\end{figure*}

Figure \ref{fig:pvals} quantifies the statistical significance of our results with the null hypothesis test described in subsection \ref{sec:null}. We find that the majority of chips do not have statistically significant alignments since $p\gg0.05$. However, there are multiple cells where the null hypothesis can be ruled out at $>95\%$ significance, including a handful at $>99\%$ significance. All of these have $\langle e\rangle\gtrsim0.1$ which, again, is at least an order of magnitude larger than the conventional weak lensing regime. Interestingly, some of these significant regions are in the southwest where we saw the clustering of tangentially-aligned galaxy-galaxy lensing candidates in Figure \ref{fig:onsky}. 

In addition to ``internal'' chip-scale significance, it is useful to think about the ``map-scale'' significance. Given that we have $80$ NIRCam chips, we expect $\sim80\times0.05\approx4$ detections by random chance at the $95\%$ confidence level assuming every bin is independent. We have more detections than this simple threshold in the combined (6), high-mass (6) and $z>2$ low-mass (8) maps. However, the low-mass map has only 3 detected chips and even though all of these are at $\gtrsim99\%$ significance, on the scale of the map this may be a statistical fluke. Formally, when we use the Benjamini-Hochberg false discovery rate and Bonferroni methods to ``correct'' the individual chip-scale $p$-values to account for the large number (80) of tests being done, none of the chips in any of the subsamples remain significant at the $>95\%$ level. However, these conservative statistical methods may not be using all of the available information (e.g., the clustering of chips with detections) so it is also useful to look at images of individual chips with alignments, which we turn to next. 

\begin{figure*}
\centering
\includegraphics[width=\hsize]{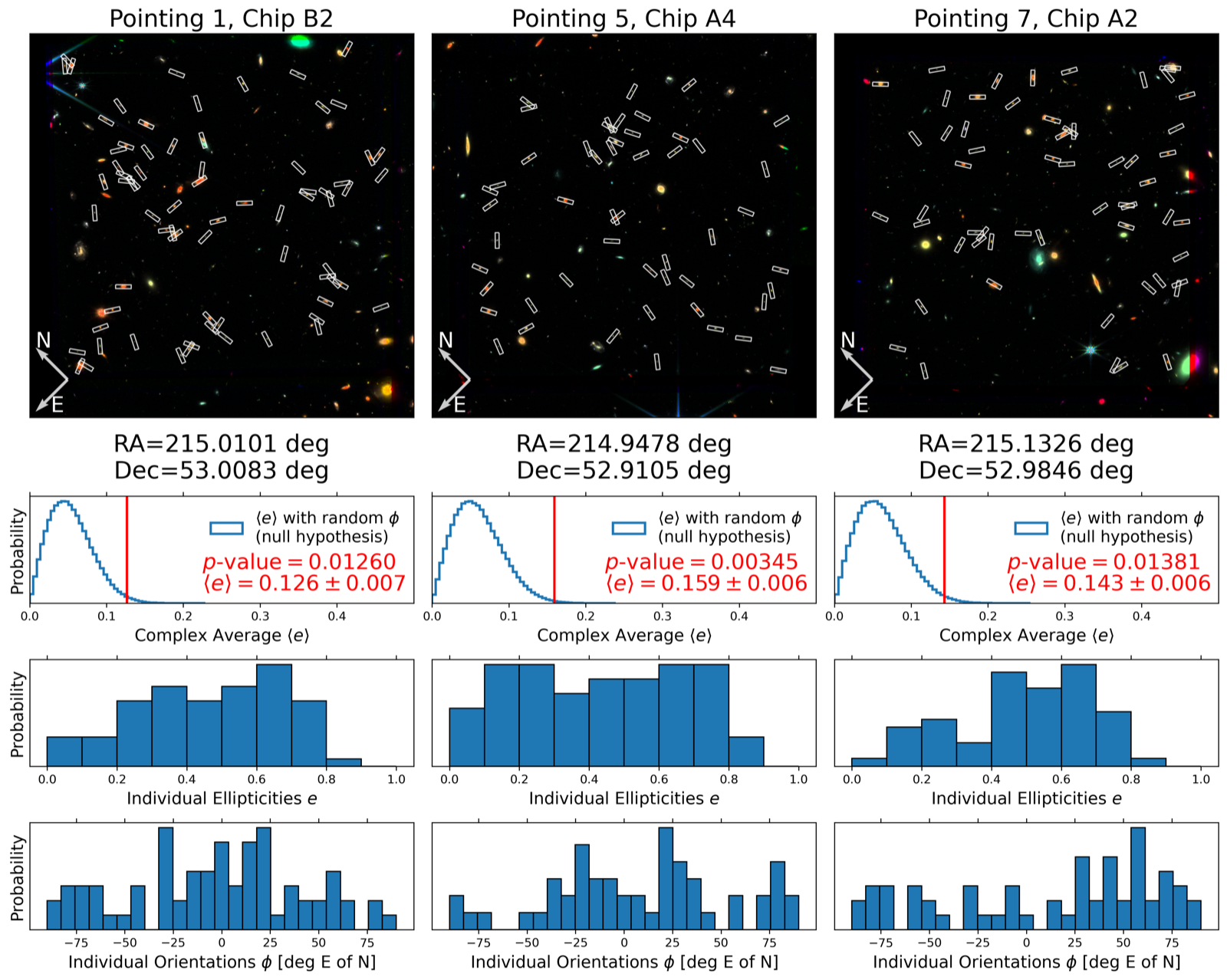}
\caption{Three example NIRCam chips in which the null hypothesis of random orientations can be ruled out at $\gtrsim99\%$ significance when averaging over the combined background sample (left-most column of Figure \ref{fig:pvals}). A two arcsec rectangle is plotted at the location of each background galaxy denoting the orientation of its major axis. The bottom three panels of each column show the distribution of $\langle e \rangle$ under the null hypothesis of random orientations (top), distribution of ellipticities (middle) and and distribution of position angles (bottom). Coherent alignments in the orientations of background galaxies can be seen.}
\label{fig:fields}
\end{figure*}

Figure \ref{fig:fields} shows three example chips that have the highest statistical significance ($p\lesssim0.01$). The distribution of $\langle e \rangle$ under the null hypothesis of random orientations can be ruled out with $\gtrsim99\%$ confidence given the large observed $\langle e \rangle$. Coherent alignments in the orientations of background galaxies can be seen with hints of circular polarization patterns. Whether these alignments are due to lensing from foreground substructure, intrinsic alignments, systematics or pure random chance remains to be seen and will require detailed follow-up spectroscopy and lens modeling. Appendix \ref{sec:psf} shows that PSF uncertainties are unlikely to explain these alignments. Another way to get a handle on such systematics is to search for alignments on even larger scales which we turn to next.

\subsection{Alignments in NIRCam modules $(2.2'\times2.2')$}
\begin{figure*}
\centering
\includegraphics[width=\hsize]{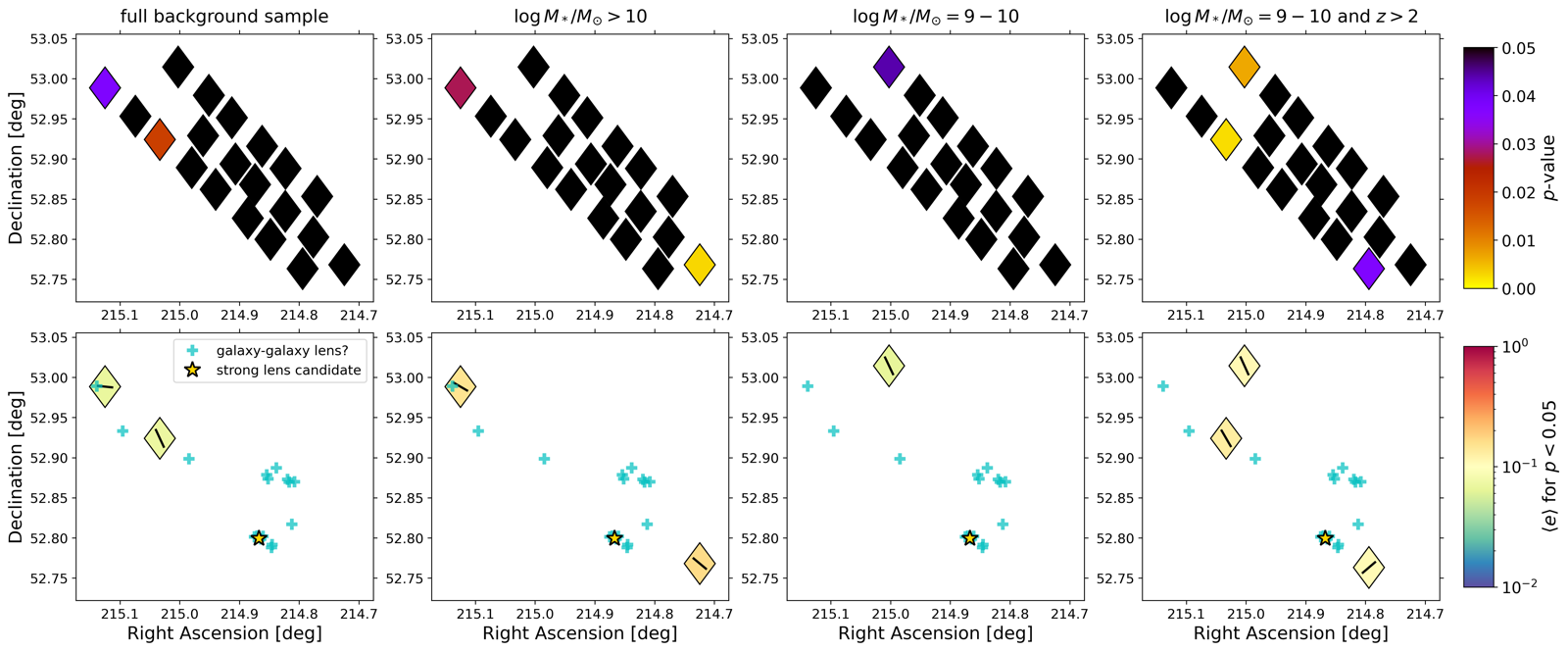}
\caption{Similar to Figure \ref{fig:pvals} but now for individual NIRCam modules ($2.2'\times2.2'$). There are a few modules with alignments at the $>95\%$ significance level for each subsample, with many showing $\langle e\rangle\sim10\%$ even on these larger scales. Those modules do not, however, typically overlap with the galaxy-galaxy lensing candidates.}
\label{fig:pvals22}
\end{figure*}

On large scales, assuming background galaxies are randomly oriented, we should find fewer regions with statistically significant alignments unless lensing or systematics are playing a role. Here we increase the map ``pixel'' size from the scale of an individual NIRCam chip ($64''\times64''$) to an individual NIRCam module ($2.2'\times2.2'$ comprising four chips).\footnote{Note that a single NIRCam module can sometimes fit a single bright foreground cluster used for conventional strong and weak lensing studies \citep[e.g.,][]{pontoppidan22} whereas here we are searching for alignments over multiple modules or pointings.} Figure \ref{fig:pvals22} identifies a handful of NIRCam modules with statistically significant alignments at the $>95\%$ level using our null test described in section \ref{sec:orientations}. A few of these have module-scale significance at the $>99\%$ level. Given that we have 20 such modules, we expect $20\times0.05\approx1$ such detection by random chance at the $95\%$ level. We have $2-3$ modules with detections for each subsample so we pass this simple threshold. Formally, the Benjamini-Hochberg and Bonferroni corrections lead to $p>0.05$ for all modules except one in the $z>2$ low-mass map which remains significant at the $\sim98\%$ level ($p=0.01882$).

\begin{figure*}
\centering
\includegraphics[width=\hsize]{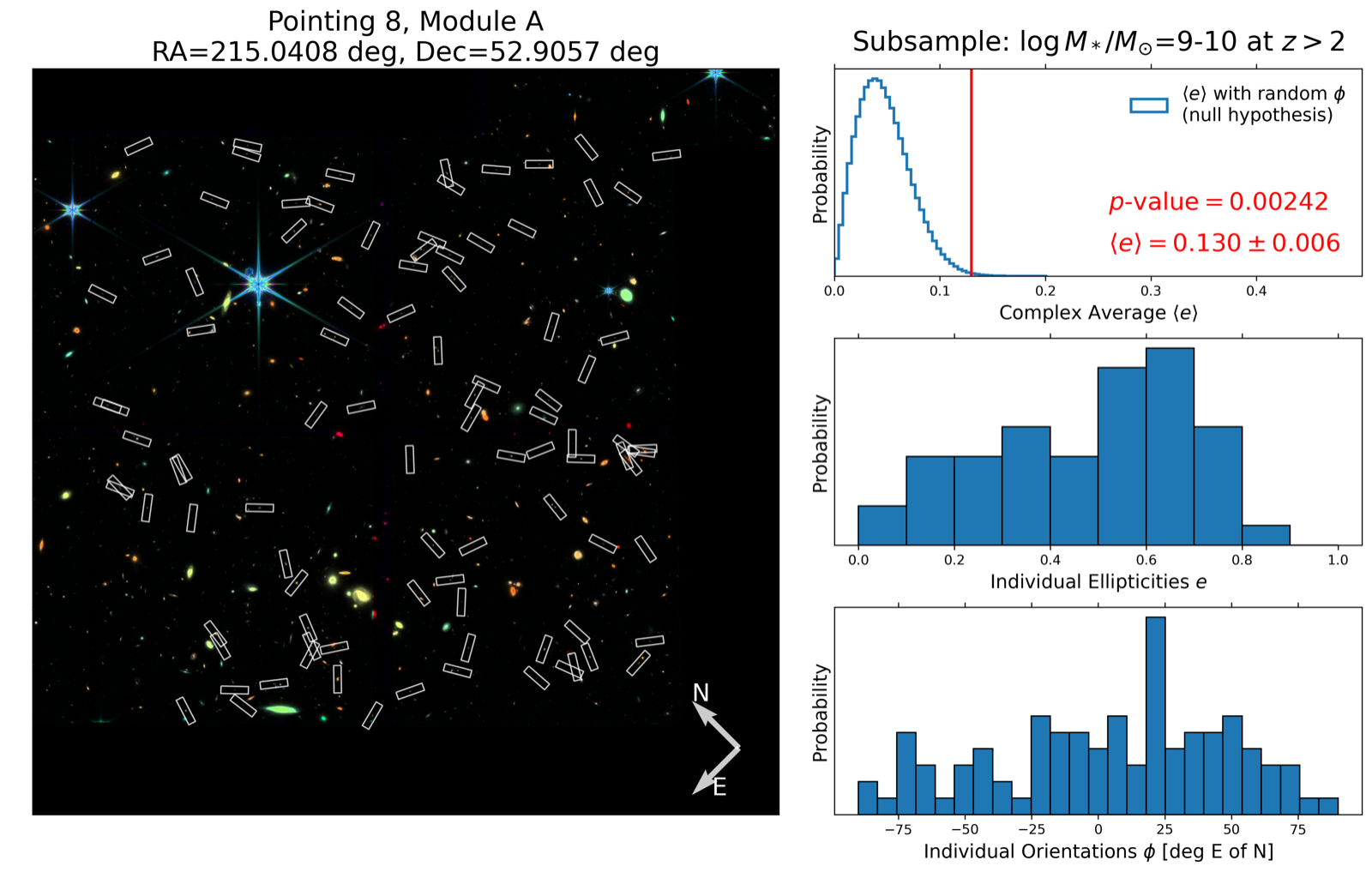}
\caption{Similar to Figure \ref{fig:fields} but for the NIRCam module with the most statistically significant alignments for the low-mass $z>2$ subsample (rightmost column of Figure \ref{fig:pvals22}). Again coherent alignments can be seen even on these larger $2.2'\times2'2$ scales, with hints of circular patterns that motivate future $E/B$-mode polarization decomposition and weak lensing mass reconstruction.}
\label{fig:fields22}
\end{figure*}

Figure \ref{fig:fields22} shows the module with the most significant alignments in the orientations of low-mass galaxies at $z>2$. Even on these larger $2.2'\times2.2'$ scales, coherent alignments can be seen. Some of these alignments appear like linear ``chains'' whereas others arise from nearly circular arrangements of background galaxies. The latter motivates future decomposition of the shear map into E- and B-mode polarizations to assess whether this signal is due to lensing, intrinsic alignments or systematics \citep[e.g.,][]{crittenden01,crittenden02}.

\subsection{Alignments in NIRCam pointings $(2.2'\times5.1')$}
\begin{figure*}
\centering
\includegraphics[width=\hsize]{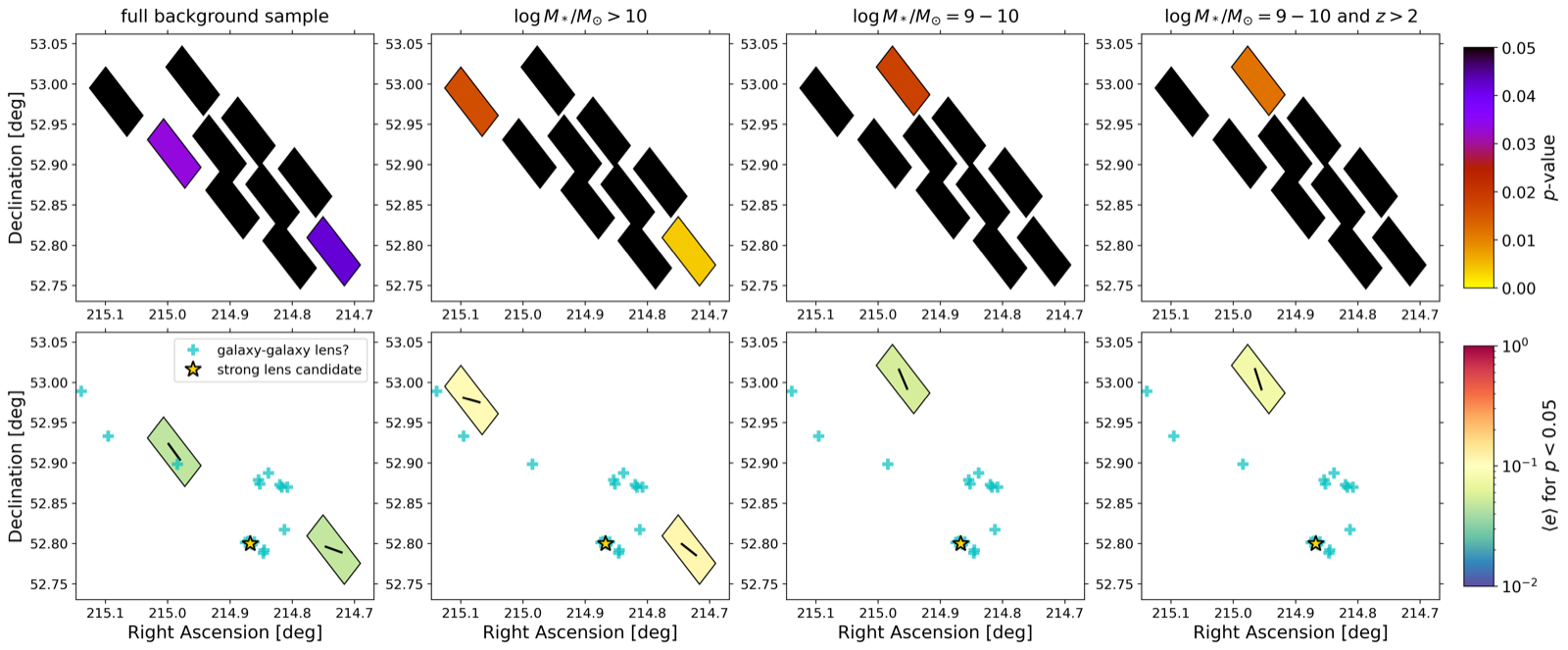}
\caption{Same as Figure \ref{fig:pvals22} but now averaging over entire NIRCam pointings (two $2.2'\times2.2'$ modules with a $\sim0.7'$ gap in between). For each subsample, there are $1-2$ pointings with alignments at $>95\%$ significance. The galaxy-galaxy lensing candidates are typically offset from such pointings by $\sim0.1$ deg.}
\label{fig:pvals44}
\end{figure*}

Here we increase the map ``pixel'' size to even larger scales of an entire NIRCam pointing which encompasses two $2.2'\times2.2'$ modules separated by a $\sim0.7'$ gap. Figure \ref{fig:pvals44} shows that we continue to find $1-2$ pointings for each subsample with significant alignments at the $>95\%$ level. Two of these are significant at the $>99\%$ level: one in the southwest corner of the high-mass map and another in the northwest corner of the $z>2$ low-mass map. Since we have 10 pointings, we expect $10\times0.05\approx0.5$ pointings with a detection by random chance alone. We are above this simple threshold for all subsamples, but the more formal Benjamini-Hochberg and Bonferroni methods lead to corrected $p$-values in excess of 0.05 for all pointings. However, in addition to these conservative statistical tests, it is still instructive to look at images of pointings with detections.

\begin{figure*}
\centering
\includegraphics[width=\hsize]{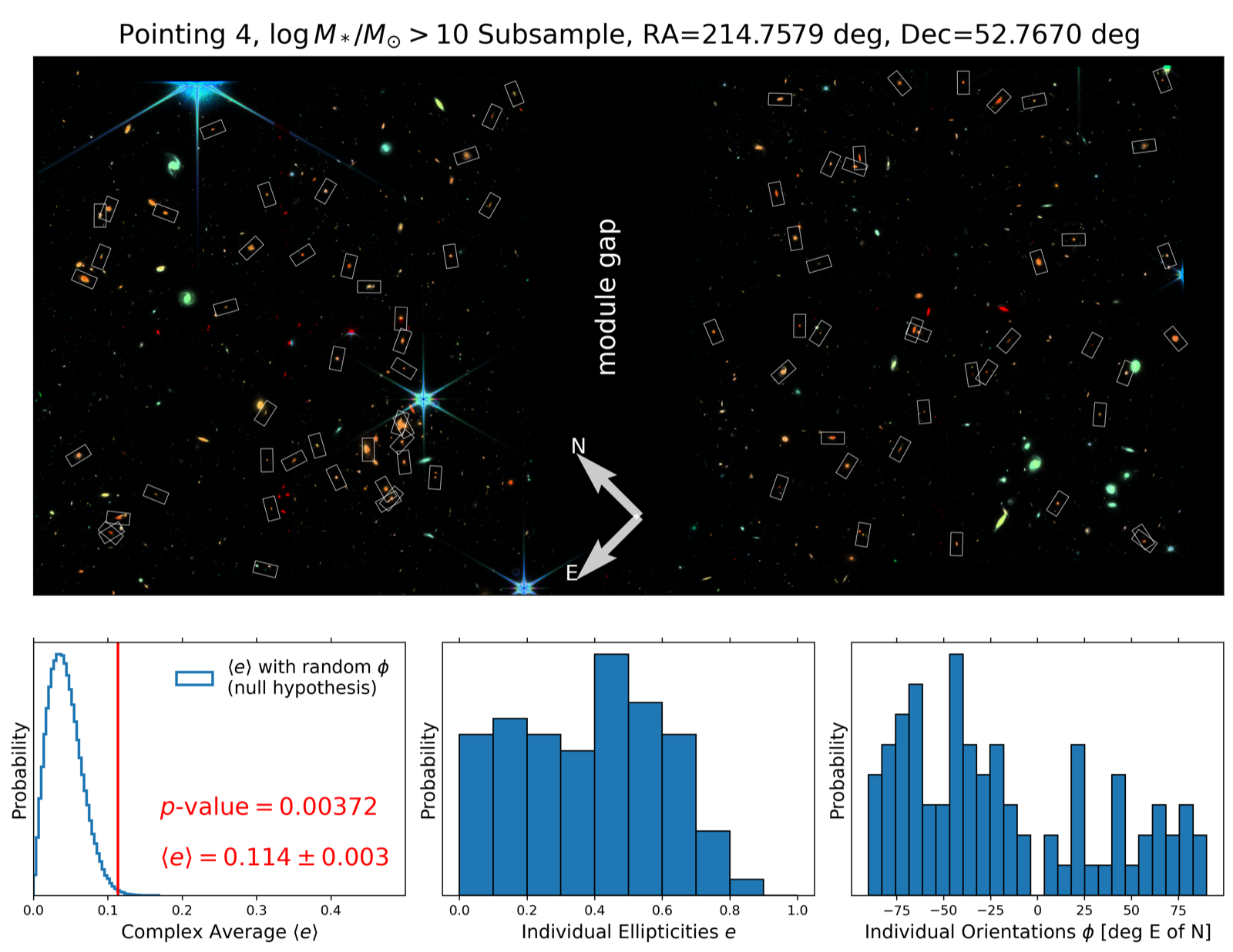}
\caption{Same as Figure \ref{fig:fields22} but now for the southwest $2.2'\times5.1'$ pointing in which our high-mass sources have significant alignments at the $>99\%$ level. Large-scale alignments are clearly present and there are hints of circular polarization patterns.}
\label{fig:fields44}
\end{figure*}

Figure \ref{fig:fields44} shows an image of the southwest pointing with significant alignments of high-mass galaxies at $z>1$. Large-scale alignments are clearly obvious by eye in both modules of this pointing. Many of these galaxies also show circular alignment patterns which reinforces the need to decompose the shear map into $E$ and $B$ mode polarizations which can help constrain whether this is due to lensing, intrinsic alignments or systematics. More generally, our results motivate computing galaxy-shear and shear-shear correlation functions in ``blank'' JWST deep fields like CEERS that are far away from obvious, bright foreground clusters. Such an analysis would naturally capture the dependence of the alignment signal as a continuous function of spatial scale and offer constraints on cosmology and the foreground lens mass. However, this is non-trivial and requires careful consideration of survey geometry so we leave it for the future.


\subsection{Alignments averaged over the entire survey}

\begin{figure*}
\centering
\includegraphics[width=\hsize]{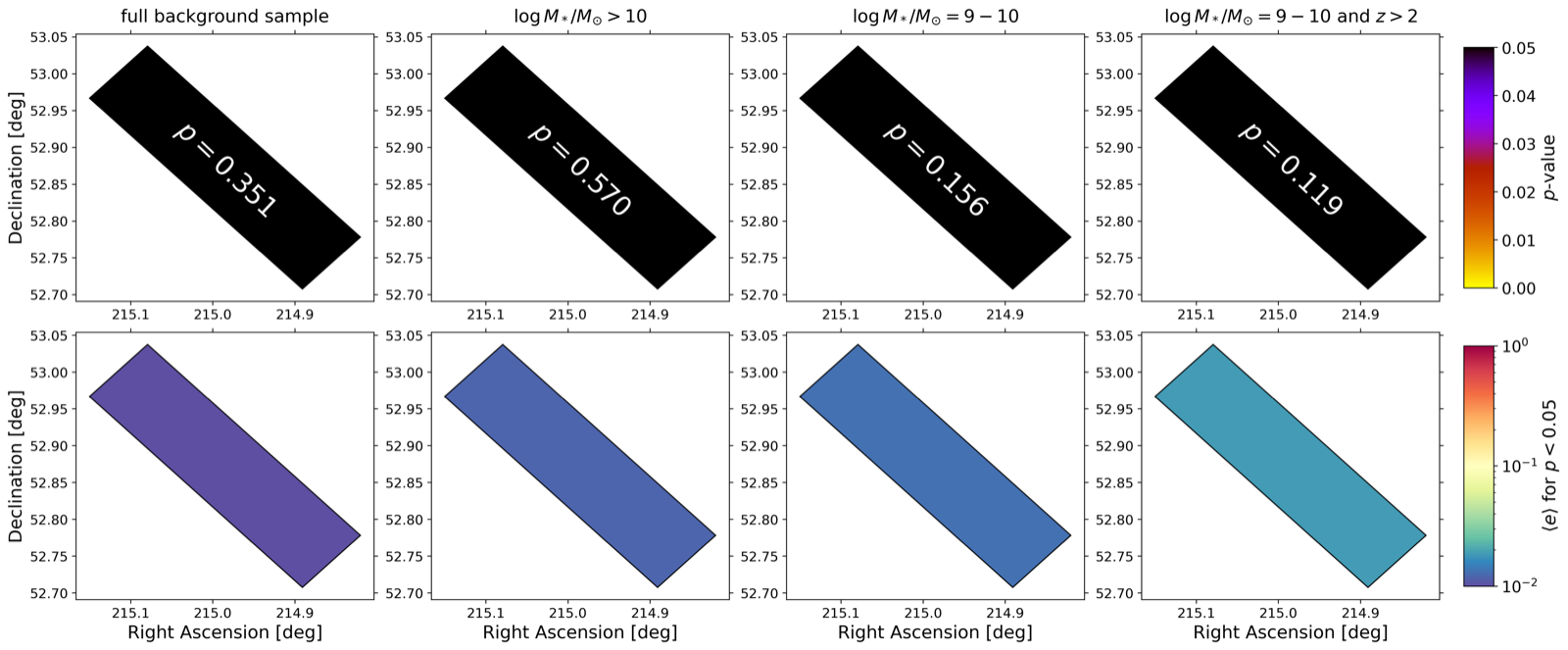}
\caption{Same as Figure \ref{fig:pvals22} but now averaging over the entire field ($\sim30'\times6'$). As expected, we do not find any statistically significant alignments implying that the background galaxies are randomly oriented over such large scales.}
\label{fig:pvalswhole}
\end{figure*}

Figure \ref{fig:pvalswhole} shows that, on the scale of the entire field ($\sim30'\times6'$), we do not find any statistically significant alignments ($p>0.05$). This makes sense since over such a large area, background galaxies should be randomly oriented and lensing must be weak. It also confirms the lack of systematics in the data and our shape measurements, at least on such large scales. Note that we also found a similar lack of net alignments in Figure \ref{fig:vectors} for galaxy-galaxy lensing pairs with negligible shear, which comprise most of our background galaxies distributed throughout the entire field.

\subsection{Background galaxy redshift distributions}
\begin{figure*}
\centering
\includegraphics[width=\hsize]{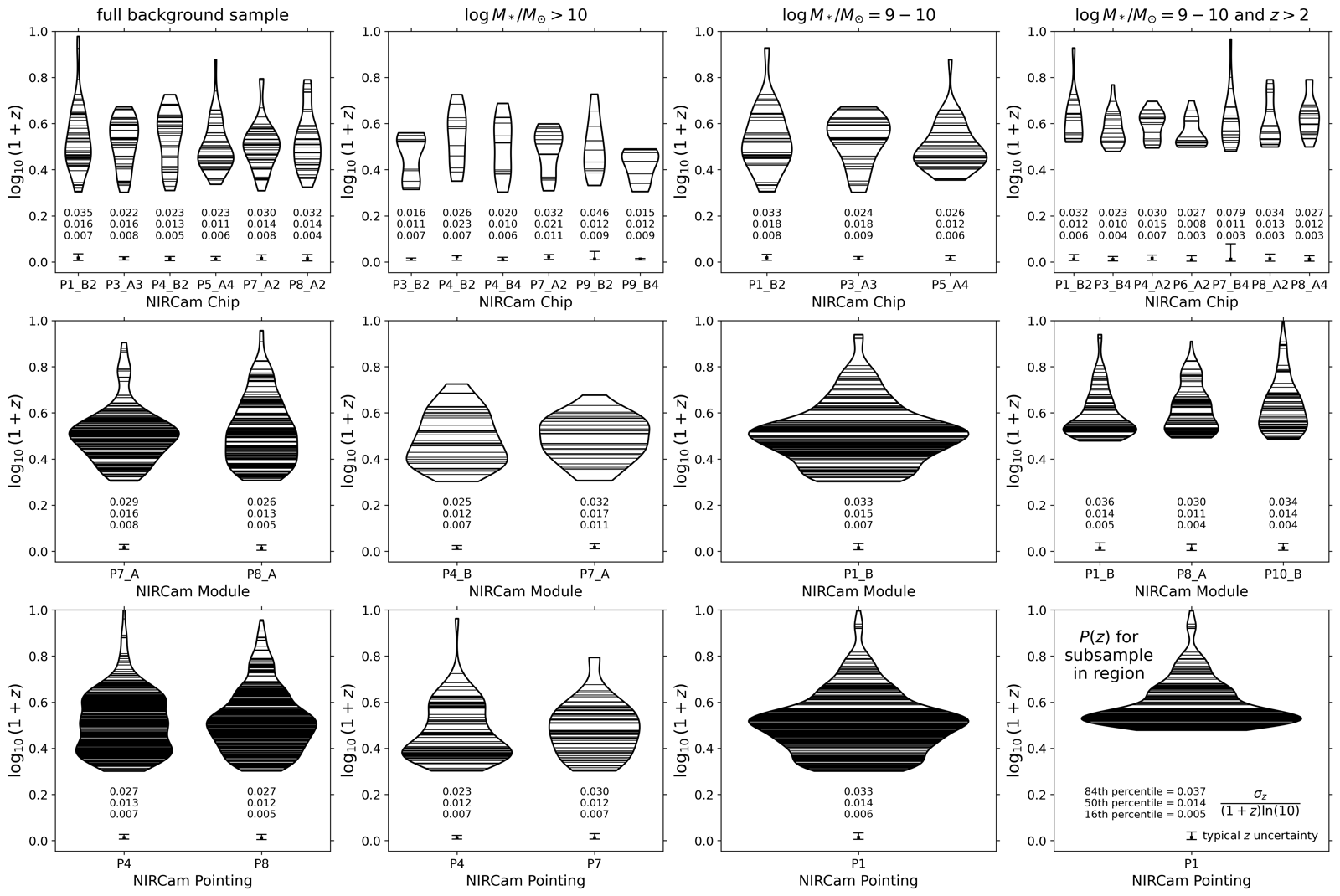}
\caption{Redshift distributions of background galaxies in the chips (top row), modules (middle row) and pointings (bottom row) where the different subsamples (from left to right) showed significant alignments ($p<0.05$ based on Figures \ref{fig:pvals}, \ref{fig:pvals22}, \ref{fig:pvals44}). The overall redshift distributions are visualized as violins and each individual galaxy redshift is denoted by a thin horizontal stripe. The thick overlapping stripes suggest that galaxies in many regions are clustered in redshift space. However, the typical fractional photometric redshift uncertainty is indicated below each violin and is generally too large to precisely identify overdensities.}
\label{fig:zdists}
\end{figure*}

Figure \ref{fig:zdists} shows the redshift distributions of background galaxies in the NIRCam chips, modules and pointings where we found significant alignments ($p<0.05$ based on Figures \ref{fig:pvals}, \ref{fig:pvals22}, \ref{fig:pvals44}). Every region has its own distinct overall redshift distribution as shown with the different shaped violins. In many regions, there are several galaxies that overlap in redshift which suggests clustering in redshift space. However, the uncertainties on $\log_{10}(1+z)$ are typically of order $\sigma_z/(1+z)/\ln10\sim0.01$ (see subsection \ref{sec:errorprop}). At $z\sim2$, this translates to $\sigma_z\sim0.07$ which implies an angular diameter distance uncertainty of $\gtrsim30$ proper Mpc. The maximum transverse (on-sky) separation probed by the overall CEERS footprint is only $\sim15$ physical Mpc at $z\sim2$ (individual chips, modules and pointings span only $\sim0.5$, $\sim1$ and $\sim2.5$ proper Mpc at $z\sim2$, respectively). Thus the redshift uncertainties are likely too large to associate galaxies along the line of sight and we have to wait for future spectroscopic follow-up \citep[as also originally argued by][]{pandya19}. Without precise redshift clustering constraints, the background galaxies in any region can be assumed to lie along widely separated large-scale structures and thus may not be expected to show strong intrinsic alignments. We thus next consider whether the alignments vary as a function of scale in the same way expected for cosmic shear.

\subsection{Shear correlations as a function of scale}\label{sec:shearvar}
\begin{figure*}
\centering
\includegraphics[width=\hsize]{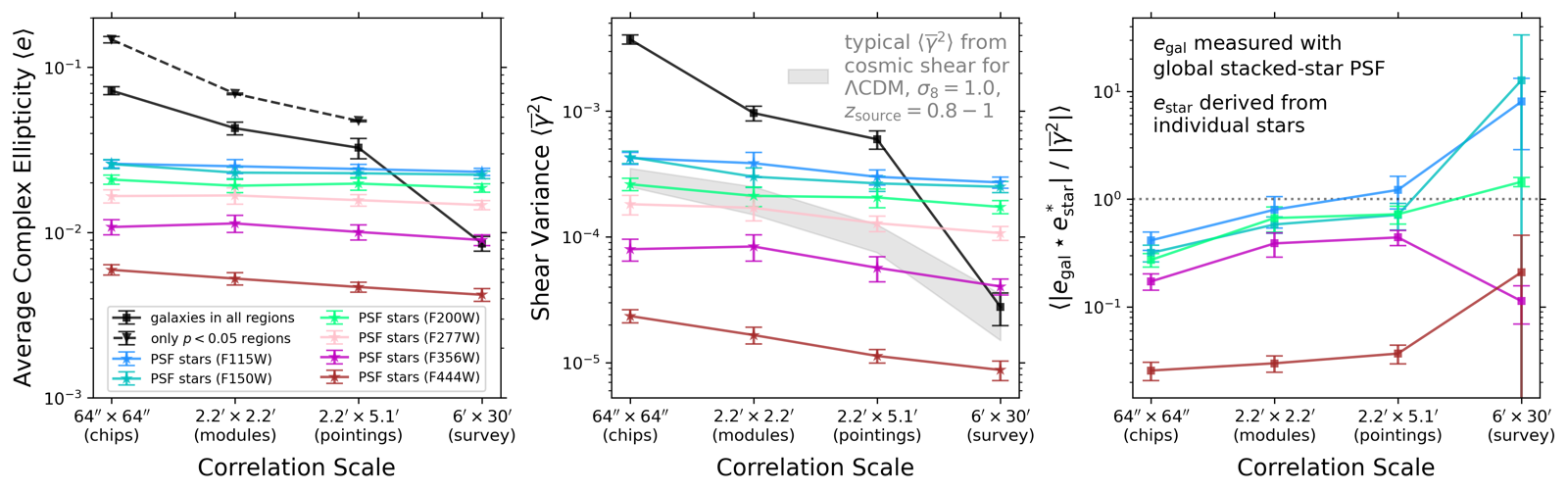}
\caption{Average complex ellipticity (left), shear variance (middle) and star-galaxy cross-correlation (right) on the scale of individual NIRCam chips, modules and pointings as well as the entire survey. All points denote averages and standard errors. Solid black lines show results for galaxies in all regions whereas dashed black lines are only based on regions where the null hypothesis of randomly oriented galaxies could be ruled out at $>95\%$ significance. Colored lines show the exact same calculations for PSF stars in each filter. Both $\langle e\rangle$ and $\langle \overline{\gamma}^2\rangle$ are above the floor set by stars on smaller scales, but drop into the systematics-dominated regime on larger scales. The shear variance in particular is several times larger than the typical value from cosmic shear (gray shaded region). The fractional contribution of the star-galaxy cross-correlation to the shear variance is sub-dominant on the chip scale but becomes significant on larger scales where the expected shear signal is lower. Note that we do not cross-correlate stars and galaxies in F277W because that filter is not used. This implies that our measured shear variance is an upper limit but even the large value on the chip scale only implies a root-mean-square shear of $\sqrt{\langle\overline{\gamma}^2\rangle}\sim5\%$ which is much smaller than the preferentially high ellipticity $e\sim0.4-0.7$ of early low-mass galaxies. Future work that corrects galaxy shapes for the local rather than global PSF may be able to measure the weak expected cosmic shear signal in ``blank'' JWST deep fields.}
\label{fig:shearvar}
\end{figure*}

Figure \ref{fig:shearvar} summarizes how the average complex ellipticity $\langle e \rangle$ and shear variance $\langle \overline{\gamma}^2\rangle$ depend on scale for both our background galaxies and PSF stars (see Appendix \ref{sec:psf} for more details about the latter). Recall that $\langle e \rangle$ gives the average orientation and ellipticity of all galaxies in some region whereas $\langle \overline{\gamma}^2\rangle$ quantifies the degree of correlation between pairwise shapes (as detailed in \ref{sec:complex}). In this context, since foreground stars should not be sheared (unlike high-redshift galaxies), they serve as a null test for PSF systematics by providing a ``floor'' on $\langle e \rangle$ and $\langle \overline{\gamma}^2\rangle$. This is complementary to our null test for statistical significance in previous subsections. 

Using our combined background galaxy sample without splitting by mass, redshift or alignment significance, we find that $\langle e \rangle$ drops from an average of $\sim8\%$ at the NIRCam chip scale to $\sim5\%$ for NIRCam modules and $\sim3\%$ for NIRCam pointings. If we restrict ourselves only to chips, modules and pointings where the null hypothesis of randomly oriented galaxies could be ruled out at $>95\%$ significance, then $\langle e\rangle$ is somewhat larger due to the alignments, as expected. The exact same calculation for the stars gives $\langle e\rangle$ that is almost constant with scale but drops monotonically from $\sim2-2.5\%$ for short-wavelength NIRCam filters (F115W, F150W, F200W) to $\sim0.5-1.5\%$ for long-wavelength NIRCam filters (F277W, F356W, F444W). This is several times lower than our $\langle e\rangle$ measured for galaxies at the chip and module scales, and to a lesser extent for pointings. This implies that the alignments we have detected on these smaller scales cannot be attributed to percent-level PSF systematics (see Appendix \ref{sec:psf} for more justification). In contrast, on the scale of the entire survey, the residual $\langle e \rangle\sim0.9\%$ of galaxies is comparable to or even exceeded by the $\langle e\rangle$ of PSF stars, which is consistent with the alignments vanishing on large scales as they should. 

The middle panel of Figure \ref{fig:shearvar} similarly shows that the shear variance drops from $\langle\overline{\gamma}^2\rangle\sim4\times10^{-3}$ at chip scales to $10^{-3}$ for modules and $6\times10^{-4}$ for pointings. The exact same calculation for the PSF stars leads to $\langle\overline{\gamma}^2\rangle$ that is almost constant with scale and drops nearly monotonically from $4\times10^{-4}$ for the bluest filter (F115W) to $\sim1-3\times10^{-5}$ for the reddest filter (F444W). Thus, pairs of galaxies have significant shear correlations above that expected from PSF systematics on the scale of chips, modules and pointings. In contrast, on the scale of the entire survey, $\langle \overline{\gamma}^2\rangle\sim3\times10^{-5}$ for galaxies, which is comparable to or even smaller than that of the stars. Thus the shear variance on the survey scale is consistent with being zero since it is so low as to be in the PSF systematics-dominated regime.

If our alignments were due to cosmic shear, we would expect a typical $\langle\overline{\gamma}^2\rangle\sim3\times10^{-4}$ on chip scales that decreases to $\sim3\times10^{-5}$ on the scale of the survey \citep[based on Figure 7 of the review by][]{refregier03}. Our measured shear variance is an order of magnitude larger at the chip scale and several times larger at the module and pointing scales, though consistent on the survey scale. The right panel of Figure \ref{fig:shearvar} shows that the contribution of the star-galaxy cross-correlation is of order $\sim10\%$ the shear variance on the chip scale, which means that star-galaxy alignments must be weaker than galaxy-galaxy alignments on these small scales. However, the fractional contribution of the star-galaxy cross-correlation becomes more significant on larger scales where the shear signal itself is lower and thus in the systematics-dominated regime. Thus our measured shear variance is an upper limit and it implies a root-mean-square shear $\sqrt{\langle\overline{\gamma}^2\rangle}\sim5\%$ which is still too small to explain the preferentially high ellipticity $e\sim0.4-0.7$ of early low-mass galaxies. The actual shear is likely even smaller and future studies that correct galaxy shapes for the local rather than global PSF may be able to overcome systematics and measure a much weaker cosmic shear signal expected in the data that we cannot.

Table \ref{tab:shear} gives alignment summary statistics for each of the four subsamples in all 111 regions (entire survey, 10 pointings, 20 modules, 80 chips).



\begin{table*}
\footnotesize 
\centering 
\begin{tabular}{|c|c|c|c|c|c|c|c|c|c|c|h|h|h|h|h|h|h|h|h|h|h|h|h|h|h|h|h|h|h|h|h|h|h|h|h|h|h|h|h|h|}\hline
Region & RA & Dec & $N^{\rm all}$ & $p^{\rm all}$ & $\langle e\rangle^{\rm all}$ & $\sigma_{\langle e \rangle}^{\rm all}$ & $\langle \phi\rangle^{\rm all}$ & $\sigma_{\langle \phi \rangle}^{\rm all}$ & $|\overline{\gamma}^2|$ & $\sigma_{|\overline{\gamma}^2|}$ & $|e_{\rm gal}\star e_{\rm star}^*|/|\overline{\gamma}^2|$ (F115W) & $\sigma_{|e_{\rm gal}\star e_{\rm star}^*|/|\overline{\gamma}^2|}$ (F115W) & $|e_{\rm gal}\star e_{\rm star}^*|/|\overline{\gamma}^2|$ (F150W) & $\sigma_{|e_{\rm gal}\star e_{\rm star}^*|/|\overline{\gamma}^2|}$ (F150W) & $|e_{\rm gal}\star e_{\rm star}^*|/|\overline{\gamma}^2|$ (F200W) & $\sigma_{|e_{\rm gal}\star e_{\rm star}^*|/|\overline{\gamma}^2|}$ (F200W) & $|e_{\rm gal}\star e_{\rm star}^*|/|\overline{\gamma}^2|$ (F277W) & $\sigma_{|e_{\rm gal}\star e_{\rm star}^*|/|\overline{\gamma}^2|}$ (F277W) & $|e_{\rm gal}\star e_{\rm star}^*|/|\overline{\gamma}^2|$ (F356W) & $\sigma_{|e_{\rm gal}\star e_{\rm star}^*|/|\overline{\gamma}^2|}$ (F356W) & $|e_{\rm gal}\star e_{\rm star}^*|/|\overline{\gamma}^2|$ (F444W) & $\sigma_{|e_{\rm gal}\star e_{\rm star}^*|/|\overline{\gamma}^2|}$ (F444W) & $N^{\rm high}$ & $p^{\rm high}$ & $\langle e\rangle^{\rm high}$ & $\sigma_{\langle e \rangle}^{\rm high}$ & $\langle \phi\rangle^{\rm high}$ & $\sigma_{\langle \phi \rangle}^{\rm high}$ & $N^{\rm low}$ & $p^{\rm low}$ & $\langle e\rangle^{\rm low}$ & $\sigma_{\langle e \rangle}^{\rm low}$ & $\langle \phi\rangle^{\rm low}$ & $\sigma_{\langle \phi \rangle}^{\rm low}$ & $N^{\rm early}$ & $p^{\rm early}$ & $\langle e\rangle^{\rm early}$ & $\sigma_{\langle e \rangle}^{\rm early}$ & $\langle \phi\rangle^{\rm early}$ & $\sigma_{\langle \phi \rangle}^{\rm early}$ \\\hline
Survey & nan & nan & 3848 & 0.351439 & 0.008596 & 0.000890 & -76.610408 & 2.856324 & 0.000028 & 0.000008 & 8.043397 & 5.160689 & 12.710213 & 20.741997 & 1.447363 & 0.145500 & nan & nan & 0.113110 & 0.043878 & 0.209793 & 0.250711 & 775 & 0.570323 & 0.012061 & 0.001081 & -7.905382 & 18.127117 & 3073 & 0.155804 & 0.013254 & 0.001070 & -80.664259 & 2.349163 & 1795 & 0.119113 & 0.018636 & 0.001606 & -80.197183 & 2.519887 \\
P1 & 214.984415 & 52.978435 & 452 & 0.066638 & 0.038976 & 0.003063 & -77.855550 & 2.039165 & 0.000576 & 0.000111 & 0.713375 & 0.127404 & 1.300976 & 0.511941 & 1.669166 & 0.447083 & nan & nan & 0.377712 & 0.134202 & 0.021795 & 0.025889 & 107 & 0.811519 & 0.020198 & 0.003630 & 35.733432 & 5.491714 & 345 & 0.018187 & 0.055699 & 0.003794 & -75.537673 & 1.768715 & 197 & 0.011739 & 0.078205 & 0.005788 & -78.680048 & 1.986709 \\
P1\_A & 214.958701 & 52.960792 & 199 & 0.720549 & 0.020503 & 0.004996 & -79.233521 & 6.838656 & 0.000463 & 0.000106 & 0.416439 & 0.068519 & 0.345581 & 0.062749 & 0.747991 & 0.244528 & nan & nan & 0.222084 & 0.161397 & 0.008705 & 0.013571 & 45 & 0.404944 & 0.061930 & 0.008421 & 11.149241 & 2.857883 & 154 & 0.329425 & 0.044365 & 0.006326 & -78.793416 & 3.526933 & 87 & 0.519956 & 0.045370 & 0.009522 & 86.437777 & 5.946934 \\
P1\_B & 215.010129 & 52.996078 & 253 & 0.053044 & 0.054112 & 0.003427 & -77.571261 & 1.917349 & 0.001001 & 0.000183 & 0.665421 & 0.184730 & 0.654481 & 0.156672 & 2.632380 & 0.960431 & nan & nan & 0.333909 & 0.046468 & 0.030292 & 0.028515 & 62 & 0.710748 & 0.034994 & 0.003156 & 76.496916 & 3.507433 & 191 & 0.044051 & 0.065597 & 0.004449 & -73.718189 & 1.987758 & 110 & 0.006606 & 0.110732 & 0.006505 & -74.277122 & 1.891703 \\
P1\_A1 & 214.940923 & 52.960792 & 36 & 0.065806 & 0.124969 & 0.009176 & 87.494525 & 2.213625 & 0.005178 & 0.001155 & nan & nan & 0.265532 & 0.080122 & nan & nan & nan & nan & nan & nan & nan & nan & 6 & 0.779766 & 0.095256 & 0.009270 & 86.258837 & 1.705034 & 30 & 0.091796 & 0.130565 & 0.011323 & 87.699022 & 2.510775 & 20 & 0.216999 & 0.137867 & 0.013514 & 75.118497 & 3.730038 \\
P1\_A2 & 214.958701 & 52.943014 & 49 & 0.813965 & 0.030689 & 0.009761 & -88.482837 & 83.918228 & 0.003371 & 0.000461 & nan & nan & nan & nan & nan & nan & nan & nan & nan & nan & nan & nan & 11 & 0.834455 & 0.043956 & 0.010735 & 32.955434 & 24.931351 & 38 & 0.831499 & 0.035383 & 0.010172 & -30.399747 & 53.268087 & 20 & 0.824904 & 0.052349 & 0.015531 & 62.230160 & 12.375861 \\
P1\_A3 & 214.958701 & 52.978570 & 53 & 0.463524 & 0.062890 & 0.013377 & -50.235209 & 5.957865 & 0.001892 & 0.000484 & 0.533952 & 0.141520 & 0.256174 & 0.247743 & nan & nan & nan & nan & nan & nan & nan & nan & 11 & 0.635722 & 0.101285 & 0.026594 & 15.370043 & 38.399175 & 42 & 0.275632 & 0.093018 & 0.016293 & -57.897640 & 4.010342 & 25 & 0.562865 & 0.076129 & 0.022612 & -58.539075 & 7.917899 \\
P1\_A4 & 214.976479 & 52.960792 & 59 & 0.745130 & 0.037742 & 0.007163 & 40.715718 & 6.197549 & 0.001791 & 0.000253 & nan & nan & nan & nan & nan & nan & nan & nan & 0.072824 & 0.109746 & nan & nan & 15 & 0.091334 & 0.165218 & 0.004144 & 14.388890 & 0.928322 & 44 & 0.728559 & 0.048309 & 0.009981 & 76.167184 & 6.054066 & 22 & 0.978232 & 0.018321 & 0.009432 & -85.628564 & 49.873509 \\\hline
\end{tabular}
\label{tab:shear}
\caption{Alignment summary statistics for all 111 regions (entire survey, 10 pointings, 20 modules and 80 chips). The central RA and Dec of each region are in decimal degrees. $N$ is the number of galaxies. We give the $p$-value from our null test, magnitude of the average complex ellipticity $\langle e\rangle$ and the associated phase angle $-90^{\circ}<\langle\phi\rangle<90^{\circ}$ defined east of north. All uncertainties are standard errors from our Monte Carlo method. We give these values for all four subsamples denoted by the superscript in the column name: all, high-mass, low-mass or ``early'' for $z>2$ low-mass galaxies. The shear variance and star-galaxy cross-correlation are given only for the combined sample. The full table is available for download from the journal.} 
\end{table*}


\section{Discussion}\label{sec:discussion}

\subsection{Evidence for cosmic shear in JWST-CEERS?}
We presented evidence for large-scale alignments of high-redshift galaxies in a ``blank'' JWST deep field far away from any obvious foreground cluster. Galaxies at $z>1$ with $\log_{10} M_*/M_{\odot}>9$ appear to be aligned on the scale of multiple NIRCam chips ($64''\times64''$), modules ($2.2'\times2.2'$) and pointings ($2.2'\times5.1'$) throughout the JWST-CEERS survey footprint. When averaging over background galaxies in these regions, the norm of the complex ellipticity vector is often $\langle e \rangle\gtrsim10\%$, which is at least an order of magnitude above the conventional weak lensing regime. Many of these chips, modules and pointings reveal circular alignment patterns that are suggestive of non-random effects and motivate follow-up $E/B$-mode polarization decomposition. In addition, our tangentially-aligned galaxy-galaxy lensing candidates, including one possible strong lens, tend to cluster in the southwest (Figure \ref{fig:onsky}). This clustering of galaxy-galaxy lensing candidates on scales of $\sim0.1$ deg as well as the detection of alignments in multiple regions on scales up to $2.2'\times5.1'$ equates to several Mpc at $z\sim0.75$ where we found an overdensity of massive foreground lens galaxies (Figure \ref{fig:pairs}). It is plausible that a cluster, proto-cluster or filament at $z\sim0.75$ may be responsible for the lensing in which case detailed follow-up modeling may help us learn something about the intrinsic 3D geometry and lensing mass profile of this overdensity \citep[e.g.,][]{kovner87,keeton97}.

On the other hand, we cannot rule out intrinsic alignments in the background galaxies \citep{kirk15,mandelbaum18,lamman24}. \citet{pandya19} showed that if early low-mass galaxies are intrinsically prolate, then they are expected to show strong intrinsic alignments with each other as they form along cosmic web filaments. Since our analysis is not done in bins of source redshift, we are presumably averaging over this and therefore our signal should not be driven by intrinsic alignments. \citet{pandya19} did not detect the expected intrinsic alignments and claimed that it was due to a lack of spectroscopically-confirmed background pairs in HST-CANDELS \citep{koekemoer11,grogin11}. However, they also did not use the complex ellipicity formalism employed here so the issue is worth revisiting in light of our results. The fact that we are detecting coherent alignments on the scale of entire NIRCam pointings ($2.2'\times5.1'$) and finding $\langle e\rangle\gtrsim10\%$ in multiple areas spread across the entire $\sim0.3$ deg extent of the survey (several Mpc at $z\sim0.75$) does point towards lensing. But because we lack spectroscopic redshifts for most of our background galaxies, we cannot precisely identify high-redshift pairs to robustly constrain intrinsic alignments (see Figure \ref{fig:zdists}).

Of course, there could be other systematics at play. It is unlikely that we are contaminated by satellite-central pair alignments \citep{hirata04} because moderate/strong galaxy-galaxy shear requires maximizing the angular diameter distance between the source and lens which implies very different photometric redshift distributions. Appendix \ref{sec:psf} shows that our PSF systematics are at the few percent-level which, while inadequate for traditional weak lensing, are an order of magnitude lower than our $\langle e\rangle\gtrsim10\%$ from averaging over the orientations of background galaxies on the chip-scale. Thus PSF uncertainties are unlikely to explain our alignments in terms of $\langle e\rangle$ but are prohibitively large for measuring the precise amplitude of the cosmic shear variance signal $\langle\overline{\gamma}^2\rangle$ (Figure \ref{fig:shearvar}). Astrometric misalignments could cause distortions that produce spurious shear signals \citep{finner23b} but the CEERS imaging we use is already tied to Gaia \citep{bagley23} and we have visually inspected all of our galaxies so this is unlikely. Future decomposition of the shear field into $E$ and $B$ mode polarizations can help constrain some of these systematics including signatures of intrinsic alignments. 

We note that JWST-CEERS is only the latest in a long line of multi-wavelength observations of EGS \citep{groth94} whose early HST-WFPC2 observations enabled the first detection of cosmic shear from space in this very same field \citep{rhodes00,rhodes01}. Those authors found mean shear values similar to ours of $\sim10\%$ using $\sim50$ background galaxies in many $1.27\times1.27$ arcmin$^2$ regions along the strip \citep[see Figure 1 of][]{rhodes01}. After carefully subtracting off systematics, they estimated the shear variance to be $\langle \overline{\gamma}^2\rangle\sim0.018^2\sim3\times10^{-4}$ on the scale of a single HST-WFPC2 chip ($1.27'\times1.27'$). This is an order of magnitude lower than our $\langle \overline{\gamma}^2\rangle\sim4\times10^{-3}$ on the slightly smaller scale of a NIRCam chip ($64''\times64''$). The discrepancy is likely due to PSF systematics on our end (Figure \ref{fig:shearvar}) which is why we consider our measurement of shear variance to be an upper limit rather than a detection. However, differences in source selection and rest-frame optical versus rest-frame ultraviolet morphology (HST-WFPC2 probed the latter) may also play a role. Future efforts to correct for the complicated spatially-varying PSF of JWST should enable precise measurements of the amplitude of the cosmic shear signal in multiple ``blank'' JWST deep fields. These can then be averaged to mitigate cosmic variance and help constrain the cosmological parameters $\sigma_8$ and $\Omega_{\rm M}$ \citep[as has been done from the ground and with HST;][]{kaiser00,bacon00,wittman00,rhodes01,pirzkal01,miralles02,hammerle02,rhodes04,schrabback07}. Indeed, the shallower but $\sim10\times$ larger COSMOS-Web survey is designed in part to enable this kind of measurement with JWST \citep{casey23}, building off of the long history of cosmic shear studies in the HST-COSMOS field \citep{rhodes07,leauthaud07,massey07b}.

EGS has benefited greatly from dedicated panchromatic imaging and spectroscopic redshift surveys such as AEGIS, DEEP2 and DEEP3 \citep{davis03,faber03,davis07,cooper07,cooper12,newman13}. This wealth of data led previous groups to identify a number of groups and clusters in EGS between $z\sim0.4-1.4$, some of which are located at $z\sim0.75$ consistent with our claimed overdensity of lenses there \citep[e.g., see Figure 6 of][]{gerke12}. The galaxy color-magnitude bimodality and color-density relations are also observed for EGS galaxies at $z\sim0.75$ which suggests the presence of rich groups \citep[][and see also later work by \citealt{pandya17} using HST-CANDELS]{faber07,cooper07}. The abundance and clustering of X-ray-selected active galactic nuclei also point towards a number of massive groups in EGS at $z\sim0.7-1.4$ \citep{coil09}. We note that our most massive lens has a spectroscopic redshift of $z\sim0.78$ and although its $\sigma_*\sim580$ km/s is very high compared to that of the most massive local galaxies \citep{ma14,pandya17b,greene19}, it seems consistent with the tail of high $\sigma_*$ LRGs out to $z\sim0.7$ \citep{thomas13}. This massive lens is also associated with our only strong lensing candidate (Figure \ref{fig:sources}). Spectroscopic constraints on the stellar kinematics and environment of this massive galaxy combined with secure redshifts for background sources would enable more sophisticated source-lens reconstruction. Finally, \citet{moustakas07} conducted a visual search for strong lenses in EGS using $\sim650$ arcmin$^2$ of HST-ACS imaging. They found three (an Einstein Cross, one with two pairs of arcs, and one with a single pair of arcs) but unfortunately those are all outside of the JWST-CEERS footprint which covers only $\sim100$ arcmin$^2$ of EGS. However, they also found that each of these lenses required $\sim10\%$ external shear which supports our claim that there may be significant cosmic shear throughout CEERS leading to $\langle e\rangle\sim0.1$.

Our study suggests that ``blank'' JWST deep fields may provide complementary constraints on lensing and intrinsic alignments beyond what is traditionally available from massive cluster surveys, provided that the systematics are better understood. For example, the high resolution of JWST imaging may allow one to study higher order distortions due to lensing such as flexion (which, to quote \citealt{schneider08}, happens when ``weak lensing goes bananas''). The prospect of searching for overdensities in ``blank'' fields is tantalizing because one of the strengths of lensing is that it is sensitive to mass concentrations even if it is ``dark.'' JWST will also enable the selection of lower mass, higher redshift sources for weak lensing which, while improving number statistics, may introduce new biases due to the unknown intrinsic shapes and intrinsic alignments of the early galaxy population. On the galaxy-galaxy lensing side, \citet{holloway23} used mock data to show that JWST deep fields should expect to have $\sim25-65$ strongly lensed systems. One new Einstein ring has already been detected in COSMOS-Web \citep{vandokkum24,mercier23}. We verified that if we had a source-lens pair in CEERS with parameters similar to the COSMOS-Web Einstein ring, we would have identified it. Assuming rings/arcs/clumps were optimally deblended, this implies there are no Einstein rings in CEERS.\footnote{We repeated our search allowing for massive lenses at any redshift and background sources down to $\log_{10} M_*/M_{\odot}=8$ but did not find any convincing strong lensing candidates in JWST-CEERS.} Given the small sizes and faint magnitudes of many of our tangentially-aligned source-lens pairs (Figure \ref{fig:demographics}), follow-up observations may be challenging but required to enable confirmation and more detailed lensing modeling.

Looking to the near future, the Roman Space Telescope, with its $\sim0.28$ deg$^2$ field of view and $\sim0.1$ arcsec resolution, has the potential to dramatically transform our understanding of lensing, intrinsic alignments and origin of low-mass elongated galaxies. A single pointing with Roman would image a region larger than the entirety of the existing HST ACS/WFC3 footprint in EGS, which required tens of tiles over only $\sim0.05-0.2$ deg$^2$ \citep{stefanon17}. A ``Roman Deep Field'' \citep{koekemoer19,drakos22} that can reach $\sim27$ AB mag at $\sim1.5\mu$m would detect and resolve elongated $\log M_*/M_{\odot}>9$ galaxies out to $z\sim3$ \citep{pandya24}. This would uniquely enable the calculation of galaxy-shear and shear-shear correlation functions over much larger areas than possible with HST or JWST. Multiple such Roman deep fields spread across the sky would help mitigate the cosmic variance that has plagued the field for decades \citep{madau14} and allow us to test whether the alignments are indeed due to lensing since the foreground mass distribution should vary across the sky. We caution that the current plan for the Roman High Latitude Wide Area Survey over $\sim2000$ deg$^2$ is only designed to reach $26.5$ AB mag for point sources \citep{troxel21,wang22,montes23}. A rule of thumb is that the completeness limit for extended high-redshift galaxies is $\sim2$ magnitudes fainter than the point source limit \citep{pandya24} so a deeper imaging component is necessary for our science case. Note that Euclid, with its sensitivity and $\sim0.2$ arcsec resolution, is not expected to be complete to such low-mass galaxies beyond $z\sim0.5$ even in its Deep Surveys spanning $\sim50$ deg$^2$ \citep{euclid22}.

\subsection{Why are early galaxies preferentially elongated?}
Despite the evidence for galaxy-galaxy lensing and large-scale alignments with $\langle e \rangle\sim10\%$, the fact that most of our background sources have much higher ellipticities of $e\sim0.4-0.7$ (top panel of Figure \ref{fig:shear}) implies that gravitational lensing cannot be primarily responsible for their elongation. Thus, we are led back to the original question that motivated this study: why are early low-mass galaxies preferentially elongated as observed by HST \citep{cowie95,elmegreen05,ravindranath06,vanderwel14,zhang19} and JWST \citep{pandya24}?

The remaining explanations for this puzzle range from the mundane to the exotic. There may still be a surface brightness detection bias against rounder, face-on disks \citep{loeb24}. \citet{pandya24} claimed that JWST-CEERS is sufficiently complete but deeper surveys have not yet been checked. Relatedly, we may be seeing elongated star-forming proto-bars at the centers of extremely low surface brightness disks. Alternatively, the elongated shapes reflect tidal debris from ongoing mergers. Finally, early low-mass galaxies may not start out as disks as commonly assumed but rather as triaxial/prolate structures that reflect the cosmic web filaments in which they form \citep{ceverino15, tomassetti16, pandya19}. \citet{pandya24} emphasized that the stellar masses of early elongated galaxies at $z\sim2-3$ are consistent with them being Milky Way progenitors, broadly defined, implying that our own Galaxy may have been elongated in its past. There are hints of this early elongation from Galactic archaeology \citep{naidu21}, and the recent discovery of a strongly lensed, elongated, clumpy Milky Way-like progenitor at $z\sim8.3$ evokes connections to chain galaxies \citep[][see also \citealt{zonoozi19}]{mowla24}.

It is imperative that we try to rule out the prolate/triaxial interpretation because so much of our current understanding of galaxy formation depends on them starting out as disks due to tidal torques and angular momentum conservation \citep{mo98}. Although prolate galaxies can form in $\Lambda$CDM, this has only been seen in some cosmological zoom-in simulations \citep{ceverino15,tomassetti16} and there is no consensus on subgrid models for uncertain baryonic physics \cite[e.g.,][]{somervilledave15}. What we really want to know is the large-scale clustering, orientation, dynamical stability and cosmological evolution of prolate galaxies but large-volume simulations typically do not have the resolution required to robustly predict the internal dynamics and therefore 3D shapes of early low-mass galaxies \citep[e.g.,][]{pillepich19}. All of this invites an exploration of more exotic scenarios such as alternatives to cold dark matter. For example, it has already been shown that warm and axionic (fuzzy) dark matter naturally leads to more elongated halos and therefore possibly more elongated galaxies at early times as observed \citep{mocz20,dome23,pozo24}. 

\citet{pandya24} claimed that stellar kinematics will be required to definitively distinguish between these interpretations. However, our analysis suggests a new test: if galaxy-shear and shear-shear correlation functions can be measured over large deep fields for $\log_{10}(M_*/M_{\odot})\sim9$ galaxies out to $z\sim2$, then we will be able to constrain whether these early elongated galaxies show strong intrinsic alignments and trace cosmic web filaments at high-redshift as predicted by \citet{pandya19}. The Euclid Deep Surveys are not expected to detect and resolve such low-mass galaxies beyond $z\sim0.5$ \citep{euclid22}, but the Roman Space Telescope may be able to do so. We therefore strongly recommend that a Roman Deep Field \citep{koekemoer19} be planned to enable this potentially transformative science: are early elongated galaxies reliable probes of lensing, do they represent an unrecognized source of significant additive bias, and can they tell us anything fundamentally new about cosmology? 

\section{Summary}\label{sec:summary}
We used data from JWST-CEERS \citep[][McGrath et al., in prep.]{finkelstein23} to explore an idea not previously considered in the literature: are early low-mass galaxies preferentially elongated because of gravitational lensing? First, we investigated galaxy-galaxy lensing for which we selected 76 massive foreground lens galaxies at $z<1$ and 3848 background sources at $z>1$ with $\log_{10} M_*/M_{\odot}>9$. Above this mass limit, CEERS has, on average, $\sim50-100$ background galaxies per arcmin$^2$ over its $\sim100$ arcmin$^2$ footprint, making it a good testbed for lensing studies in ``blank'' JWST deep fields. For each source, we assigned the nearest on-sky massive lens and estimated the shear $\gamma$ assuming a singular isothermal sphere lens model as well as the tangential alignment angle. Second, we looked for alignments on larger scales by averaging over the orientations of background galaxies on the scale of individual NIRcam chips ($64''\times64''$), modules ($2.2'\times2.2'$) and pointings ($2.2'\times5.1'$) across the survey. Our results are summarized as follows.

\begin{enumerate}

\item Galaxy-galaxy lensing cannot be the primary driver for the elongated appearance of most early low-mass galaxies because our typical predicted shear is of order only $\gamma\approx1\%$. This is much lower than the typical ellipticity $e\sim0.4-0.7$ of high-redshift galaxies (Figure \ref{fig:shear}). However, there is a broad tail towards larger shears in excess of $10\%$. We use non-parametric quantile regression with Bayesian Additive Regression Trees to show that there is a hint of an excess of tangentially-aligned source-lens pairs with such high shear (Figure \ref{fig:alignments}). We identify 17 tangentially-aligned lensing candidates, only one of which may truly be in the strong lensing regime (Figure \ref{fig:sources} and Table \ref{tab:pairs}). These galaxy-galaxy lensing candidates are clustered in the southwest region of the survey (Figure \ref{fig:onsky}) and show evidence that their orientations are correlated (Figure \ref{fig:vectors}).

\item We find evidence for coherent large-scale alignments when averaging over the orientations of background galaxies in multiple NIRCam chips ($64''\times64''$), modules ($2.2'\times2.2'$) and pointings ($2.2'\times5.1'$). We can rule out the null hypothesis of random orientations at the $\gtrsim99\%$ confidence level for multiple individual regions (Figures \ref{fig:pvals}, \ref{fig:pvals22} and \ref{fig:pvals44}, and Table \ref{tab:shear}). The number of such regions is small on the scale of the survey so formally the detections may be due to random chance, but the imaging clearly reveals non-random alignment patterns (Figures \ref{fig:fields}, \ref{fig:fields22} and \ref{fig:fields44}). On the chip scale, the average complex ellipticity $\langle e \rangle\gtrsim10\%$ and shear variance $\langle\overline{\gamma}^2\rangle\sim4\times10^{-3}$ are an order of magnitude above the conventional weak lensing regime (Figure \ref{fig:shearvar}). We consider these to be upper limits due to PSF uncertainties (Appendix \ref{sec:psf}), intrinsic alignments, cosmic variance and other lensing systematics. The maximum implied ``cosmic shear'' is still only $\sqrt{\langle\overline{\gamma}^2\rangle}\sim5\%$, much lower than the ellipticity of early low-mass galaxies.

\item We speculate that the large-scale alignments may be due to lensing from a foreground protocluster or filament spanning several Mpc at $z\sim0.75$ where we found an overabundance of massive lens galaxies (Figure \ref{fig:pairs}). Future spectroscopic surveys of the regions with net alignments are needed to confirm whether galaxies are clustered in redshift space and intrinsically aligned along the same high-redshift large-scale structures (Figure \ref{fig:zdists}). We also strongly recommend the planning of a deep field with the forthcoming Roman Space Telescope whose sensitivity, resolution and $0.28$ deg$^2$ field of view will uniquely enable the calculation of galaxy-shear and shear-shear correlation functions for elongated galaxies with $\log_{10}(M_*/M_{\odot})\sim9$ out to $z\sim2$. This will provide a novel way to constrain the intrinsic shapes of early elongated galaxies which, if they are forming along filaments, are expected to show strong intrinsic alignments \citep{pandya19}. This would also help calibrate any bias from this population for weak lensing analyses.

\end{enumerate}

\begin{acknowledgements}
This paper is the result of an initial scientific disagreement between VP and AL about whether ``galaxies are going bananas'' \citep{pandya24,loeb24}. Thanks to an invitation from AL, VP gave a colloquium at Harvard's ITC and that visit inspired conversations about the enigmatic nature of early elongated galaxies and novel ways to follow them up. We thank Haowen Zhang, Raphael Gavazzi and Greg Bryan for helpful suggestions. VP thanks Jo Bovy for writing the interactive online ``Dynamics and Astrophysics of Galaxies'' textbook at \url{https://galaxiesbook.org/}. VP also thanks Joel Primack, Avishai Dekel, Sandy Faber and David Koo for getting him interested in early elongated galaxies. We thank the anonymous referee for helpful suggestions. Support for VP was provided by NASA through: (1) the NASA Hubble Fellowship grant HST-HF2-51489 awarded by the Space Telescope Science Institute, which is operated by the Association of Universities for Research in Astronomy, Inc., for NASA, under contract NAS5-26555, and (2) NSF Astronomy \& Astrophysics Grant \#2307419.

\end{acknowledgements}

\bibliographystyle{aasjournal}
\bibliography{references}

\appendix 

\section{Investigation of Possible PSF Systematics}\label{sec:psf}

In this Appendix we show that PSF systematics are unlikely to explain our results, which are at least an order of magnitude above the conventional weak lensing regime. Our galaxy shape measurements were done with \texttt{Galfit} \citep[][McGrath et al., in prep.]{peng02} assuming a single-component S\'ersic model convolved with an empirical NIRCam PSF. The PSFs are ``global'' averages based on stacking all stars that fall within the CEERS footprint in each filter as described in section 3.2 of \citet{finkelstein23}. It is customary to summarize a PSF by only its FWHM, enclosed flux within some aperture and corresponding limiting magnitude for a point source, but for our purposes, we also need to estimate its complex ellipticity. Furthermore, in addition to the global stacked PSF, we also need to consider spatial variations by fitting individual stars throughout the survey. We use a similar list of stars in each filter throughout the CEERS footprint as \citet{finkelstein23}. These stars were selected via custom cuts on half-light radius and magnitude. For each star in each filter, we made a $101\times101$ pixel cutout and discarded objects with obvious issues (neighbor, detector edge, etc.). This leaves us with $111-134$ stars per filter, i.e., $\sim1$ star arcmin$^{-2}$.

Following standard practice in weak lensing studies \citep[e.g.,][]{kaiser95,schneider05,heymans05,heymans06,massey07,jee07,bridle10,mandelbaum18,finner23,finner23b,cha24}, we measure the ``quadrupole moments'' of every individual star as well as the global stacked PSF image from \citet{finkelstein23}. These quadrupole moments characterize the variance and covariance of the light intensity $I(x,y)$ along the $x$ and $y$ directions. To suppress noise and ensure that only the ``core'' of the PSF contributes to the measurements, we adopt the common choice of a circular Gaussian with $\sigma=3$ pixels as a weight function $\mathcal{W}(x,y)$ (our results are insensitive to reasonable variations in $\sigma$). This lets us compute the first moment of the 2D light distribution which is the centroid ($\bar{x}$,$\bar{y}$):

\begin{equation}
\bar{x} = \frac{\sum x \mathcal{W}(x,y) I(x,y)}{\sum \mathcal{W}(x,y)I(x,y)}\;, 
\end{equation}

\begin{equation}
\bar{y} = \frac{\sum y \mathcal{W}(x,y)I(x,y)}{\sum \mathcal{W}(x,y)I(x,y)}\;.
\end{equation}

Then we compute the second moments using the pixel indices shifted relative to the centroid: 

\begin{equation}
Q_{xx} = \frac{\sum (x-\bar{x})^2 \mathcal{W}(x,y) I(x,y)}{\sum \mathcal{W}(x,y)I(x,y)}\;, 
\end{equation}

\begin{equation}
Q_{yy} = \frac{\sum (y-\bar{y})^2 \mathcal{W}(x,y) I(x,y)}{\sum \mathcal{W}(x,y)I(x,y)}\;, 
\end{equation}

\begin{equation}
Q_{xy} = \frac{\sum (x-\bar{x})(y-\bar{y}) \mathcal{W}(x,y) I(x,y)}{\sum \mathcal{W}(x,y)I(x,y)}\;.
\end{equation}

Finally, this lets us assign to every star (i.e., PSF) a complex ellipticity
\begin{equation}
e_1 + i e_2 = \frac{Q_{xx}-Q_{yy}+2iQ_{xy}}{Q_{xx}+Q_{yy}+2\sqrt{Q_{xx}Q_{yy}-Q_{xy}^2}} \;. 
\end{equation}
This is analogous to the complex ellipticity that we assigned to every galaxy in Equation \ref{eqn:complex} except here we are using quadrupole moments. The complex magnitude $e=\sqrt{e_1^2+e_2^2}$ gives us the PSF ellipticity and the complex phase $\phi=0.5\rm{arctan2}(e_2,e_1)$ gives us the orientation. The size is defined as $R=\sqrt{Q_{xx}+Q_{yy}}$.

\begin{figure*}
\centering
\includegraphics[width=\hsize]{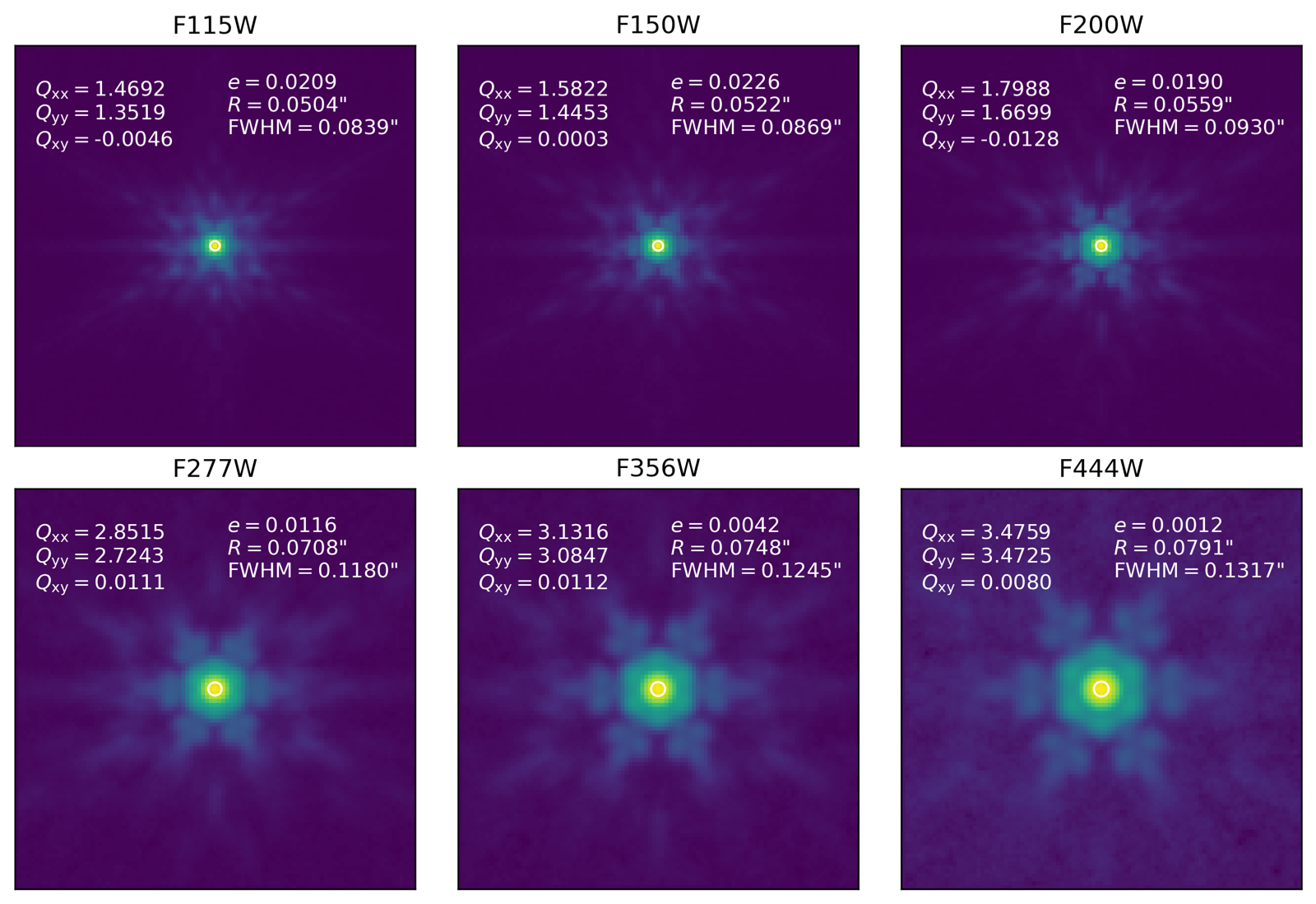}
\caption{The global empirical PSF from stacking stars in each filter from \citet{finkelstein23} along with our quadrupole moment measurements. The white ellipse illustrates the ellipticity, orientation and size of the quadrupole moments that characterize the core. Naturally, the ellipticities are small with $e\lesssim0.02$ so these ellipses are nearly circular. All images are on a log scale to emphasize JWST's unique diffraction features outside the ``core'' of the PSF. On a linear scale, the size $R$ derived from quadrupole moments adequately reflects the extent of the core. Results for individual stars are similar.}
\label{fig:psf_fits}
\end{figure*}

As an example, Figure \ref{fig:psf_fits} shows the global empirical PSF in each filter from \citet{finkelstein23} along with our quadrupole moments. As expected, the ellipticities are very small with $e\lesssim0.02$. The size $R$ translated to a Gaussian FWHM, as is common practice, yields similar values as \citet{finkelstein23} with the resolution decreasing from $\sim0.08"$ in F115W to $\sim0.13"$ in F444W. Results for individual stars are similar. Before proceeding, we caution that although this kind of quadrupole analysis is standard in weak lensing, the NIRCam PSF is clearly very complicated (i.e., not a simple Gaussian or Airy disk). The features outside the ``core'' of the PSF are due to JWST's hexagonally segmented primary mirror and the ``tripod'' structure that supports the secondary mirror \citep{rigby23}. Other recent JWST studies \citep{finner23,cha24,zhuang24} also use quadrupole moments or fit an elliptical 2D Gaussian to the ``core'' of the PSF and arrive at similar conclusions as us, but we recommend more detailed follow-up analysis.

\begin{figure*}
\centering
\includegraphics[width=\hsize]{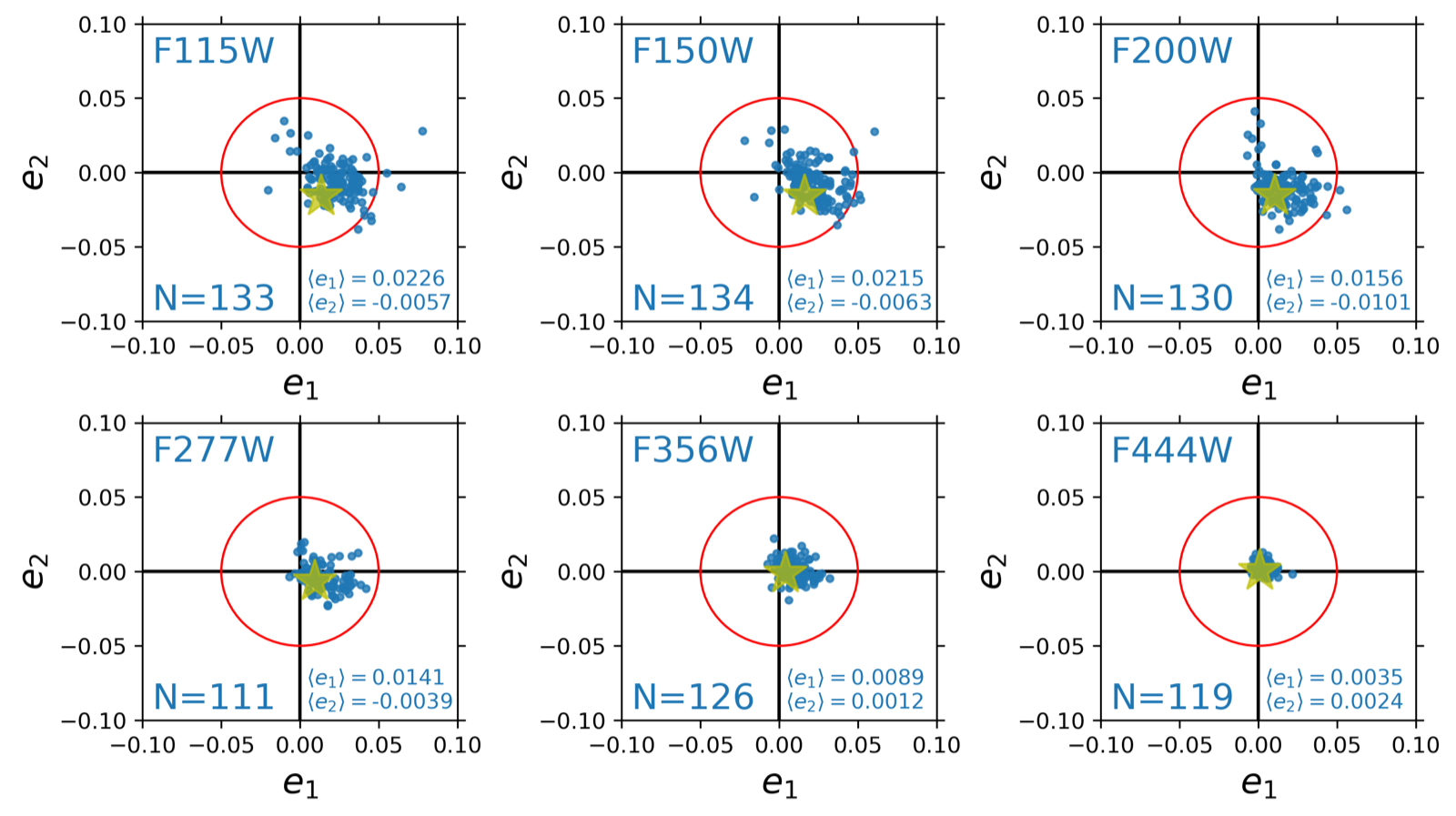}
\caption{Distribution of our individual stars (blue points) and global stacked PSF (yellow star) in the complex $(e_1,e_2)$ plane. This is in the standard ``north up'' frame. Nearly all stars in every filter have $e<0.05$, i.e., the PSF is very round. It is slightly more elliptical in the short-wavelength filters (F115W, F150W, F200W) with $\langle e_1\rangle\sim0.02$ and $\langle e_2 \rangle\sim-0.01$ and there is clearly a bias towards the lower-right quadrant. The long-wavelength filters (F277W, F356W, F444W) have even rounder PSFs as expected. These percent-level PSF systematics cannot explain our much larger $\langle e \rangle\gtrsim10\%$ when averaging over background galaxies (Figure \ref{fig:shearvar}) but must be corrected in any future analysis of the conventional weak lensing regime.}
\label{fig:psf_e1e2}
\end{figure*}

Figure \ref{fig:psf_e1e2} shows the distribution of our individual stars and the global PSF in the complex $(e_1,e_2)$ plane. Nearly all stars in every filter have $e<0.05$, i.e., the PSF is very round as expected. The short-wavelength NIRCam filters (F115W, F150W, F200W) have average components $\langle e_1\rangle\sim0.02$ and $\langle e_2 \rangle\sim-0.01$ and there is clearly a bias towards the lower-right quadrant. The long-wavelength NIRCam filters (F277W, F356W, F444W) have substantially rounder PSFs as evidenced by their smaller $\langle e_1\rangle\lesssim0.01$ and $\langle e_2 \rangle\sim\pm0.001$. These percent-level PSF systematics are unlikely to explain our much larger observed $\langle e \rangle\gtrsim0.1$ when averaging over galaxies in individual NIRCam chips, modules and pointings (Figure \ref{fig:shearvar}). At most, they would lead to an additive bias in $\langle e\rangle$ of a few percent, not tens of percent.

\begin{figure*}
\centering
\includegraphics[width=\hsize]{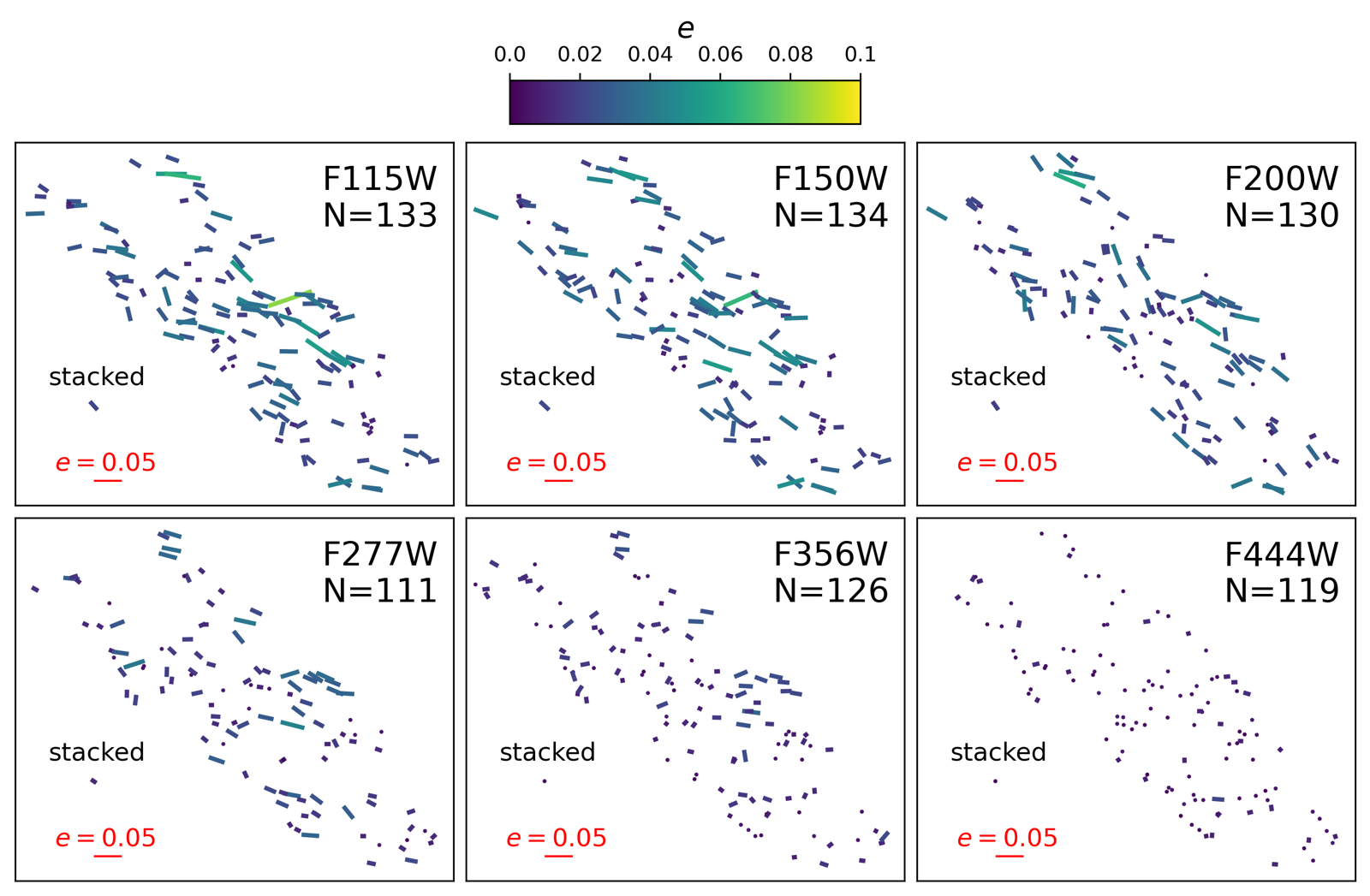}
\caption{Ellipticity and orientation of the PSF at the positions of individual stars across the entire CEERS footprint. For convenience, the ellipticity is indicated both with the line length (scale bar in lower left) and with the colorbar. The ellipticity and orientation of the stacked PSF is shown in the left center of each panel. This is in the standard ``north up'' frame. The ellipticities are very small and appear random in the long-wavelength filters. In contrast, for the short-wavelength filters, there are systematic PSF alignments towards $45^{\circ}$ east of north but the small ellipticities mean that these will cause an additive bias of at most a few percent which is an order of magnitude lower than our $\langle e\rangle\gtrsim10\%$ from averaging background galaxies (Figure \ref{fig:shearvar}). However, future studies in the conventional weak lensing regime must correct for this bias.}
\label{fig:psf_maps}
\end{figure*}

Figure \ref{fig:psf_maps} shows a map of the PSF ellipticity and orientation across the entire CEERS footprint using our individual stars. Most stars have very small ellipticity but there are a handful with $e\sim0.05-0.08$. The cores of these stars do look slightly elliptical and it is not clear whether they are outliers since other stars near them show smaller ellipticities. The orientations of PSF stars in the long-wavelength filters appear random due to their very small ellipticities. In contrast, for the short-wavelength filters, there is an overall pattern of many stars pointing towards the upper-left of each panel ($45^{\circ}$ east of north). However, again, since the individual PSF ellipticities are very small, their additive bias due to these alignments cannot contribute more than a few percent to our much larger observed $\langle e\rangle>10\%$ from averaging over background galaxies. 

\begin{figure}
\centering
\includegraphics[width=\hsize]{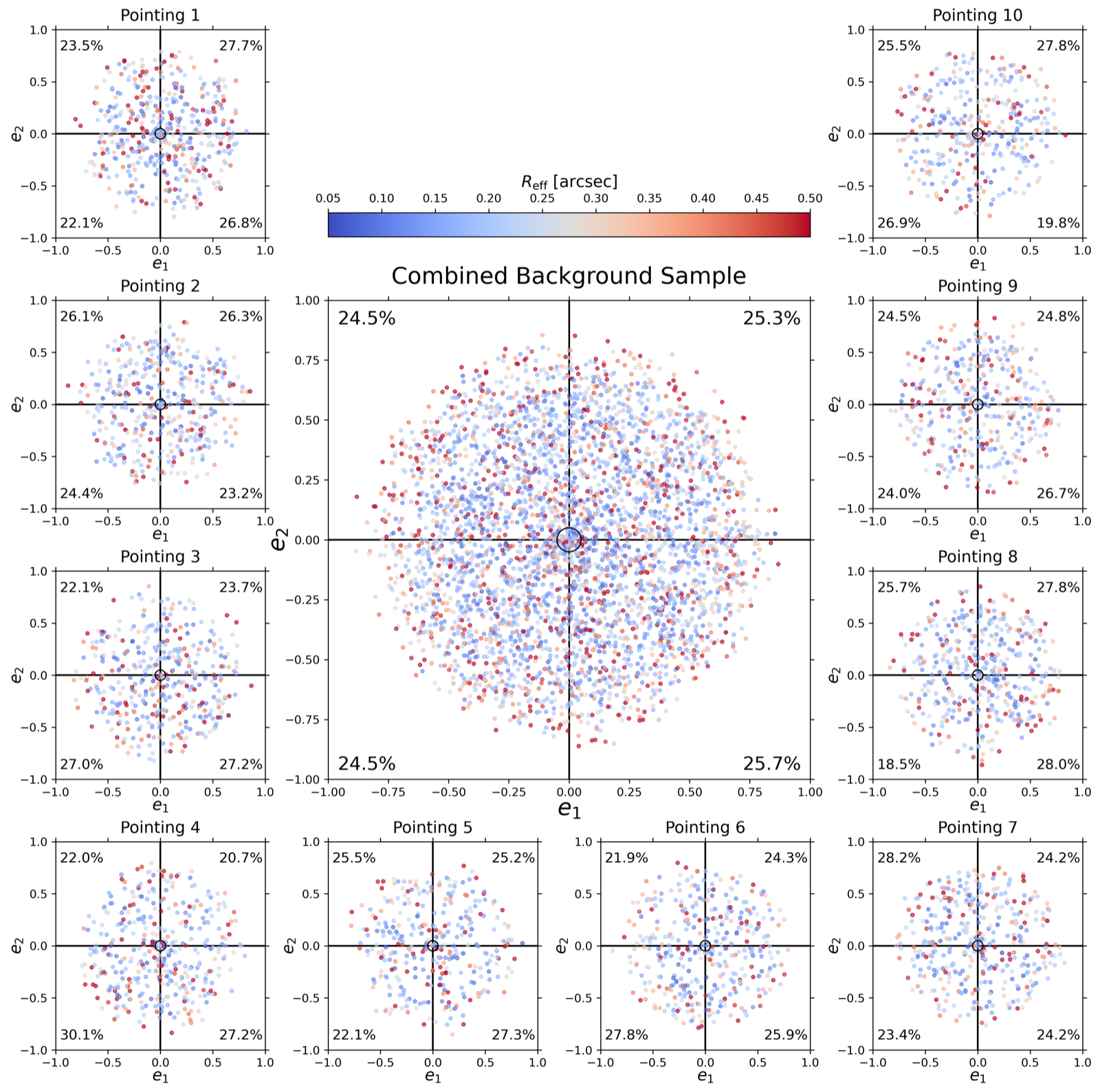}
\caption{Distribution of our background galaxy sample in the complex $(e_1,e_2)$ plane (standard ``north-up'' frame). The large center panel shows our combined background sample whereas the smaller panels show the distributions in individual pointings. In all cases, $\sim2\%$ of our galaxies have $e<0.05$ which is the PSF-dominated regime (within the central black circle). Our galaxies also tend to be roughly uniformly distributed between the four quadrants. This is particularly true for our smaller-size galaxies (bluer colored points) which would be most susceptible to PSF effects. The fact that most of our galaxies have intrinsic ellipticities that are much larger than that of our stars implies that the small bias towards the lower-right quadrant in Figure \ref{fig:psf_e1e2} is unlikely to dictate the orientations and hence alignments of our galaxies.}
\label{fig:psf_galaxies}
\end{figure}

Finally, Figure \ref{fig:psf_galaxies} shows that our actual background galaxy sample spans a much larger range in this same complex $(e_1,e_2)$ plane. Only $\sim2\%$ of our galaxies have $e<0.05$ similar to our PSF stars and they are roughly uniformly split between the different quadrants (phase angles). Note that the majority of our galaxies ($2972/3848\sim77\%$) are assigned to a short-wavelength NIRCam filter (F115W, F150W, F200W) to obtain the rest-frame morphology corresponding to their redshift. Moreover, our smallest size galaxies (which would be preferentially subject to PSF systematics) are also roughly uniformly spread out in this plane, i.e., not biased towards any one quadrant. This is true regardless of whether we look at the combined sample or only galaxies in individual pointings. Given that the PSF tends to round out objects, the fact that our galaxies have intrinsic ellipticities that are much larger than that of the stars implies that correcting for the small bias towards the lower-right quadrant in Figure \ref{fig:psf_e1e2} would not cause most of our galaxies to change quadrants. Thus we do not believe that the percent-level PSF systematics remaining in the data can explain our observed alignments in regions with $\langle e\rangle\gtrsim10\%$. However, measuring the precise amplitude of the shear variance $\langle\overline{\gamma}^2\rangle$ will require correcting galaxy shapes for this spatially-varying PSF because the expected signal is weak, especially on larger scales (Figure \ref{fig:shearvar}). It is unclear whether CEERS itself has enough stars to interpolate the PSF shape to the position of every galaxy so a more sophisticated strategy may be required. We defer that to a future more detailed weak lensing analysis of ``blank'' JWST deep fields.

\end{document}